\documentclass[final,3p,times,compress]{elsarticle}

\usepackage{amssymb}
\usepackage{lipsum}
\usepackage{color}
\usepackage{amsthm}
\usepackage{braket}
\usepackage{mathrsfs}
\usepackage{nccmath}
\usepackage{bm}
\usepackage{multirow}
\usepackage[colorlinks=true]{hyperref}
\usepackage{comment}

\journal{Annals of Physics}

\begin{document}

\begin{frontmatter}



\title{Path-Integral Formulation of Bosonic Markovian Open Quantum Dynamics with Monte Carlo stochastic trajectories using the Glauber-Sudarshan P, Wigner, and Husimi Q Functions and Hybrids}


\author[first]{Toma~Yoneya}
\author[second]{Kazuya~Fujimoto}
\author[first,third]{Yuki~Kawaguchi}
\affiliation[first]{organization={Department of Applied Physics, Nagoya University},
            city={Nagoya},
            state={464-8603},
            country={Japan}}

\affiliation[second]{organization={Department of Physics, Institute of Science Tokyo},
            city={Tokyo},
            state={152-8551},
            country={Japan}}

\affiliation[third]{organization={Research Center for Crystalline Materials Engineering, Nagoya University},
            city={Nagoya},
            state={464-8603},
            country={Japan}}
\begin{abstract}
The Monte Carlo trajectory sampling of stochastic differential equations based on the quasiprobability distribution functions, such as the Glauber-Sudarshan P, Wigner, and Husimi Q functions, enables us to investigate bosonic open quantum many-body dynamics described by the Gorini-Kossakowski-Sudarshan-Lindblad (GKSL) equation.
In this method, the Monte Carlo samplings for the initial distribution and stochastic noises incorporate quantum fluctuations, and thus, we can go beyond the mean-field approximation.
However, description using stochastic differential equations is possible only when the corresponding Fokker-Planck equation has a positive-semidefinite diffusion matrix.
In this work, we analytically derive the stochastic differential equations for arbitrary Hamiltonian and jump operators based on the path-integral formula, independently of the derivation of the Fokker-Planck equation.
In the course of the derivation, we formulate the path-integral representation of the GKSL equation by using the $s$-ordered quasiprobability distribution function, which systematically describes the aforementioned quasiprobability distribution functions by changing the real parameter $s$.
The essential point of this derivation is that we employ the Hubbard-Stratonovich transformation in the path integral, and its application is not always feasible.
We find that the feasible condition of the Hubbard-Stratonovich transformation is identical to the positive-semidefiniteness condition of the diffusion matrix in the Fokker-Planck equation.
In the benchmark calculations, we confirm that the Monte Carlo simulations of the obtained stochastic differential equations well reproduce the exact dynamics of physical quantities and non-equal time correlation functions of numerically solvable models, including the Bose-Hubbard model.
This work clarifies the applicability of the approximation and gives systematic and simplified procedures to obtain the stochastic differential equations to be numerically solved.
\end{abstract}



\begin{keyword}
Open quantum dynamics \sep Path integral \sep Phase-space method



\end{keyword}

\end{frontmatter}

\tableofcontents



\section{Introduction}
\label{introduction}

The phase-space formulation of quantum mechanics gives us a physical interpretation of quantum many-body states and phenomena \cite{Hillery,Lee}.
It also enables us to investigate bosonic quantum many-body dynamics while considering the effects of quantum fluctuations and has been developed to investigate open quantum many-body dynamics \cite{Gardiner,Milburn,Carmichael1,Carmichael2} described by the Gorini-Kossakowski-Sudarshan-Lindblad (GKSL) equation \cite{Gorini,Lindblad}.
In the phase-space formulation, operators are mapped into $c$-number functions, and the density operator is represented as the quasiprobability distribution function, such as the Glauber-Sudarshan P \cite{Glauber,Sudarshan}, Wigner \cite{Wigner}, and Husimi Q \cite{Husimi,GlauberQ} functions.
When one applies the formulation to an open quantum system and neglects higher-order fluctuations, the GKSL equation is
approximated into the Fokker-Planck equation for the quasiprobability distribution function.
We usually investigate the dynamics following the Fokker-Planck equation by
a Monte Carlo simulation of
corresponding stochastic differential equations.
However, the Fokker-Planck equation does not always reduce to the stochastic differential equations because the diffusion matrix is not always positive-semidefinite depending on details of the Hamiltonian and jump operators \cite{Gardiner,Milburn,Carmichael1,Carmichael2}.

The choice of the quasiprobability distribution function is also crucial for obtaining the Fokker-Planck equation with a positive-semidefinite diffusion matrix.
The approximation using the Wigner function is particularly referred to as the truncated Wigner approximation (TWA) \cite{Gardiner,Milburn,Carmichael1,Carmichael2}.
In isolated systems, the Fokker-Planck equation for the Wigner function does not involve diffusion terms and thus we can always simulate it by iteratively solving classical equations of motion.
For this tractability, the TWA has been widely utilized \cite{Steel,Alice,Blakie,Polkovnikov2010} and generalized to describe the many-body dynamics of spins \cite{Schachenmayer,Davidson2015,Wurtz,Zhu} and fermions \cite{Davidson2017}.
The performance of the TWA, including these generalizations, has been investigated through comparisons with experiments \cite{Orioli,Fersterer,Takasu,Nagao2021,Christopher2022,Christopher2023,Nagao2024}.
Recently, the TWA has been applied to investigate open quantum many-body dynamics such as the dissipative Bose-Hubbard model \cite{kordas2015,Vicentini,PRXQuantumDeuar2021},  cavities \cite{Iacopo2005,Dagvadorj,KeBler2020,Seibold}, and dissipative spin systems \cite{Huber,Singh,Mink,Huber2022}.
In open quantum systems, the TWA is not always applicable depending on the details of the jump operators \cite{Gardiner,Milburn,Carmichael1,Carmichael2,Huber,Plimak,Yoneya2025}. 
On the other hand, although the use of the Glauber-Sudarshan P and Husimi Q functions are necessary for calculating non-equal time correlation functions of normally- and antinormally-ordered operators \cite{Gardiner,Milburn,Carmichael1,Carmichael2,Deuar2021},
the Monte Carlo simulation for these quasiprobability distribution functions is unfeasible in isolated systems because the diffusion matrix of the Fokker-Planck equation always has at least one negative eigenvalue.
The only exception is a non-interacting system.
However, when we consider open quantum systems, the effects of couplings with environments can make the diffusion matrix positive-semidefinite even if the Hamiltonian involves many-body interactions \cite{Gardiner,Milburn,Carmichael1,Carmichael2}.

Considering these facts, a question naturally arises as to when the diffusion matrix of the GKSL equation becomes positive-semidefinite depending on details of the Hamiltonian, jump operators, and the choice of the quasiprobability distribution function.
To the best of our knowledge, the general description of the diffusion matrix is absent.
For the Wigner function, we have analytically derived the positive-semidefiniteness condition of the diffusion matrix under the restriction that the jump operators do not couple different degrees of freedom \cite{Yoneya2025}.

In this work, we analytically obtain the diffusion matrix for the Glauber-Sudarshan P, Wigner, and Husimi Q functions and a hybrid of them for the GKSL equation with an arbitrary Hamiltonian and jump operators.
In the course of the derivation, we formulate the path-integral representation of the GKSL equation by using the $s$-ordered quasiprobability distribution function \cite{Cahill1,Cahill2}, which systematically describes the aforementioned quasiprobability distribution functions by changing the real parameter $s$.
For a system with multiple degrees of freedom, we can use a different $s$ for a different internal degree of freedom, namely, we can hybridize the different quasiprobability distribution functions.
The action in the resulting path-integral representation involves classical and quantum fields, which respectively characterize the classical motion and quantum fluctuations of the system.
Taking the perturbations up to the second-order terms of the action with respect to the quantum fields, we can derive the Fokker-Planck equation, whose diffusion matrix is composed of the second-order terms of the action.
On the other hand, the Hubbard-Stratonovich transformation of the second-order terms of the action leads us to obtain the stochastic differential equation independently of the Fokker-Planck equation.
Here, the Hubbard-Stratonovich transformation is not always feasible, and we confirm that the feasible condition of the Hubbard-Stratonovich transformation is identical to the positive-semidefiniteness condition of the diffusion matrix in the Fokker-Planck equation.
Furthermore, the analytical expression of the action for Markovian open quantum systems obtained in this paper will enable us to clarify the effects of higher-order quantum fluctuations beyond the Fokker-Planck equation, as has been done for isolated systems \cite{Polkovnikov2010,Polkovnikov2003,plimak2009}.
In the benchmark calculations, we investigate the relaxation dynamics of numerically solvable models including Bose-Hubbard model with various jump operators.
By comparing the exact dynamics of physical quantities including non-equal time correlation functions with the ones obtained from the Monte Carlo dynamics of the derived stochastic differential equations, we confirm that our results well reproduce the exact dynamics.

This paper is organized as follows.
In Sec.~\ref{sec:Target of this paper}, we introduce the GKSL equation and the systems we deal with.
In Sec.~\ref{sec:phase-space mapping}, we briefly review the phase-space mapping of bosonic systems and quasiprobability distribution functions.
In Sec.~\ref{sec:Functional representation of the GKSL equation}, we formulate the path-integral representation of the GKSL equation based on the $s$-ordered phase-space mapping and derive the stochastic differential equations and the Fokker-Planck equation.
The formula for calculating the non-equal time correlation functions is also in this section.
We show some benchmark calculations in Sec.~\ref{sec:Benchmark calculations}.
The summary and conclusions are in Sec.~\ref{sec:Summary and conclusions}.


\section{\label{sec:Target of this paper}Target of this paper}
Our formulation is applicable to an arbitrary bosonic open quantum system obeying the GKSL equation.
Below, after introducing the GKSL equation, we summarize the system addressed in this work.


\subsection{\label{subsec:Gorini-Kossakowski-Sudarshan-Lindblad equation}Gorini-Kossakowski-Sudarshan-Lindblad equation}
We first introduce the general description of a quantum system interacting with environments.
Following the conventional literature \cite{Breuer}, we divide the total system into a system we focus on and environments coupling with the system and assume that the total system is isolated from other systems such that its density operator $\hat{\rho}_{\rm tot}$ obeys the von Neumann equation.
Suppose there is no entanglement between the system and the environment in the initial state, i.e., $\hat{\rho}_{\rm tot}(t_0) = \hat{\rho}(t_0)\otimes\hat{\rho}_{\rm B}(t_0)$ with $\rho_{\rm B}(t_0)$ being the initial density operator of the environment, the dynamical map $\hat{\mathcal{V}}(t,t_0)$ that propagates the system's density operator as $\hat{\rho}(t_0)\to\hat{\rho}(t)$ is given by
\begin{align}
    \label{eq:Kraus_representation}
    \hat{\rho}(t) = \hat{\mathcal{V}}(t,t_0)\left[\hat{\rho}(t_0)\right] = \sum_k\hat{M}_k(t,t_0)\hat{\rho}(t_0)\hat{M}^{\dagger}_k(t,t_0),
\end{align}
where $\hat{M}_k(t,t_0)$ is the Kraus operator satisfying $\sum_k\hat{M}_k^{\dagger}(t,t_0)\hat{M}_k(t,t_0) = \hat{1}$ with $\hat{1}$ being the identity operator.
The map \eqref{eq:Kraus_representation} is a completely positive and trace-preserving (CPTP) map that guarantees the trace-preserving property of the density operator in the time evolution and the positive-semidefiniteness of $\hat{\rho}(t)$ and $\hat{\rho}_{\rm tot}(t)$ \cite{Choi}.

Eq.~\eqref{eq:Kraus_representation} generally exhibits a non-Markovian dynamics. In this work, however, we restrict ourselves to considering a Markovian open quantum system.
Under the assumption that the dynamical map is a CPTP map satisfying the Markov condition $\hat{\rho}(t) = \hat{\mathcal{V}}(t,t_0)[\hat{\rho}(t_0)] = \hat{\mathcal{V}}(t,t_j)[\hat{\mathcal{V}}(t_j,t_0)[\hat{\rho}(t_0)]]$ for $t \geq t_j \geq t_0$, the equation of motion of the system's density operator reduces to the GKSL equation \cite{Gorini,Lindblad}:
\begin{align}
    \label{eq:def of GKSL equation}
    \frac{d\hat{\rho}(t)}{dt} = -\frac{i}{\hbar}\left[\hat{H},\hat{\rho}(t)\right]_- + \sum_k\gamma_k\left(\hat{L}_k\hat{\rho}(t)\hat{L}^{\dagger}_k - \frac{1}{2}\left[\hat{L}^{\dagger}_k\hat{L}_k,\hat{\rho}(t)\right]_+\right),
\end{align}
where $[\cdots]_{\mp}$ denote the commutator $(-)$ and anti-commutator $(+)$.
The first term of the right-hand side describes unitary dynamics generated by the system's Hamiltonian $\hat{H}$, and the second term describes non-unitary dynamics, where the jump operator $\hat{L}_k$ characterizes the interaction between the system and the environment, $\gamma_k$ represents the strength, and the subscript $k$ distinguishes a variety of couplings with environments.


\subsection{Setup}
In this paper, we consider a bosonic system with total $M$ degrees of freedom which we identify by using subscripts $m,n \in \{1,2,\dots,M\}$.
The Hamiltonian $\hat{H}$ and the jump operators $\hat{L}_k$ for $\forall k$ are composed of bosonic creation and annihilation operators $\hat{a}_m$ and $\hat{a}^{\dagger}_m$, which satisfy the commutation relation $[\hat{a}_m,\hat{a}^{\dagger}_n]_- = \delta_{mn}$.
Here, $\hat{H}$ and $\hat{L}_k$ for $\forall k$ can include higher-body interactions and couple different degrees of freedom.


\section{\label{sec:phase-space mapping}Review of phase-space mapping of bosonic operators}
In the phase space, a bosonic operator is mapped into a $c$-number function, and the density operator is expressed as a quasiprobability distribution function.
Here, the way of the mapping is not unique, and the most general and comprehensive description has been established in Refs.~\cite{Agarwal1,Agarwal2,Agarwal3}, where the quasiprobability distribution function generally takes complex values depending on the mapping.
In this work, we utilize the phase-space mapping that leads to a real-valued quasiprobability distribution function \cite{Cahill1,Cahill2}.
This condition is necessary for performing the phase-space calculation using a classical computer.  
In this mapping, the phase-space representation is characterized by a real parameter $s$.
Below, we introduce the phase-space mapping and the resulting quasiprobability distribution function with focusing on the relation between the mapping and the operator ordering.
In Secs.~\ref{sec:Mapping to phase space} and \ref{sec:Quasiprobability distribution functions}, we consider a system with a single degree of freedom and extend to the result to a system with multiple degrees of freedom in Sec.~\ref{sec:Extension to multiple degrees of freedom}.


\subsection{\label{sec:Mapping to phase space}Mapping to phase space}
An an arbitrary bosonic operator $\hat{A}$ is mapped into a $c$-number function on the phase space, $\hat{A}\mapsto A_s(\alpha,\alpha^*)$, via
\begin{gather}
    \label{eq:s-parametrized mapping of A single}
    A_s(\alpha,\alpha^*) = \int\frac{d^2\eta}{\pi}\chi_A(\eta,s)e^{\alpha^*\eta - \alpha\eta^*}, \\
    \label{eq:definition of the characteristic function single}
    \chi_A(\eta,s) = {\rm Tr}\left[\hat{A}\hat{D}^{\dagger}(\eta,-s)\right],
\end{gather}
where $-1 \leq s \leq 1$, $\alpha = \alpha^{\rm re} + i\alpha^{\rm im} \in \mathbb{C}$ ($\alpha^{\rm re},\alpha^{\rm im}\in\mathbb{R}$), $\eta = \eta^{\rm re} + i\eta^{\rm im} \in \mathbb{C}$ ($\eta^{\rm re},\eta^{\rm im} \in \mathbb{R}$), $\int d^2\eta = \int_{-\infty}^\infty d\eta^{\rm re} \int_{-\infty}^\infty d\eta^{\rm im}$, and $\chi_{A}(\eta,s)$ is the characteristic function of $\hat{A}$.
Here, $\hat{D}(\eta,s)$ is defined by using the displacement operator $\hat{D}(\eta)=e^{\eta\hat{a}^\dagger - \eta^*\hat{a}}$ as
\begin{align}
    \label{eq:definition of s-ordered displacement operator}
    \hat{D}(\eta,s) = \hat{D}(\eta)e^{s|\eta|^2/2}=e^{\eta \hat{a}^\dagger - \eta^*\hat{a}}e^{s|\eta|^2/2},
\end{align}
where $\hat{a}^\dagger$ and $\hat{a}$ are the creation and annihilation operators of bosons.
We can show that $A_s^*(\alpha,\alpha^*)$ is the phase-space representation of $\hat{A}^{\dagger}$, i.e., $A_s^*(\alpha,\alpha^*) = [\hat{A}^{\dagger}]_s(\alpha,\alpha^*)$, by taking the complex conjugate of Eq.~\eqref{eq:s-parametrized mapping of A single} and transforming $\eta$ to $-\eta$.

In the original papers \cite{Cahill1,Cahill2}, the parameter $s$ can take complex values, and the corresponding quasiprobability distribution function \eqref{eq:definition of the s-ordered quasiprobability distribution function single}, which we will introduce in the next section, becomes a complex function.
However, when $-1 \leq s \leq 1$, the quasiprobability distribution function always takes real values.
Although the basic formulation in this section applies to $-1 \leq s \leq 1$, we focus on integer $s$ ($=1,0,-1$) in our formulation in the subsequent sections.

The phase-space mapping Eq.~\eqref{eq:s-parametrized mapping of A single} transforms a set of ordered products of bosonic creation and annihilation operators into a product of $c$-numbers.
The operator ordering is characterized by the parameter $s$, and is referred to as the $s$-ordering, which is defined by
\begin{align}
    \label{eq:definition of s-ordering}
    \left\{\hat{a}^{\dagger p}\hat{a}^{q}\right\}_{s} = \left.(-1)^{q}\frac{\partial^{p+q}}{\partial\alpha^{p}\partial\alpha^{*q}}\hat{D}(\alpha,s)\right|_{\alpha = 0},
\end{align}
where $p,q\in\mathbb{Z}_{\geq 0}$.
In particular, the $s$-ordering with an integer $s$ gives the widely used ordered product:
$s=1$ provides the normal ordering $\{\hat{a}^{\dagger p}\hat{a}^{q}\}_1 = \hat{a}^{\dagger p}\hat{a}^{q}$,
$s=0$ corresponds to the Weyl (symmetric) ordering,
and $s=-1$ gives the anti-normal ordering $\{\hat{a}^{\dagger p}\hat{a}^{q}\}_{-1} = \hat{a}^{q}\hat{a}^{\dagger p}$.
The Weyl-ordering ($s=0$) consists of all possible ordering products of $\hat{a}$ and $\hat{a}^{\dagger}$, e.g., $\{\hat{a}^{\dagger}\hat{a}\}_0 = (\hat{a}^{\dagger}\hat{a} + \hat{a}\hat{a}^{\dagger})/2$ and $\{\hat{a}^{\dagger 2}\hat{a}^{2}\}_0 = (\hat{a}^{\dagger 2}\hat{a}^{2} + \hat{a}^{\dagger}\hat{a}\hat{a}^{\dagger}\hat{a} + \hat{a}^{\dagger}\hat{a}^2\hat{a}^{\dagger} + \hat{a}\hat{a}^{\dagger 2}\hat{a} + \hat{a}\hat{a}^{\dagger}\hat{a}\hat{a}^{\dagger} + \hat{a}^{2}\hat{a}^{\dagger 2})/6$.
The phase-space mapping Eq.~\eqref{eq:s-parametrized mapping of A single} maps $s$-ordered operators as
\begin{align}
    \label{eq:phase-space transformation of s-ordered operators single}
    \left\{\hat{a}^{\dagger p}\hat{a}^{q}\right\}_s\mapsto\alpha^{*p}\alpha^{q}.
\end{align}
Thus, by expanding $\hat{A}$ in the $s$-ordered form as $\hat{A} = \sum_{p,q}A^{p}_{q}(s)\{\hat{a}^{\dagger p}\hat{a}^{q}\}_{s}$ with $A^{p}_{q}(s) \in \mathbb{C}$ being an expansion coefficient,
we obtain $A_s(\alpha,\alpha^*) = \sum_{p,q}A^{p}_{q}(s)\alpha^{*p}\alpha^{q}$.
Below, we refer to $A_s(\alpha,\alpha^*)$ and $\hat{D}(\alpha,s)$ as the $s$-ordered phase-space representation of $\hat{A}$ and the $s$-ordered displacement operator, respectively.

We further introduce the function $A^e_s(\alpha + \zeta,\alpha^* + \xi^*)$, whose arguments are not in the complex conjugated pairs, as 
\begin{align}
    \label{eq:definition of extended s-ordered phase-space representation of A single}
    A^e_s(\alpha + \zeta,\alpha^* + \xi^*) = {\rm exp}\left(\zeta\frac{\partial}{\partial\alpha} + \xi^*\frac{\partial}{\partial\alpha^*}\right)A_s(\alpha,\alpha^*).
\end{align}
This is equivalent to the one obtained by formally replacing the arguments $\alpha$ and $\alpha^*$ with $\alpha + \zeta$ and $\alpha^* + \xi^*$, respectively, in $A_s(\alpha,\alpha^*)$.
Accordingly, $A^e_s(\alpha+\zeta,\alpha^*+\zeta^*)=A_s(\alpha+\zeta,\alpha^*+\zeta^*)$ holds.
Here, we make two remarks about Eq.~\eqref{eq:definition of extended s-ordered phase-space representation of A single}.
First, $A^e_s(\alpha + \zeta,\alpha^* + \xi^*)$ is not the same as the one defined in the doubled phase-space representation, such as the positive-P representation \cite{Gardiner,Plimak,Drummond1980,Oliveira}.
To avoid confusion, in this paper, we refer to the function $A^e_s(\alpha + \zeta,\alpha^* + \xi^*)$ as the extended $s$-ordered phase-space representation of $\hat{A}$.
Second, $[A^{e}_s(\alpha + \zeta,\alpha^* + \xi^*)]^*$ is not the extended $s$-ordered phase-space representation of $\hat{A}^{\dagger}$, where the latter is obtained by replacing $\alpha$ and $\alpha^*$ in $A_s^*(\alpha,\alpha^*)$ with $\alpha+\zeta$ and $\alpha^*+\xi^*$, respectively, and is defined as
\begin{align}
    \label{eq:definition of extended s-ordered phase-space representation of A dagger single}
    \bar{A}^e_s(\alpha + \zeta,\alpha^* + \xi^*) = [\hat{A}^{\dagger}]^e_s(\alpha + \zeta,\alpha^* + \xi^*) = {\rm exp}\left(\zeta\frac{\partial}{\partial\alpha} + \xi^*\frac{\partial}{\partial\alpha^*}\right)A^*_s(\alpha,\alpha^*).
\end{align}

\subsection{\label{sec:Quasiprobability distribution functions}Quasiprobability distribution functions}

According to the original paper, Ref.~\cite{Cahill2},
we specifically refer to the $(-s)$-ordered phase-space representation of the density operator $\hat{\rho}(t)$ as the $s$-ordered quasiprobability distribution function $W_s(\alpha,\alpha^*,t) \in \mathbb{R}$, which is defined by
\begin{gather}
    \label{eq:definition of the s-ordered quasiprobability distribution function single}
    W_s(\alpha,\alpha^*,t) = \int\frac{d^2\eta}{\pi}\chi_{\rho}(\eta,-s)e^{\alpha^*\eta - \alpha\eta^*}, \\
    \label{eq:definition of the characteristic function of the s-ordered quasiprobability distribution function single}
    \chi_{\rho}(\eta,-s) = {\rm Tr}\left[\hat{\rho}(t)\hat{D}^{\dagger}(\eta,s)\right].
\end{gather}
Here, $W_s(\alpha,\alpha^*,t)$ with $s=1,0,$ and $-1$ correspond to the Glauber-Sudarshan P function, the Wigner function, and the Husimi Q function, respectively.
By using the relation \cite{Cahill1,Cahill2}
\begin{align}
    \label{eq:s-parametrized representation of TrAB single}
    {\rm Tr}\left[\hat{A}\hat{B}\right] = \int\frac{d^2\alpha}{\pi}A_s(\alpha,\alpha^*)B_{-s}(\alpha,\alpha^*)
\end{align}
with $\hat{B} = \hat{\rho}(t)$, we can evaluate the expectation value of a physical quantity $\braket{\hat{A}(t)} = {\rm Tr}[\hat{A}\hat{\rho}(t)]$ as
\begin{align}
    \label{eq:physical quantity in the s-ordered phase space single}
    \braket{\hat{A}(t)} = \int\frac{d^2\alpha}{\pi}A_s(\alpha,\alpha^*)W_s(\alpha,\alpha^*,t).
\end{align}
When we choose $\hat{A}$ as the identity operator $\hat{1}$ and use the normalization property of the density operator ${\rm Tr}[\hat{\rho}(t)] = 1$, we obtain the normalization condition for $W_s(\alpha,\alpha^*,t)$:
\begin{align}
    \label{eq:normalization condition for the quasiprobability distribution function single}
    \int\frac{d^2\alpha}{\pi}W_s(\alpha,\alpha^*,t) = 1.
\end{align}
The quasiprobability distribution function can generally take negative values except for the Husimi Q function $W_{s=-1}(\alpha,\alpha^*,t)$, which can only take non-negative values.


\subsection{\label{sec:Extension to multiple degrees of freedom}Extension to multiple degrees of freedom}
The extension of the phase-space mapping in the previous sections to a system with multiple degrees of freedom is straightforward.
We consider a bosonic operator $\hat{A}$ consisting of $\hat{a}^{\dagger}_m$ and $\hat{a}_m$ with various $m$.
In the phase space, the operator $\hat{A}$ is mapped into a $c$-number function $A_{\vec{s}}(\vec{\alpha},\vec{\alpha}^*)$ through
\begin{gather}
    \label{eq:s-parametrized mapping of A}
    A_{\vec{s}}(\vec{\alpha},\vec{\alpha}^*) = \int\frac{d^2\vec{\eta}}{\pi^M}\chi_A(\vec{\eta},\vec{s})e^{\vec{\alpha}^*\cdot\vec{\eta} - \vec{\alpha}\cdot\vec{\eta}^*},\\
    \label{eq:definition of the characteristic function}
    \chi_A(\vec{\eta},\vec{s}) = {\rm Tr}\left[\hat{A}\hat{D}^{\dagger}(\vec{\eta},-\vec{s})\right],
\end{gather}
where $\vec{\alpha}=(\alpha_1, \alpha_2, \cdots, \alpha_M)^{\rm T}$ with ${\rm T}$ being the transposition and $\alpha_m = \alpha^{\rm re}_m + i\alpha_m^{\rm im}$ ($\alpha^{\rm re}_m,\alpha^{\rm im}_m \in \mathbb{R}$) for $\forall m$, $\vec{\eta}=(\eta_1, \eta_2, \cdots, \eta_M)^{\rm T}$, $\int d^2\vec{\eta} = \prod_{m=1}^M\int d^2\eta_m = \prod_{m}\int_{-\infty}^\infty d\eta_m^{\rm re} \int_{-\infty}^\infty d\eta_m^{\rm im}$ with $\eta_m=\eta_m^{\rm re} + i\eta_m^{\rm im}\in\mathbb{C}$ ($\eta_m^{\rm re}, \eta_m^{\rm im}\in\mathbb{R}$) for $\forall m$, $\cdot$ indicates the inner product, $\vec{s} = (s_1,s_2,\cdots,s_M)^{\rm T}$ with $-1 \leq s_m \leq 1$ for $\forall m$, and $\chi_{A}(\vec{\eta},\vec{s})$ is a characteristic function with $\hat{D}(\vec{\alpha},\vec{s})$ defined by
\begin{align}
    \label{eq:s-ordered displacememt operator}
    \hat{D}(\vec{\eta},\vec{s}) = \bigotimes_{m=1}^M\hat{D}(\eta_m,s_m),
\end{align}
where $\hat{D}(\alpha_m,s_m)$ is given by Eq.~\eqref{eq:definition of s-ordered displacement operator}.
In Eq.~\eqref{eq:s-parametrized mapping of A}, the parameters $s_1,s_2,\cdots, s_M$ can take different values for each degree of freedom.
Below, we refer to $A_{\vec{s}}(\vec{\alpha},\vec{\alpha}^*)$ as the $\vec{s}$-ordered phase-space representation of $\hat{A}$ and $\hat{D}(\vec{\alpha},\vec{s})$ as the $\vec{s}$-ordered displacement operator.

The phase-space mapping Eq.~\eqref{eq:s-parametrized mapping of A} transforms a product of bosonic creation and annihilation operators as follows:
\begin{align}
    \label{eq:phase-space transformation of s-ordered operators}
    \left\{\hat{a}_m^{\dagger p_m}\hat{a}_m^{q_m}\right\}_{s_m}\mapsto\alpha_m^{*p_m}\alpha_m^{q_m},\qquad\prod_{m=1}^M\left\{\hat{a}_m^{\dagger p_m}\hat{a}_m^{q_m}\right\}_{s_m}\mapsto\prod_{m=1}^M \alpha_m^{*p_m}\alpha_m^{q_m},
\end{align}
where $p_m,q_m\in\mathbb{Z}_{\geq 0}$ for $\forall m$.
In order to obtain $A_{\vec{s}}(\vec{\alpha},\vec{\alpha}^*)$, we first expand $\hat{A}$ as $\hat{A} = \sum_{\{p_m\},\{q_m\}}A^{p_1\cdots p_M}_{q_1\cdots q_M}(\vec{s})\prod_m\{\hat{a}_m^{\dagger p_m}\hat{a}_m^{q_m}\}_{s_m}$ with an expansion coefficient $A^{p_1\cdots p_M}_{q_1\cdots q_M}(\vec{s}) \in \mathbb{C}$, and replace the operators by the $c$-numbers according to Eq.~\eqref{eq:phase-space transformation of s-ordered operators}, obtaining $A_{\vec{s}}(\vec{\alpha},\vec{\alpha}^*) = \sum_{\{p_m\},\{q_m\}}A^{p_1\cdots p_M}_{q_1\cdots q_M}(\vec{s})\prod_m\alpha_m^{*p_m}\alpha_m^{q_m}$.
Here, we need to appropriately reorder the creation and annihilation operators for each degree of freedom before mapping to the phase space.

We also introduce the extended $\vec{s}$-ordered phase-space representation of $\hat{A}$ and $\hat{A}^{\dagger}$ as
\begin{gather}
    \label{eq:definition of extended s-ordered phase-space representation of A}
    A^e_{\vec{s}}(\vec{\alpha} + \vec{\zeta},\vec{\alpha}^* + \vec{\xi}^*) = {\rm exp}\left\{\sum_{m=1}^M\left(\zeta_{m}\frac{\partial}{\partial\alpha_{m}} + \xi^*_{m}\frac{\partial}{\partial\alpha^*_{m}}\right)\right\}A_{\vec{s}}(\vec{\alpha},\vec{\alpha}^*),\\
    \label{eq:definition of extended s-ordered phase-space representation of A dagger}
    \bar{A}^e_{\vec{s}}(\vec{\alpha} + \vec{\zeta},\vec{\alpha}^* + \vec{\xi}^*) = {\rm exp}\left\{\sum_{m=1}^M\left(\zeta_{m}\frac{\partial}{\partial\alpha_{m}} + \xi^*_{m}\frac{\partial}{\partial\alpha^*_{m}}\right)\right\}A^*_{\vec{s}}(\vec{\alpha},\vec{\alpha}^*),
\end{gather}
which are respectively obtained by replacing $\vec{\alpha}$ with $\vec{\alpha} + \vec{\zeta}$ and $\vec{\alpha}^*$ with $\vec{\alpha}^* + \vec{\xi}^*$ in $A_{\vec{s}}(\vec{\alpha},\vec{\alpha}^*)$ and $A_{\vec{s}}^*(\vec{\alpha},\vec{\alpha}^*)$.

The $\vec{s}$-ordered quasiprobability distribution function $W_{\vec{s}}(\vec{\alpha},\vec{\alpha}^*,t)$ is defined as the $(-\vec{s})$-ordered phase-space representation of the density operator $\hat{\rho}(t)$:
\begin{gather}
    \label{eq:definition of the s-ordered quasiprobability distribution function}
    W_{\vec{s}}(\vec{\alpha},\vec{\alpha}^*,t) = \int\frac{d^2\vec{\eta}}{\pi^M}\chi_{\rho}(\vec{\eta},-\vec{s})e^{\vec{\alpha}^*\cdot\vec{\eta} - \vec{\alpha}\cdot\vec{\eta}^*}, \\
    \label{eq:definition of the characteristic function of the s-ordered quasiprobability distribution function}
    \chi_{\rho}(\vec{\eta},-\vec{s}) = {\rm Tr}\left[\hat{\rho}(t)\hat{D}^{\dagger}(\vec{\eta},\vec{s})\right].
\end{gather}
With the normalized quasiprobability distribution function $W_{\vec{s}}(\vec{\alpha},\vec{\alpha}^*,t)$,
\begin{align}
    \label{eq:normalization condition for the quasiprobability distribution function}
    \int\frac{d^2\vec{\alpha}}{\pi^M}W_{\vec{s}}(\vec{\alpha},\vec{\alpha}^*,t) = 1,
\end{align}
we can calculate the expectation value of a physical quantity $\hat{A}(t)$ as
\begin{align}
    \label{eq:physical quantity in the s-ordered phase space}
    \braket{\hat{A}(t)} = \int\frac{d^2\vec{\alpha}}{\pi^M}A_{\vec{s}}(\vec{\alpha},\vec{\alpha}^*)W_{\vec{s}}(\vec{\alpha},\vec{\alpha}^*,t),
\end{align}
where Eqs.~\eqref{eq:normalization condition for the quasiprobability distribution function} and \eqref{eq:physical quantity in the s-ordered phase space} are respectively obtained by using the relation:
\begin{align}
    \label{eq:s-parametrized representation of TrAB}
    {\rm Tr}\left[\hat{A}\hat{B}\right] = \int\frac{d^2\vec{\alpha}}{\pi^M}A_{\vec{s}}(\vec{\alpha},\vec{\alpha}^*)B_{-\vec{s}}(\vec{\alpha},\vec{\alpha}^*)
\end{align}
with $(\hat{A},\hat{B})=(\hat{1},\hat{\rho}(t))$ and $(\hat{A},\hat{\rho}(t))$.
When we choose $s_m = s = 1,0$, or $-1$ for $\forall m$ in Eq.~\eqref{eq:definition of the s-ordered quasiprobability distribution function}, the quasiprobability distribution function reduces to the Glauber-Sudarshan P function ($s=1$), the Wigner function ($s=0$), and the Husimi Q function ($s=-1$).
If a system is in a product state $\hat{\rho}(t) = \bigotimes_m\hat{\rho}_m(t)$ with $\rho_m(t)$ being a reduced density operator for the $m$th degree of freedom, the corresponding $\vec{s}$-ordered quasiprobability distribution function becomes $W_{\vec{s}}(\vec{\alpha},\vec{\alpha}^*,t) = \prod_{m=1}^MW_{s_m}(\alpha_m,\alpha^*_m,t)$, where $W_{s_m}(\alpha_m,\alpha^*_m,t)$ is given by Eq.~\eqref{eq:definition of the s-ordered quasiprobability distribution function single}.


\section{\label{sec:Functional representation of the GKSL equation}Functional representation of Markovian open quantum systems in the phase space}
In the phase space, the GKSL equation can be approximated into the Fokker-Planck equation for the $\vec{s}$-ordered quasiprobability distribution function, which does not always reduce to stochastic differential equations depending on details of the Hamiltonian, jump operators, and parameters $s_m$.
Below, we derive the stochastic differential equations and the condition for obtaining them by using the path-integral approach.
In Secs.\ref{subsec:Markov condition in the phase space} and \ref{subsec:Path-integral representation}, we first derive the path-integral representation of the GKSL equation for the $\vec{s}$-ordered quasiprobability distribution function.
Based on this representation, in Sec.~\ref{eq:Equation of motion in the phase space}, we obtain the stochastic differential equation and discuss its relation to the Fokker–Planck equation, whose detailed derivation is given separately in \ref{appendix:Equations of motion in the phase space}.


\subsection{\label{subsec:Markov condition in the phase space}Markov condition in the phase space}
We first present an integral representation of an open quantum system using the propagator of the $\vec{s}$-ordered quasiprobability distribution function, which is a starting point for formulating the path-integral representation of the GKSL equation in the next section.
Using Eq.~\eqref{eq:definition of the s-ordered quasiprobability distribution function}, we obtain the $(-\vec{s})$-ordered phase-space representation of the Kraus representation \eqref{eq:Kraus_representation} as
\begin{align}
    \label{eq:Kraus representation in the phase space}
    W_{\vec{s}}(\vec{\alpha}_{\rm f},\vec{\alpha}^*_{\rm f},t) = \int\frac{d^2\vec{\alpha}_0}{\pi^M}\varUpsilon_{\vec{s}}(\vec{\alpha}_{\rm f},t;\vec{\alpha}_0,t_0)W_{\vec{s}}(\vec{\alpha}_0,\vec{\alpha}^*_0,t_0),
\end{align}
where $\varUpsilon_{\vec{s}}(\vec{\alpha}_{\rm f},t;\vec{\alpha}_0,t_0)$ is the propagator of the $\vec{s}$-ordered quasiprobability distribution function given by
\begin{align}
    \label{eq: phase space representation of non Markov propagator}
    \varUpsilon_{\vec{s}}(\vec{\alpha}_{\rm f},t;\vec{\alpha}_0,t_0) = \int\frac{d^2\vec{\xi} d^2\vec{\eta} }{\pi^{2M}}\sum_k{\rm Tr}\left[\hat{D}^{\dagger}(\vec{\xi},\vec{s})\hat{M}_k(t,t_0)\hat{D}^{\dagger}(\vec{\eta},-\vec{s})\hat{M}^{\dagger}_k(t,t_0)\right]e^{\vec{\alpha}_{\rm f}^*\cdot\vec{\xi} - \vec{\alpha}_{\rm f}\cdot\vec{\xi}^*}e^{\vec{\alpha}^*_0\cdot\vec{\eta} - \vec{\alpha}_0\cdot\vec{\eta}^*}.
\end{align}
The detailed derivation of Eq.~\eqref{eq: phase space representation of non Markov propagator} is given in \ref{appendix:The propagator in the Kraus representation}.
When the dynamical map \eqref{eq:Kraus_representation} satisfies the Markov condition, the propagator satisfies the following condition:
\begin{align}
    \label{eq:Markov conditoin in the phase space}
    \varUpsilon_{\vec{s}}(\vec{\alpha}_{\rm f},t;\vec{\alpha}_0,t_0) = \int\frac{d^2\vec{\alpha}_j}{\pi^M}\varUpsilon_{\vec{s}}(\vec{\alpha}_{\rm f},t;\vec{\alpha}_j,t_j)\varUpsilon_{\vec{s}}(\vec{\alpha}_j,t_j;\vec{\alpha}_0,t_0).
\end{align}
This is the Markov condition in the phase space.
We provide the derivation of Eq.~\eqref{eq:Markov conditoin in the phase space} in \ref{appendix:Markov condition for the propagator}.


\subsection{\label{subsec:Path-integral representation}Path-integral representation}
Eq.~\eqref{eq:Markov conditoin in the phase space} enables us to write the time-evolved $\vec{s}$-ordered quasiprobability distribution function $W_{\vec{s}}(\vec{\alpha}_{\rm f},\vec{\alpha}^*_{\rm f},t)$ as an infinite product of infinitesimal time propagators as
\begin{align}
    \label{eq:s-ordered quasiprobability distribution functoin with infinite products of infinitesimal time propagator}
    W_{\vec{s}}(\vec{\alpha}_{\rm f},\vec{\alpha}^*_{\rm f},t) = \lim_{\Delta t\to 0}\prod_{j=0}^{N_t - 1}\int\frac{d^2\vec{\alpha}_j}{\pi^M}\varUpsilon_{\vec{s}}(\vec{\alpha}_{j+1},t_{j+1};\vec{\alpha}_j,t_j)W_{\vec{s}}(\vec{\alpha}_0,\vec{\alpha}^*_0,t_0),
\end{align}
where we discretize the time interval $[t,t_0]$ into $N_t$ with width $\Delta t$:
\begin{align}
    N_t=\frac{t-t_0}{\Delta t},\quad
    t_j = t_0 + j\Delta t,\quad
    t_{N_t} = t,\quad
    \alpha_{N_t} = \alpha_{\rm f}.
\end{align}
Below, after deriving the explicit form of the infinitesimal time propagator $\varUpsilon_{\vec{s}}(\vec{\alpha}_{j+1},t_{j+1};\vec{\alpha}_j,t_j)$ from the GKSL equation, we obtain the path-integral representation of the $\vec{s}$-ordered quasiprobability distribution function by substituting the obtained propagator into Eq.~\eqref{eq:s-ordered quasiprobability distribution functoin with infinite products of infinitesimal time propagator}.

Substituting $\vec{\alpha}_{\rm f} = \vec{\alpha}_{j+1}$, $\vec{\alpha}_0 = \vec{\alpha}_j$, $t = t_{j+1}$, $t_0 = t_j$, $\hat{D}(\vec{\xi},\vec{s}) = \hat{D}(\vec{\xi},\vec{0})e^{\sum_m s_m|\xi_m|^2/2}$, and $\hat{D}(\vec{\eta},-\vec{s}) = \hat{D}(\vec{\eta},\vec{0})e^{-\sum_m s_m|\eta_m|^2/2}$ into Eq.~\eqref{eq: phase space representation of non Markov propagator}, we obtain
\begin{align}
    \varUpsilon_{\vec{s}}(\vec{\alpha}_{j+1},t_{j+1};\vec{\alpha}_j,t_j) &= \int\frac{d^2\vec{\xi} d^2\vec{\eta} }{\pi^{2M}}\sum_k{\rm Tr}\left[\hat{D}^{\dagger}(\vec{\xi},\vec{0})\hat{M}_k(t_{j+1},t_j)\hat{D}^{\dagger}(\vec{\eta},\vec{0} )\hat{M}^{\dagger}_k(t_{j+1},t_j)\right]e^{\vec{\alpha}_{j+1}^*\cdot\vec{\xi} - \vec{\alpha}_{j+1}\cdot\vec{\xi}^*}e^{\vec{\alpha}^*_j\cdot\vec{\eta} - \vec{\alpha}_j\cdot\vec{\eta}^*}e^{\sum_ms_m(|\xi_m|^2 - |\eta_m|^2)/2} \nonumber\\
    \label{eq:relation between s-ordered propagator and s=0 propagator}
    &= {\rm exp}\left\{\sum_{m=1}^{M}\frac{s_m}{2}\left(\frac{\partial^2}{\partial\alpha_{m,j}\partial\alpha^*_{m,j}} - \frac{\partial^2}{\partial\alpha_{m,j+1}\partial\alpha^*_{m,j+1}}\right)\right\}\varUpsilon_{\vec{0}}(\vec{\alpha}_{j+1},t_{j+1};\vec{\alpha}_j,t_j),
\end{align}
where $\vec{0}$ is the zero vector of dimension $M$.
In Ref.~\cite{Yoneya2025}, we have derived the infinitesimal time propagator for the Wigner function $\varUpsilon_{\vec{0}}(\vec{\alpha}_{j+1},t_{j+1};\vec{\alpha}_j,t_j)$ as
\begin{align}
    \label{eq:propagato for the Wigner function}
    \varUpsilon_{\vec{0}}(\vec{\alpha}_{j+1},t_{j+1};\vec{\alpha}_j,t_j) =  \int\frac{d^2\vec{\eta}_{j+1}}{\pi^M}&e^{\vec{\eta}_{j+1}^*\cdot(\vec{\alpha}_{j+1} - \vec{\alpha}_j) - \vec{\eta}_{j+1}\cdot(\vec{\alpha}_{j+1}^* - \vec{\alpha}_j^*)}\nonumber \\
    \times&\left[1 + \frac{i\Delta t}{\hbar}\right. \left\{\sum_{n=0,1}(-1)^nH_{\vec{0}}\left(\vec{\alpha}_j + \frac{(-1)^n}{2}\vec{\eta}_{j+1},\vec{\alpha}^*_j + \frac{(-1)^n}{2}\vec{\eta}^*_{j+1}\right)\right.\nonumber\\
    &\hphantom{\times[1 + \frac{i\Delta t}{\hbar}\{}\left.\left.-  i\hbar\mathcal{D}_{\vec{0}}\left(\vec{\alpha}_j + \frac{1}{2}\vec{\eta}_{j+1},\vec{\alpha}^*_j + \frac{1}{2}\vec{\eta}^*_{j+1},\vec{\alpha}_j - \frac{1}{2}\vec{\eta}_{j+1},\vec{\alpha}^*_j - \frac{1}{2}\vec{\eta}^*_{j+1}\right)\right\}\right],
\end{align}
where $\mathcal{D}_{\vec{0}}$ is given by
\begin{align}
    \label{eq:definition of the non-unitary term of the propagator of the Wigner function}
    \mathcal{D}_{\vec{0}}(\vec{\alpha},\vec{\alpha}^*,\vec{\beta},\vec{\beta}^*) = \sum_k\gamma_k\left\{L^*_{k\vec{0}}(\vec{\alpha},\vec{\alpha}^*)\star^e L_{k\vec{0}}(\vec{\beta},\vec{\beta}^*) - \frac{1}{2}L^*_{k\vec{0}}(\vec{\alpha},\vec{\alpha}^*)\star L_{k\vec{0}}(\vec{\alpha},\vec{\alpha}^*) - \frac{1}{2}L^*_{k\vec{0}}(\vec{\beta},\vec{\beta}^*)\star L_{k\vec{0}}(\vec{\beta},\vec{\beta}^*)\right\}
\end{align}
with $H_{\vec{0}}(\vec{\alpha},\vec{\alpha}^*)$ and $L_{k\vec{0}}(\vec{\alpha},\vec{\alpha}^*)$ being the $\vec{s}$-ordered phase-space representation of $\hat{H}$ and $\hat{L}_k$ with $\vec{s} = \vec{0}$, respectively.
In Eq.~\eqref{eq:definition of the non-unitary term of the propagator of the Wigner function}, $\star^e$ is the extended Moyal product defined by
\begin{align}
    \label{eq:definition of the extended Moyal product}
    A_{\vec{0}}(\vec{\alpha},\vec{\alpha}^*)\star^e B_{\vec{0}}(\vec{\beta},\vec{\beta}^*) = A_{\vec{0}}(\vec{\alpha},\vec{\alpha}^*){\rm exp}\left\{\sum_{m=1}^{M}\left(\frac{1}{2}\frac{\overleftarrow{\partial}}{\partial \alpha_m}\frac{\overrightarrow{\partial}}{\partial \beta^*_m} - \frac{1}{2}\frac{\overleftarrow{\partial}}{\partial \alpha^*_m}\frac{\overrightarrow{\partial}}{\partial \beta_m}\right)\right\}B_{\vec{0}}(\vec{\beta},\vec{\beta}^*)
\end{align}
with the arrows above the derivative symbols indicating which function, left or right, is to be differentiated.
When we choose $\vec{\beta} = \vec{\alpha}$, the extended Moyal product $\star^e$ reduces to the Moyal product $\star$:
\begin{align}
    \label{eq:definition of the Moyal product}
    A_{\vec{0}}(\vec{\alpha},\vec{\alpha}^*)\star B_{\vec{0}}(\vec{\alpha},\vec{\alpha}^*) = A_{\vec{0}}(\vec{\alpha},\vec{\alpha}^*){\rm exp}\left\{\sum_{m=1}^{M}\left(\frac{1}{2}\frac{\overleftarrow{\partial}}{\partial \alpha_m}\frac{\overrightarrow{\partial}}{\partial \alpha^*_m} - \frac{1}{2}\frac{\overleftarrow{\partial}}{\partial \alpha^*_m}\frac{\overrightarrow{\partial}}{\partial \alpha_m}\right)\right\}B_{\vec{0}}(\vec{\alpha},\vec{\alpha}^*).
\end{align}
Substituting Eq.~\eqref{eq:propagato for the Wigner function} into the right-hand side of Eq.~\eqref{eq:relation between s-ordered propagator and s=0 propagator}, we obtain $\varUpsilon_{\vec{s}}(\vec{\alpha}_{j+1},t_{j+1};\vec{\alpha}_j,t_j)$ as
\begin{align}
    \label{eq:s-parametrized phase space representation of Markov propagator}
    \varUpsilon_{\vec{s}}(\vec{\alpha}_{j+1},t_{j+1};\vec{\alpha}_j,t_j) = \int\frac{d^2\vec{\eta}_{j+1}}{\pi^M}e^{\vec{\eta}_{j+1}^*\cdot(\vec{\alpha}_{j+1} - \vec{\alpha}_j) - \vec{\eta}_{j+1}\cdot(\vec{\alpha}_{j+1}^* - \vec{\alpha}_j^*)}\left[1 + \frac{i\Delta t}{\hbar}\left\{H^e_{\vec{s}}(\vec{\psi}^+_{\vec{s},j},\vec{\psi}^{+*}_{-\vec{s},j}) - H^e_{\vec{s}}(\vec{\psi}^-_{-\vec{s},j},\vec{\psi}^{-*}_{\vec{s},j}) - i\hbar\mathcal{D}_{\vec{s}}(\vec{\psi}^+_{\vec{s},j},\vec{\psi}^{+*}_{-\vec{s},j},\vec{\psi}^-_{-\vec{s},j},\vec{\psi}^{-*}_{\vec{s},j})\right\}\right].
\end{align}
See \ref{appendix:Infinitesimal time propagator} for the detailed derivation.
In Eq.~\eqref{eq:s-parametrized phase space representation of Markov propagator}, we have introduced the vectors
\begin{gather}
    \label{eq:+ vectors contains alpha and eta discrete}
    \vec{\psi}^+_{\vec{s},j} = \left(\alpha_{1,j} + \frac{1+s_1}{2}\eta_{1,j+1},\alpha_{2,j} + \frac{1+s_2}{2}\eta_{2,j+1},\cdots,\alpha_{M,j} + \frac{1+s_M}{2}\eta_{M,j+1}\right), \\
    \label{eq:- vectors contains alpha and eta discrete}
    \vec{\psi}^-_{\vec{s},j} = \left(\alpha_{1,j} - \frac{1+s_1}{2}\eta_{1,j+1},\alpha_{2,j} - \frac{1+s_2}{2}\eta_{2,j+1},\cdots,\alpha_{M,j} - \frac{1+s_M}{2}\eta_{M,j+1}\right),
\end{gather}
and define $\mathcal{D}_{\vec{s}}$ as
\begin{align}
    \label{eq:definition of the non-unitary term of the propagator of the s-ordered quasiprobability distributino function}
    \mathcal{D}_{\vec{s}}(\vec{\alpha},\vec{\beta},\vec{\gamma},\vec{\delta}) = \sum_k\gamma_k\left\{\bar{L}^e_{k\vec{s}}(\vec{\alpha},\vec{\beta})\star_{\vec{s}} L^e_{k\vec{s}}(\vec{\gamma},\vec{\delta}) - \frac{1}{2}\bar{L}^e_{k\vec{s}}(\vec{\alpha},\vec{\beta})\star_{\vec{s}} L^e_{k\vec{s}}(\vec{\alpha},\vec{\beta}) - \frac{1}{2}\bar{L}^e_{k\vec{s}}(\vec{\gamma},\vec{\delta})\star_{\vec{s}} L^e_{k\vec{s}}(\vec{\gamma},\vec{\delta})\right\},
\end{align}
where $H^e_{\vec{s}}(\vec{\alpha},\vec{\beta})$, $L^e_{k\vec{s}}(\vec{\alpha},\vec{\beta})$, and $\bar{L}^e_{k\vec{s}}(\vec{\alpha},\vec{\beta})$ are, respectively, the extended $\vec{s}$-ordered phase-space representation of $\hat{H}$, $\hat{L}_k$, and $\hat{L}^{\dagger}_k$ for defined by Eqs.~\eqref{eq:definition of extended s-ordered phase-space representation of A} and \eqref{eq:definition of extended s-ordered phase-space representation of A dagger}, and we have introduced the differential operator $\star_{\vec{s}}$ as
\begin{align}
    \label{eq:definition of the s-ordered Moyal product}
    A^e_{\vec{s}}(\vec{\alpha},\vec{\gamma})\star_{\vec{s}} B^e_{\vec{s}}(\vec{\beta},\vec{\delta}) = A^e_{\vec{s}}(\vec{\alpha},\vec{\gamma}){\rm exp}\left\{\sum_{m=1}^{M}\left(\frac{1+s_m}{2}\frac{\overleftarrow{\partial}}{\partial \alpha_m}\frac{\overrightarrow{\partial}}{\partial \delta_m} - \frac{1-s_m}{2}\frac{\overleftarrow{\partial}}{\partial \gamma_m}\frac{\overrightarrow{\partial}}{\partial \beta_m}\right)\right\}B^e_{\vec{s}}(\vec{\beta},\vec{\delta}).
\end{align}
When we choose $\vec{\gamma} = \vec{\alpha}^*$, $\vec{\delta} = \vec{\beta}^*$, and $s_m=0$ for $\forall m$ in Eq.~\eqref{eq:definition of the s-ordered Moyal product}, the differential operator $\star_{\vec{s}}$ reduces to the extended Moyal product \eqref{eq:definition of the extended Moyal product}.
Finally, rewriting Eq.~\eqref{eq:s-parametrized phase space representation of Markov propagator} as
\begin{align}
    \label{eq:s-parametrized phase space representation of Markov propagator exp}
    \varUpsilon_{\vec{s}}(\vec{\alpha}_{j+1},t_{j+1};\vec{\alpha}_j,t_j) = \int\frac{d^2\vec{\eta}_{j+1}}{\pi^M}&e^{\vec{\eta}_{j+1}^*\cdot(\vec{\alpha}_{j+1} - \vec{\alpha}_j) - \vec{\eta}_{j+1}\cdot(\vec{\alpha}_{j+1}^* - \vec{\alpha}_j^*)}\nonumber \\
    \times &{\rm exp}\left[\frac{i\Delta t}{\hbar}\left\{H^e_{\vec{s}}(\vec{\psi}^+_{\vec{s},j},\vec{\psi}^{+*}_{-\vec{s},j}) - H^e_{\vec{s}}(\vec{\psi}^-_{-\vec{s},j},\vec{\psi}^{-*}_{\vec{s},j}) - i\hbar\mathcal{D}_{\vec{s}}(\vec{\psi}^+_{\vec{s},j},\vec{\psi}^{+*}_{-\vec{s},j},\vec{\psi}^-_{-\vec{s},j},\vec{\psi}^{-*}_{\vec{s},j})\right\} + o(\Delta t)\right],
\end{align}
substituting Eq.~\eqref{eq:s-parametrized phase space representation of Markov propagator exp} into Eq.~\eqref{eq:s-ordered quasiprobability distribution functoin with infinite products of infinitesimal time propagator}, and ignoring the terms of order $o(\Delta t)$, we obtain the path-integral representation for the $\vec{s}$-ordered quasiprobability distribution function:
\begin{gather}
    \label{eq:path-integral representaiton discrete}
    W_{\vec{s}}(\vec{\alpha}_{\rm f},\vec{\alpha}^*_{\rm f},t) = \lim_{\Delta t \to 0}\prod_{j=0}^{N_t-1}\int\frac{d^2\vec{\alpha}_j d^2\vec{\eta}_{j+1}}{\pi^{2M}}e^{i\Delta t \mathcal{L}^{\vec{s}}_j/\hbar} W_{\vec{s}}(\vec{\alpha}_0,\vec{\alpha}^*_0,t_0), \\
    \label{eq:action discrete}
    \mathcal{L}^{\vec{s}}_j = i\hbar\left\{\vec{\eta}_{j+1}\cdot\left(\frac{\vec{\alpha}_{j+1}^* - \vec{\alpha}_j^*}{\Delta t}\right) - \vec{\eta}_{j+1}^*\cdot\left(\frac{\vec{\alpha}_{j+1} - \vec{\alpha}_j}{\Delta t}\right)\right\} + H^e_{\vec{s}}(\vec{\psi}^+_{\vec{s},j},\vec{\psi}^{+*}_{-\vec{s},j}) - H^e_{\vec{s}}(\vec{\psi}^-_{-\vec{s},j},\vec{\psi}^{-*}_{\vec{s},j}) - i\hbar\mathcal{D}_{\vec{s}}(\vec{\psi}^+_{\vec{s},j},\vec{\psi}^{+*}_{-\vec{s},j},\vec{\psi}^-_{-\vec{s},j},\vec{\psi}^{-*}_{\vec{s},j}),
\end{gather}
where $\mathcal{L}^{\vec{s}}_j$ is the Lagrangian of the system.
Eqs.~\eqref{eq:path-integral representaiton discrete} and \eqref{eq:action discrete} reduce to the path-integral representation for the $\vec{s}$-ordered quasiprobability distribution function for an isolated system \cite{plimak2009} when we choose $\gamma_k=0$ for $\forall k$, and the one for the Wigner function \cite{Polkovnikov2010,Yoneya2025,Polkovnikov2009} when we choose $s_m=0$ for $\forall m$.
From these correspondences, we can respectively regard the fields $\vec{\alpha}_j$ and $\vec{\eta}_{j+1}$ as classical and quantum fields,
where the classical fields describe the classical motion of the system and the quantum fields characterize quantum fluctuations around the classical motion \cite{Polkovnikov2010,Polkovnikov2003,plimak2009,Polkovnikov2009,Marinov,Dittrich2006,Dittrich2010,Gozzi,Pagani}.

In the continuous limit, we formally represent Eqs.~\eqref{eq:path-integral representaiton discrete} and \eqref{eq:action discrete} as 
\begin{gather}
    \label{eq:path-integral representaiton continuous}
    W_{\vec{s}}(\vec{\alpha}_{\rm  f},\vec{\alpha}_{\rm f}^*,t) = \int\mathscr{D}^2\vec{\alpha}\mathscr{D}^2\vec{\eta} e^{i\mathcal{S}[\vec{\alpha},\vec{\eta}]/\hbar}W_{\vec{s}}(\vec{\alpha}_0,\vec{\alpha}^*_0,t_0), \\
    \label{eq:action continuous}
    \mathcal{S}[\vec{\alpha},\vec{\eta}] = \int^t_{t_0}d\tau \left\{i\hbar\left(\vec{\eta}\cdot\frac{d\vec{\alpha}^*}{d\tau} - \vec{\eta}^*\cdot\frac{d\vec{\alpha}}{d\tau}\right) + H^e_{\vec{s}}(\vec{\psi}^+_{\vec{s}},\vec{\psi}^{+*}_{-\vec{s}}) - H^e_{\vec{s}}(\vec{\psi}^-_{-\vec{s}},\vec{\psi}^{-*}_{\vec{s}}) - i\hbar\mathcal{D}_{\vec{s}}(\vec{\psi}^+_{\vec{s}},\vec{\psi}^{+*}_{-\vec{s}},\vec{\psi}^-_{-\vec{s}},\vec{\psi}^{-*}_{\vec{s}})\right\},
\end{gather}
where $\mathcal{S}[\vec{\alpha},\vec{\eta}]$ is the action of the system.
At the boundaries, while the classical fields take $\vec{\alpha}(t_0) = \vec{\alpha}_0$ and $\vec{\alpha}(t) = \vec{\alpha}_{\rm f}$, the quantum fields are unconstrained.
Fig.~\ref{fig:Path integral short summary}(a) displays a schematic illustration for the path-integral representation for a system with a single degree of freedom.
When we choose a point in the phase space as an initial state, the point moves along infinite paths in the time evolution.
Eq.~\eqref{eq:path-integral representaiton continuous} says that we need to sum up all of the paths with multiplying the appropriate phase factor $e^{i\mathcal{S}[\vec{\alpha},\vec{\eta}]/\hbar}$.
Then, we can obtain the time-evolved $\vec{s}$-ordered quasiprobability distribution function $W_{\vec{s}}(\vec{\alpha}_{\rm f},\vec{\alpha}^*_{\rm f},t)$ by applying the same procedure for all initial points in the phase space and taking the ensemble average of the results weighted by $W_{\vec{s}}(\vec{\alpha}_0,\vec{\alpha}^*_0,t_0)$.


\subsection{\label{eq:Equation of motion in the phase space}Equation of motion in the phase space}
\begin{figure}[t]
	\centering 
	\includegraphics[width = \linewidth]{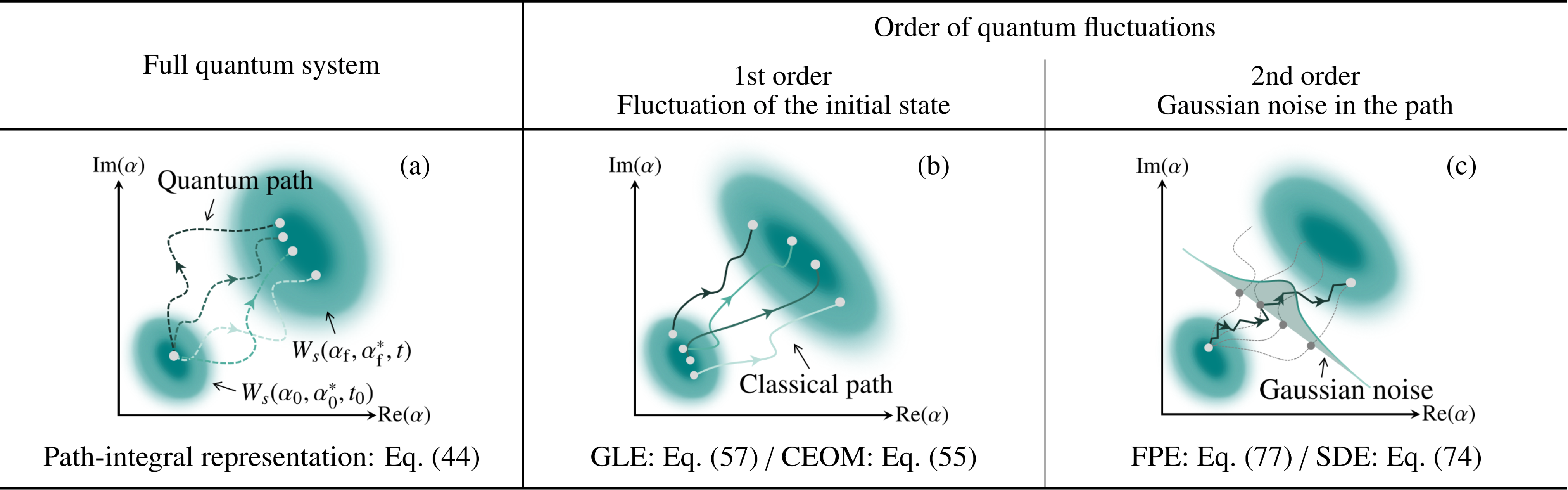}
	\caption{Schematic images of (a) the path-integral representation \eqref{eq:path-integral representaiton continuous} and (b)-(c) the approximations of the GKSL equation in the phase space.
(b) Within the first order of quantum fluctuations, the GKSL equation is approximated into the generalized Liouville equation (GLE) \cite{Gerlich,Steeb}, where each point distributed by the initial $\vec{s}$-ordered quasiprobability distribution function $W_{\vec{s}}(\vec{\alpha}_0,\vec{\alpha}^*_0,t_0)$ follows the classical equation of motion (CEOM), the equation of motion of the classical path.
(c) The effects of the second order of the quantum fluctuations are incorporated into the classical path as Gaussian noises,
where each point follows the stochastic differential equation (SDE).
Here, the GKSL equation is approximated into the Fokker-Planck
equation (FPE).
}
	\label{fig:Path integral short summary}
\end{figure}

By assuming small quantum fluctuations, we can expand the Lagrangian Eq.~\eqref{eq:action discrete} with respect to the quantum fields $\vec{\eta}_{j+1}$ order by order up to second order and derive the equations of motion within each order of quantum fluctuations.
We summarize the results of this section in Fig.~\ref{fig:Path integral short summary}(b) and (c).
Within the first order of quantum fluctuations, the integration with respect to the quantum fields gives rise to a classical path starting from each point in the phase space.
This is equivalent to approximate the GKSL equation into the generalized Liouville equation \cite{Gerlich,Steeb} [Fig.~\ref{fig:Path integral short summary}(b), Sec.~\ref{subsec:First order of quantum fluctuations}].
By taking the effects of the second order of quantum fluctuations, we obtain stochastic differential equations that the points in the phase space follow.
The same equation is obtained from the Fokker-Planck equation, which approximates the GKSL equation [Fig.~\ref{fig:Path integral short summary}(c), Sec.~\ref{subsec:Second order of quantum fluctuations}].
By analytically deriving these equations for general Hamiltonian and jump operators, we get the condition for the stochastic differential equation to be available in terms of the parameters in the Hamiltonian, jump operators, and $\vec{s}$.


\subsubsection{\label{subsec:First order of quantum fluctuations}First order of quantum fluctuations}
We derive the classical equations of motion, or the generalized Liouville equation, by expanding the Lagrangian Eq.~\eqref{eq:action discrete} with respect to the quantum fields up to first order.
The integration of the phase factor over the quantum fields leads to the Dirac delta function in the phase space, whose argument gives the trajectory of the classical fields.
We can perform the procedure in this section for arbitrary Hamiltonian, jump operators, and the parameters $s_m$ for $\forall m$.

By expanding $\mathcal{L}^{\vec{s}}_j$ in Eq.~\eqref{eq:action discrete} with respect to the quantum fields up to first order, we obtain
\begin{gather}
    \mathcal{L}^{\vec{s}}_j = \mathcal{L}^{\vec{s}(1)}_j + o(\vec{\eta}_{j+1}), \\
    \label{eq:First order of the action}
    \mathcal{L}^{\vec{s}(1)}_j = -\sum_{m=1}^{M}\eta^*_{m,j+1}\left\{i\hbar\left(\frac{\alpha_{m,j+1} - \alpha_{m,j}}{\Delta t}\right) - \frac{\partial H_{\vec{s}}(\vec{\alpha}_j,\vec{\alpha}^*_j)}{\partial\alpha^*_{m,j}} + i\hbar \mathcal{K}^{\vec{s}}_m(\vec{\alpha}_j,\vec{\alpha}^*_j)\right\} + {\rm c.c.},
\end{gather}
where ${\rm c.c.}$ denotes the complex conjugation of the proceeding term, $\mathcal{L}^{\vec{s}(1)}_j$ denotes the first-order contribution of the quantum fields in $\mathcal{L}^{\vec{s}}_j$, and $\mathcal{K}^{\vec{s}}_m$ is given by
\begin{align}
    \label{eq:definition of Ksm}
    \mathcal{K}^{\vec{s}}_m(\vec{\alpha},\vec{\alpha}^*) = - \frac{1}{2}\sum_k\gamma_k\left\{L^*_{k\vec{s}}(\vec{\alpha},\vec{\alpha}^*)\star_{\vec{s}}\frac{\partial L_{k\vec{s}}(\vec{\alpha},\vec{\alpha}^*)}{\partial\alpha^*_{m}} - \frac{\partial L^*_{k\vec{s}}(\vec{\alpha},\vec{\alpha}^*)}{\partial\alpha^*_{m}}\star_{\vec{s}}L_{k\vec{s}}(\vec{\alpha},\vec{\alpha}^*)\right\}.
\end{align}
Approximating $\mathcal{L}^{\vec{s}}_j$ in Eq.~\eqref{eq:path-integral representaiton discrete} with $\mathcal{L}^{\vec{s}(1)}_j$, we obtain
\begin{align}
    W_{\vec{s}}(\vec{\alpha}_{\rm f},\vec{\alpha}^*_{\rm f},t) &\approx \lim_{\Delta t \to 0}\prod_{j=0}^{N_t-1}\int\frac{d^2\vec{\alpha}_j d^2\vec{\eta}_{j+1}}{\pi^{2M}}e^{i\Delta t \mathcal{L}^{\vec{s}(1)}_j/\hbar} W_{\vec{s}}(\vec{\alpha}_0,\vec{\alpha}^*_0,t_0), \\
    \label{eq:calculation in the first order of quantum fluctuations}
    &= \lim_{\Delta t \to 0}\prod_{j=0}^{N_t-1}\int d^2\vec{\alpha}_j\prod_{m=1}^{M}\int\frac{ d^2\eta_{m,j+1}}{\pi^{2}}{\rm exp}\left\{\eta^*_{m,j+1}\left(\alpha_{m,j+1} - \alpha_{m,j} - \frac{\Delta t}{i\hbar}\frac{\partial H_{\vec{s}}}{\partial\alpha^*_{m,j}} + \Delta t \mathcal{K}^{\vec{s}}_m\right) - {\rm c.c.}\right\}W_{\vec{s}}(\vec{\alpha}_0,\vec{\alpha}^*_0,t_0).
\end{align}
Considering that the integration over $\vec{\eta}_{j+1}$ leads to the Dirac delta function
\begin{align}
    \label{eq:definitnio of the Dirac delta function}
    \int \frac{d^2\vec{\eta}}{\pi^{2M}}e^{\vec{\eta}^*\cdot\vec{\alpha}-\vec{\eta}\cdot\vec{\alpha}^*} = \prod_{m=1}^M\int \frac{d^2\eta_m}{\pi^{2}}e^{\eta_m^*\alpha_m-{\rm c.c.}} =\prod_{m=1}^M\delta^{(2)}(\alpha_m)=\prod_{m=1}^M\delta(\alpha_m^{\rm re})\delta(\alpha_m^{\rm im}),
\end{align}
we can rewrite Eq.~\eqref{eq:calculation in the first order of quantum fluctuations} as
\begin{align}
    \label{eq:formal solution of the generalized Liouville equation}
    W_{\vec{s}}(\vec{\alpha}_{\rm f},\vec{\alpha}^*_{\rm f},t) = \lim_{\Delta t\to 0}\prod_{j=0}^{N_t - 1}\int\frac{d^2\vec{\alpha}_j}{\pi^M}\varUpsilon^{(1)}_{\vec{s}}(\vec{\alpha}_{j+1},t_{j+1};\vec{\alpha}_j,t_j)W_{\vec{s}}(\vec{\alpha}_0,\vec{\alpha}^*_0,t_0),
\end{align}
where $\varUpsilon^{(1)}_{\vec{s}}(\vec{\alpha}_{j+1},t_{j+1};\vec{\alpha}_j,t_j)$ is the first-order propagator given by
\begin{align}
    \label{eq:first-order propagator}
    \varUpsilon^{(1)}_{\vec{s}}(\vec{\alpha}_{j+1},t_{j+1};\vec{\alpha}_j,t_j) = \prod_{m=1}^{M}\pi\delta^{(2)}\left(\alpha_{m,j+1} - \alpha_{m,j} - \frac{\Delta t}{i\hbar}\frac{\partial H_{\vec{s}}}{\partial\alpha^*_{m,j}} - \frac{\Delta t }{2}\sum_k\gamma_k\left(L^*_{k\vec{s}}\star_{\vec{s}}\frac{\partial L_{k\vec{s}}}{\partial\alpha^*_{m,j}} - \frac{\partial L^*_{k\vec{s}}}{\partial\alpha^*_{m,j}}\star_{\vec{s}}L_{k\vec{s}}\right) \right).
\end{align}
Eq.~\eqref{eq:formal solution of the generalized Liouville equation} is a formal solution of the GKSL equation within the first order of quantum fluctuations.
We display the schematic illustration of Eq.~\eqref{eq:formal solution of the generalized Liouville equation} in Fig.~\ref{fig:Path integral short summary}(b).
The points in the phase space, which are initially distributed according to the $\vec{s}$-ordered quasiprobability distribution function $W_{\vec{s}}(\vec{\alpha}_0,\vec{\alpha}^*_0,t_0)$, follow the classical path determined by the argument of the Dirac delta function in the right-hand side of Eq.~\eqref{eq:first-order propagator}: 
\begin{align}
    \label{eq:clasical equation of motion discrete}
    \alpha_{m,j+1} - \alpha_{m,j} = \frac{\Delta t}{i\hbar}\frac{\partial H_{\vec{s}}(\vec{\alpha}_j,\vec{\alpha}^*_j)}{\partial\alpha^*_{m,j}} + \frac{\Delta t }{2}\sum_k\gamma_k\left\{L^*_{k\vec{s}}(\vec{\alpha}_j,\vec{\alpha}^*_j)\star_{\vec{s}}\frac{\partial L_{k\vec{s}}(\vec{\alpha}_j,\vec{\alpha}^*_j)}{\partial\alpha^*_{m,j}} - \frac{\partial L^*_{k\vec{s}}(\vec{\alpha}_j,\vec{\alpha}^*_j)}{\partial\alpha^*_{m,j}}\star_{\vec{s}}L_{k\vec{s}}(\vec{\alpha}_j,\vec{\alpha}^*_j)\right\}.
\end{align}
By taking the continuous limit of Eq.~\eqref{eq:clasical equation of motion discrete}, we obtain the classical equation of motion for $\alpha_m$:
\begin{align}
    \label{eq:classical equation of motion}
    i\hbar\frac{d\alpha_m}{dt} = \frac{\partial H_{\vec{s}}(\vec{\alpha},\vec{\alpha}^*)}{\partial\alpha^*_{m}} + \frac{i\hbar}{2}\sum_k\gamma_k\left\{L^*_{k\vec{s}}(\vec{\alpha},\vec{\alpha}^*)\star_{\vec{s}}\frac{\partial L_{k\vec{s}}(\vec{\alpha},\vec{\alpha}^*)}{\partial\alpha^*_m} - \frac{\partial L^*_{k\vec{s}}(\vec{\alpha},\vec{\alpha}^*)}{\partial\alpha^*_m}\star_{\vec{s}}L_{k\vec{s}}(\vec{\alpha},\vec{\alpha}^*)\right\}.
\end{align}
From Eqs.~\eqref{eq:physical quantity in the s-ordered phase space} and \eqref{eq:formal solution of the generalized Liouville equation}, we obtain the path-integral representation of the physical quantity within the first order of quantum fluctuations as
\begin{align}
    \label{eq:physical quantity first order}
    \braket{\hat{A}(t)} = \int\frac{d^2\vec{\alpha}_{\rm f}}{\pi^M}A_{\vec{s}}(\vec{\alpha}_{\rm f},\vec{\alpha}^*_{\rm f})\lim_{\Delta t\to 0}\prod_{j=0}^{N_t - 1}\int\frac{d^2\vec{\alpha}_j}{\pi^M}\varUpsilon^{(1)}_{\vec{s}}(\vec{\alpha}_{j+1},t_{j+1};\vec{\alpha}_j,t_j)W_{\vec{s}}(\vec{\alpha}_0,\vec{\alpha}^*_0,t_0).
\end{align}
This means that we can calculate $\braket{\hat{A}(t)}$ by a Monte Carlo simulation:
We iteratively solve the classical equation of motion Eq.~\eqref{eq:classical equation of motion} for $\forall m$ with various initial conditions stochastically sampled from $W_{\vec{s}}(\vec{\alpha}_0,\vec{\alpha}^*_0,t_0)$, calculate $A_{\vec{s}}(\vec{\alpha}_{\rm f},\vec{\alpha}^*_{\rm f})$, and take the ensemble average over the results.

We can also show that within the first order of quantum fluctuations, the GKSL equation is transformed into the following generalized Liouville equation:
\begin{align}
    \label{eq:generalized Liouville equation}
    i\hbar\frac{dW_{\vec{s}}(\vec{\alpha},\vec{\alpha}^*,t)}{dt} = -\sum_{m=1}^{M}\left[\frac{\partial}{\partial\alpha_m} \left\{\frac{\partial H_{\vec{s}}}{\partial\alpha^*_m} + \frac{i\hbar}{2}\sum_k\gamma_k\left(L^*_{k\vec{s}}\star_{\vec{s}}\frac{\partial L_{k\vec{s}}}{\partial\alpha^*_m} - \frac{\partial L^*_{k\vec{s}}}{\partial\alpha^*_m}\star_{\vec{s}}L_{k\vec{s}}\right) \right\}W_{\vec{s}}(\vec{\alpha},\vec{\alpha}^*,t)\right] - {\rm c.c.}
\end{align}
See \ref{appendix:Generalized Liouville equation} for the derivation.


\subsubsection{\label{subsec:Second order of quantum fluctuations}Second order of quantum fluctuations}
Next, we expand the Lagrangian Eq.~\eqref{eq:action discrete} with respect to the quantum fields up to second order.
In order to perform the integration with respect to the quantum fields, we perform the Hubbard-Stratonovich transformation with introducing auxiliary fields, which give the stochastic process in the equation of motion in the phase space.
Here, the Hubbard-Stratonovich transformation is not always feasible.
We show that the feasible condition of the Hubbard-Stratonovich transformation is equivalent to the positive-semidefiniteness condition of the diffusion matrix in the Fokker-Planck equation.

Expansion of $\mathcal{L}^{\vec{s}}_j$ up to the second order of the quantum fields reads
\begin{gather}
    \mathcal{L}^{\vec{s}}_j = \mathcal{L}^{\vec{s}(1)}_j + \mathcal{L}^{\vec{s}(2)}_j + o(\vec{\eta}^2_{j+1}), \\
    \mathcal{L}^{\vec{s}(2)}_j = i\hbar\sum_{m,n = 1}^M\left\{\lambda^{\vec{s}*}_{mn}(\vec{\alpha}_j,\vec{\alpha}^*_j)\eta_{m,j+1}\eta_{n,j+1} + 2\Lambda^{\vec{s}}_{mn}(\vec{\alpha}_j,\vec{\alpha}^*_j)\eta^*_{m,j+1}\eta_{n,j+1} + \lambda^{\vec{s}}_{mn}(\vec{\alpha}_j,\vec{\alpha}^*_j)\eta^*_{m,j+1}\eta^*_{n,j+1}\right\},
\end{gather}
where $\mathcal{L}^{\vec{s}(1)}_j$ is given by Eq.~\eqref{eq:First order of the action}, $\mathcal{L}^{\vec{s}(2)}_j$ denotes the second-order contribution of the quantum fields in $\mathcal{L}^{\vec{s}}_j$, and $\lambda^{\vec{s}}_{mn}(\vec{\alpha},\vec{\alpha}^*) \in \mathbb{C}$ and $\Lambda^{\vec{s}}_{mn}(\vec{\alpha},\vec{\alpha}^*) \in \mathbb{C}$ are defined as
\begin{align}
    \label{eq:definitnion of lambdamn}
    &\lambda^{\vec{s}}_{mn}(\vec{\alpha},\vec{\alpha}^*) = \sum_k\frac{\gamma_k}{4}\left\{\left(1 - \frac{s_m - s_n}{2}\right)\frac{\partial L^*_{k\vec{s}}(\vec{\alpha},\vec{\alpha}^*)}{\partial\alpha^*_{m}}\star_{\vec{s}} \frac{\partial L_{k\vec{s}}(\vec{\alpha},\vec{\alpha}^*)}{\partial\alpha^*_{n}} + \left(1 + \frac{s_m - s_n}{2}\right)\frac{\partial L^*_{k\vec{s}}(\vec{\alpha},\vec{\alpha}^*)}{\partial\alpha^*_{n}}\star_{\vec{s}} \frac{\partial L_{k\vec{s}}(\vec{\alpha},\vec{\alpha}^*)}{\partial\alpha^*_{m}}\right\} \nonumber\\
    &\hphantom{\lambda^{\vec{s}}_{mn}(\vec{\alpha},\vec{\alpha}^*) =}- \frac{s_m + s_n}{2}\left[\sum_k\frac{\gamma_k}{4}\left\{L^*_{k\vec{s}}(\vec{\alpha},\vec{\alpha}^*)\star_{\vec{s}}\frac{\partial^2 L_{k\vec{s}}(\vec{\alpha},\vec{\alpha}^*)}{\partial\alpha^*_{m}\partial\alpha^*_{n}} - \frac{\partial^2 L^*_{k\vec{s}}(\vec{\alpha},\vec{\alpha}^*)}{\partial\alpha^*_{m}\partial\alpha^*_{n}}\star_{\vec{s}} L_{k\vec{s}}(\vec{\alpha},\vec{\alpha}^*)\right\} - \frac{i}{2\hbar}\frac{\partial^2 H_{\vec{s}}(\vec{\alpha},\vec{\alpha}^*)}{\partial\alpha^*_{m}\partial\alpha^*_{n}}\right], \\
    \label{eq:definitnion of Lambdamn}
    &\Lambda^{\vec{s}}_{mn}(\vec{\alpha},\vec{\alpha}^*) = \sum_k\frac{\gamma_k}{4}\left\{\left(1-\frac{s_m + s_n}{2}\right)\frac{\partial L^*_{k\vec{s}}(\vec{\alpha},\vec{\alpha}^*)}{\partial\alpha^*_{m}}\star_{\vec{s}} \frac{\partial L_{k\vec{s}}(\vec{\alpha},\vec{\alpha}^*)}{\partial\alpha_{n}} + \left(1+\frac{s_m + s_n}{2}\right)\frac{\partial L^*_{k\vec{s}}(\vec{\alpha},\vec{\alpha}^*)}{\partial\alpha_{n}}\star_{\vec{s}} \frac{\partial L_{k\vec{s}}(\vec{\alpha},\vec{\alpha}^*)}{\partial\alpha^*_{m}}\right\} \nonumber\\
    &\hphantom{\Lambda^{\vec{s}}_{mn}(\vec{\alpha},\vec{\alpha}^*) =} - \frac{s_m - s_n}{2}\left[\sum_k\frac{\gamma_k}{4}\left\{L^*_{k\vec{s}}(\vec{\alpha},\vec{\alpha}^*)\star_{\vec{s}}\frac{\partial^2 L_{k\vec{s}}(\vec{\alpha},\vec{\alpha}^*)}{\partial\alpha^*_{m}\partial\alpha_{n}} - \frac{\partial^2 L^*_{k\vec{s}}(\vec{\alpha},\vec{\alpha}^*)}{\partial\alpha^*_{m}\partial\alpha_{n}}\star_{\vec{s}} L_{k\vec{s}}(\vec{\alpha},\vec{\alpha}^*)\right\} - \frac{i}{2\hbar}\frac{\partial^2 H_{\vec{s}}(\vec{\alpha},\vec{\alpha}^*)}{\partial\alpha^*_{m}\partial\alpha_{n}}\right].
\end{align}
Here, when we refer to the terms $\gamma_k{\hat{L}}_k\hat{\rho}(t){\hat{L}}_k^\dagger$ and $(\gamma_k/2)[{\hat{L}}_k^\dagger{\hat{L}}_k,\hat{\rho}\left(t\right)]_+$ in the GKSL equation~\eqref{eq:def of GKSL equation} as the quantum-jump and quantum-diffusion terms, respectively,
the both terms equally contribute to $\mathcal{L}^{\vec{s}(2)}_j$.
On the other hand, only the quantum-jump term affect $\mathcal{L}^{\vec{s}(1)}_j$.

Approximating $\mathcal{L}^{\vec{s}}_j$ in Eq.~\eqref{eq:path-integral representaiton discrete} with $\mathcal{L}^{\vec{s}(1)}_j + \mathcal{L}^{\vec{s}(2)}_j$, we obtain
\begin{align}
    W_{\vec{s}}(\vec{\alpha}_{\rm f},\vec{\alpha}^*_{\rm f},t) &\approx \lim_{\Delta t \to 0}\prod_{j=0}^{N_t-1}\int\frac{d^2\vec{\alpha}_j d^2\vec{\eta}_{j+1}}{\pi^{2M}}e^{i\Delta t \mathcal{L}^{\vec{s}(1)}_j/\hbar} e^{i\Delta t \mathcal{L}^{\vec{s}(2)}_j/\hbar}W_{\vec{s}}(\vec{\alpha}_0,\vec{\alpha}^*_0,t_0) \\
    \label{eq:calculation in the second order of quantum fluctuations 1}
    &= \lim_{\Delta t \to 0}\prod_{j=0}^{N_t-1}\int\frac{d^2\vec{\alpha}_j d^2\vec{\eta}_{j+1}}{\pi^{2M}}e^{i\Delta t \mathcal{L}^{\vec{s}(1)}_j/\hbar}{\rm exp}\left\{-\frac{\Delta t}{2}
    \begin{bmatrix}
        \vec{\eta}_{j+1}^{*{\rm T}},\vec{\eta}^{\rm T}_{j+1}
    \end{bmatrix}
    \bm{\mathcal{A}}^{\vec{s}}(\vec{\alpha}_j,\vec{\alpha}^*_j)
    \begin{bmatrix}
        \vec{\eta}_{j+1} \\
        \vec{\eta}_{j+1}^{*}
    \end{bmatrix}
    \right\}W_{\vec{s}}(\vec{\alpha}_0,\vec{\alpha}^*_0,t_0),
\end{align}
where $\bm{\mathcal{A}}^{\vec{s}}(\vec{\alpha},\vec{\alpha}^*)$ is a $2M\times 2M$ Hermitian matrix given by
\begin{align}
    \label{eq:definitnioa of the diffusion matrix A}
    \bm{\mathcal{A}}^{\vec{s}}(\vec{\alpha},\vec{\alpha}^*) = 2
    \begin{bmatrix}
        \bm{\Lambda}^{\vec{s}}(\vec{\alpha},\vec{\alpha}^*) & \bm{\lambda}^{\vec{s}}(\vec{\alpha},\vec{\alpha}^*) \\
        \bm{\lambda}^{\vec{s}*}(\vec{\alpha},\vec{\alpha}^*) & \bm{\Lambda}^{\vec{s}*}(\vec{\alpha},\vec{\alpha}^*)
    \end{bmatrix}.
\end{align}
Here, $\bm{\Lambda}^{\vec{s}}(\vec{\alpha},\vec{\alpha}^*)$ and $\bm{\lambda}^{\vec{s}}(\vec{\alpha},\vec{\alpha}^*)$ are $M\times M$ Hermitian and symmetric matrices whose matrix elements are $\Lambda^{\vec{s}}_{mn}(\vec{\alpha},\vec{\alpha}^*)$ and $\lambda^{\vec{s}}_{mn}(\vec{\alpha},\vec{\alpha}^*)$, respectively.

In order to perform the integration with respect to the quantum fields in Eq.~\eqref{eq:calculation in the second order of quantum fluctuations 1}, we perform the Hubbard-Stratonovich transformation by introducing auxiliary fields $\Delta\overrightarrow{\mathcal{W}} \in \mathbb{R}^{2M}$ as
\begin{align}
    \label{eq:Hubbard-Stratonovich transformation}
    {\rm exp}\left\{-\frac{\Delta t}{2} 
    \begin{bmatrix}
        \vec{\eta}_{j+1}^{*{\rm T}},\vec{\eta}^{\rm T}_{j+1}
    \end{bmatrix}
    \bm{\mathcal{A}}^{\vec{s}}(\vec{\alpha}_j,\vec{\alpha}^*_j)
    \begin{bmatrix}
        \vec{\eta}_{j+1} \\
        \vec{\eta}_{j+1}^*
    \end{bmatrix}
    \right\}
    =
    \prod_{\mu =1}^{2M}\int_{-\infty}^{\infty} d\Delta\mathcal{W}_{\mu}\frac{e^{-\Delta\mathcal{W}^2_{\mu}/(2\Delta t)}}{\sqrt{2\pi\Delta t}}\prod_{m=1}^{M}{\rm exp}\left(\eta^*_{m,j+1}\left[i\bm{\mathcal{U}}^{\vec{s}}(\vec{\alpha}_j,\vec{\alpha}^*_j)\sqrt{\bm{\mathcal{A}}^{\vec{s}}_{\rm diag}(\vec{\alpha}_j,\vec{\alpha}^*_j)}\bm{\mathcal{Q}}\Delta \overrightarrow{\mathcal{W}}\right]_m - {\rm c.c.}\right),
\end{align}
which is feasible when
\begin{align}
    \label{eq:feasible condition of the Hubbard-Stratonovich transformation}
    \bm{\mathcal{A}}^{\vec{s}}(\vec{\alpha}_j,\vec{\alpha}^*_j) \succeq 0,
\end{align}
i.e., $\bm{\mathcal{A}}^{\vec{s}}(\vec{\alpha}_j,\vec{\alpha}^*_j)$ is a positive-semidefinite matrix.
The derivation of Eq.~\eqref{eq:Hubbard-Stratonovich transformation} is given in \ref{appendix:Hubbard-Stratonovich transformation}.
In Eq.~\eqref{eq:Hubbard-Stratonovich transformation},
$\bm{\mathcal{Q}}$ is an arbitrary $2M\times 2M$ orthogonal matrix \cite{Gardinerstochas},
and $\bm{\mathcal{U}}^{\vec{s}}(\vec{\alpha},\vec{\alpha}^*)$ is a $2M\times 2M$ unitary matrix that diagonalizes $\bm{\mathcal{A}}^{\vec{s}}$:
\begin{align}
\bm{\mathcal{U}}^{\vec{s}\dagger} \bm{\mathcal{A}}^{\vec{s}} \bm{\mathcal{U}}^{\vec{s}} = \bm{\mathcal{A}}^{\vec{s}}_{\rm diag},
\end{align}
where $\bm{\mathcal{A}}^{\vec{s}}_{\rm diag}$ is a $2M \times 2M$ diagonal matrix having the eigenvalues of $\bm{\mathcal{A}}^{\vec{s}}$ on its diagonal entries.
Here, we restrict $\bm{\mathcal{U}}^{\vec{s}}$ to those that can be decomposed into 
\begin{align}
    \bm{\mathcal{U}}^{\vec{s}}(\vec{\alpha},\vec{\alpha}^*) = \bm{\mathcal{P}}\bm{\mathcal{V}}^{\vec{s}}(\vec{\alpha},\vec{\alpha}^*),
\end{align}
where $\bm{\mathcal{P}}$ is a $2M\times 2M$ unitary matrix:
\begin{align}
    \bm{\mathcal{P}} = \frac{1}{\sqrt{2}}
    \begin{bmatrix}
        \bm{1} & i\bm{1} \\
        \bm{1} & -i\bm{1}
    \end{bmatrix}
\end{align}
with $\bm{1}$ being the $M\times M$ identity matrix and $\bm{\mathcal{V}}^{\vec{s}}(\vec{\alpha},\vec{\alpha}^*)$ a $2M\times 2M$ orthogonal matrix: Such an orthogonal matrix $\bm{\mathcal{V}}^{\vec{s}}$ always exists because $\bm{\mathcal{P}}^{\dagger}\bm{\mathcal{A}}^{\vec{s}}\bm{\mathcal{P}}$ is a real symmetric matrix (see \ref{appendix:Hubbard-Stratonovich transformation} for details).

Substituting Eqs.~\eqref{eq:First order of the action} and \eqref{eq:Hubbard-Stratonovich transformation} into Eq.~\eqref{eq:calculation in the second order of quantum fluctuations 1}, we obtain
\begin{align}
    \label{eq:calculation in the second order of quantum fluctuations 2}
    W_{\vec{s}}(\vec{\alpha}_{\rm f},\vec{\alpha}^*_{\rm f},t) =& \lim_{\Delta t \to 0}\prod_{j=0}^{N_t-1}\int d^2\vec{\alpha}_j
    \prod_{\mu = 1}^{2M}\int_{-\infty}^{\infty} d\Delta\mathcal{W}_{\mu}\frac{e^{-\Delta\mathcal{W}^2_{\mu}/(2\Delta t)}}{\sqrt{2\pi\Delta t}}\nonumber \\
    &\times\prod_{m=1}^{M}\int\frac{ d^2\eta_{m,j+1}}{\pi^{2}}{\rm exp}\left\{\eta^*_{m,j+1}\left(\alpha_{m,j+1} - \alpha_{m,j} - \frac{\Delta t}{i\hbar}\frac{\partial H_{\vec{s}}}{\partial\alpha^*_{m,j}} + \Delta t \mathcal{K}^{\vec{s}}_m + \left[i\bm{\mathcal{U}}^{\vec{s}}\sqrt{\bm{\mathcal{A}}^{\vec{s}}_{\rm diag}}\bm{\mathcal{Q}}\Delta \overrightarrow{\mathcal{W}}\right]_m\right) - {\rm c.c.}\right\}W_{\vec{s}}(\vec{\alpha}_0,\vec{\alpha}^*_0,t_0).
\end{align}
Integrating out the quantum fields by using Eq.~\eqref{eq:definitnio of the Dirac delta function}, we can rewrite Eq.~\eqref{eq:calculation in the second order of quantum fluctuations 2} as
\begin{align}
    \label{eq:formal solution of the Fokker-Planck equation}
    W_{\vec{s}}(\vec{\alpha}_{\rm f},\vec{\alpha}^*_{\rm f},t) = \lim_{\Delta t\to 0}\prod_{j=0}^{N_t - 1}\int\frac{d^2\vec{\alpha}_j}{\pi^M}\varUpsilon^{(2)}_{\vec{s}}(\vec{\alpha}_{j+1},t_{j+1};\vec{\alpha}_j,t_j)W_{\vec{s}}(\vec{\alpha}_0,\vec{\alpha}^*_0,t_0),
\end{align}
where$\varUpsilon^{(2)}_{\vec{s}}(\vec{\alpha}_{j+1},t_{j+1};\vec{\alpha}_j,t_j)$ is a second-order propagator given by
\begin{align}
    \label{eq:second-order propagator}
    \varUpsilon^{(2)}_{\vec{s}}(\vec{\alpha}_{j+1},t_{j+1};\vec{\alpha}_j,t_j) = \prod_{\mu = 1}^{2M}\int_{-\infty}^{\infty} d\Delta\mathcal{W}_{\mu}\frac{e^{-\Delta\mathcal{W}^2_{\mu}/(2\Delta t)}}{\sqrt{2\pi\Delta t}}\prod_{m=1}^{M}\pi\delta^{(2)}\left(\alpha_{m,j+1} - \alpha_{m,j} - \frac{\Delta t}{i\hbar}\frac{\partial H_{\vec{s}}}{\partial\alpha^*_{m,j}} + \Delta t \mathcal{K}^{\vec{s}}_m + \left[i\bm{\mathcal{U}}^{\vec{s}}\sqrt{\bm{\mathcal{A}}^{\vec{s}}_{\rm diag}}\bm{\mathcal{Q}}\Delta \overrightarrow{\mathcal{W}}\right]_m \right).
\end{align}
Eq.~\eqref{eq:formal solution of the Fokker-Planck equation} is a formal solution of the GKSL equation within the second-order approximation.
We illustrate the schematic image of Eq.~\eqref{eq:formal solution of the Fokker-Planck equation} in Fig.~\ref{fig:Path integral short summary}(c):
Initial points distributed according to $W_{\vec{s}}(\vec{\alpha}_0,\vec{\alpha}^*_0,t_0)$ move in the phase space in time along the path obtained by the stochastic differential equation given as the argument of the Dirac delta function in the right-hand side of Eq.~\eqref{eq:second-order propagator}:
\begin{align}
    \label{eq:stochastic differential equation discrete}
    \alpha_{m,j+1} - \alpha_{m,j} = \frac{\Delta t}{i\hbar}\frac{\partial H_{\vec{s}}(\vec{\alpha}_j,\vec{\alpha}^*_j)}{\partial\alpha^*_{m,j}} - \Delta t \mathcal{K}^{\vec{s}}_m(\vec{\alpha}_j,\vec{\alpha}^*_j) + \left[i\bm{\mathcal{U}}^{\vec{s}}(\vec{\alpha}_j,\vec{\alpha}^*_j)\sqrt{\bm{\mathcal{A}}^{\vec{s}}_{\rm diag}(\vec{\alpha}_j,\vec{\alpha}^*_j)}\bm{\mathcal{Q}}\Delta \overrightarrow{\mathcal{W}}\right]_m.
\end{align}
Substituting Eq.~\eqref{eq:definition of Ksm} and taking the continuous limit of Eq.~\eqref{eq:stochastic differential equation discrete}, we obtain
\begin{align}
    \label{eq:stochastic differential equation}
    i\hbar d\alpha_m = \left[\frac{\partial H_{\vec{s}}}{\partial\alpha^*_{m}} + \frac{i\hbar}{2}\sum_k\gamma_k\left(L^*_{k\vec{s}}\star_{\vec{s}}\frac{\partial L_{k\vec{s}}}{\partial\alpha^*_m} - \frac{\partial L^*_{k\vec{s}}}{\partial\alpha^*_m}\star_{\vec{s}}L_{k\vec{s}}\right)\right]dt + i\hbar\left[i\bm{\mathcal{U}}^{\vec{s}}(\vec{\alpha},\vec{\alpha}^*)\sqrt{\bm{\mathcal{A}}^{\vec{s}}_{\rm diag}(\vec{\alpha},\vec{\alpha}^*)}\bm{\mathcal{Q}}\cdot d\overrightarrow{\mathcal{W}}(t)\right]_m,
\end{align}
where $\cdot$ denotes the Ito product \cite{Risken} and $\overrightarrow{\mathcal{W}}(t) \in \mathbb{R}^{2M}$ is a real stochastic process vector whose components are Wiener processes and independent of each other, i.e., the changes of $\Delta\mathcal{W}_{\mu} = \mathcal{W}_{\mu}(t + \Delta t) - \mathcal{W}_{\mu}(t)$ in the time interval $\Delta t$ obey the following Gaussian distribution function:
\begin{align}
    \label{eq:complex_stochastic_process_real}
    P\left[\Delta \mathcal{W}_{\mu}\right] &= \frac{1}{\sqrt{2\pi\Delta t}}{\rm exp}\left(-\frac{\Delta \mathcal{W}_{\mu}^2}{2\Delta t}\right).
\end{align}
Here, we note that the stochastic term in Eq.~\eqref{eq:stochastic differential equation} is not uniquely determined because there is an arbitrariness in the choice of $\bm{\mathcal{U}}^{\vec{s}}$, $\bm{\mathcal{A}}^{\vec{s}}_{\rm diag}$ (the order of the eigenvalues), and $\bm{\mathcal{Q}}$.
We can choose them at our convenience.
Below, we choose $\bm{\mathcal{Q}}$ as the identity matrix.
Using Eqs.~\eqref{eq:physical quantity in the s-ordered phase space} and \eqref{eq:formal solution of the Fokker-Planck equation}, we obtain the path-integral representation of the physical quantity within the second order of quantum fluctuations:
\begin{align}
    \label{eq:physical quantity second order}
    \braket{\hat{A}(t)} = \int\frac{d^2\vec{\alpha}_{\rm f}}{\pi^M}A_{\vec{s}}(\vec{\alpha}_{\rm f},\vec{\alpha}^*_{\rm f})\lim_{\Delta t\to 0}\prod_{j=0}^{N_t - 1}\int\frac{d^2\vec{\alpha}_j}{\pi^M}\varUpsilon^{(2)}_{\vec{s}}(\vec{\alpha}_{j+1},t_{j+1};\vec{\alpha}_j,t_j)W_{\vec{s}}(\vec{\alpha}_0,\vec{\alpha}^*_0,t_0),
\end{align}
from which we can calculate $\braket{\hat{A}(t)}$ by the Monte Carlo simulation of the stochastic differential equation~\eqref{eq:stochastic differential equation}.
Here, we note that in the calculation of the stochastic differential equations~\eqref{eq:stochastic differential equation}, we need to numerically diagonalize the matrix $\bm{\mathcal{P}}^{\dagger}\bm{\mathcal{A}}^{\vec{s}}\bm{\mathcal{P}}$ in each time steps to obtain $\bm{\mathcal{U}}^{\vec{s}}$ and $\bm{\mathcal{A}}^{\vec{s}}_{\rm diag}$.

We can also see that the expansion of the GKSL equation up to second-order quantum fluctuations leads to the Fokker-Planck equation, which is given by
\begin{align}
    \label{eq:Fokker-Planck equation}
    i\hbar\frac{dW_{\vec{s}}(\vec{\alpha},\vec{\alpha}^*,t)}{dt} =& -\sum_{m=1}^{M}\frac{\partial}{\partial\alpha_m}\left[ \left\{\frac{\partial H_{\vec{s}}}{\partial\alpha^*_m} + \frac{i\hbar}{2}\sum_k\gamma_k\left(L^*_{k\vec{s}}\star_{\vec{s}}\frac{\partial L_{k\vec{s}}}{\partial\alpha^*_m} - \frac{\partial L^*_{k\vec{s}}}{\partial\alpha^*_m}\star_{\vec{s}}L_{k\vec{s}}\right) \right\}W_{\vec{s}}(\vec{\alpha},\vec{\alpha}^*,t)\right] \nonumber\\
    & -i\hbar\sum_{m,n=1}^M\frac{\partial^2}{\partial\alpha_m\partial\alpha_n}\left[\lambda^{\vec{s}}_{mn}W_{\vec{s}}(\vec{\alpha},\vec{\alpha}^*,t)\right] + i\hbar\sum_{m,n=1}^M\frac{\partial^2}{\partial\alpha_m\partial\alpha^*_n}\left[\Lambda^{\vec{s}}_{mn}W_{\vec{s}}(\vec{\alpha},\vec{\alpha}^*,t)\right] - {\rm c.c.}
\end{align}
We provide the derivation of Eq.~\eqref{eq:Fokker-Planck equation} in \ref{appendix:Fokker-Planck equation}.
After some calculation, the diffusion matrix in Eq.~\eqref{eq:Fokker-Planck equation} is found to be $\bm{\mathcal{P}}^{\dagger}\bm{\mathcal{A}}^{\vec{s}}\bm{\mathcal{P}}$.
Since $\bm{\mathcal{P}}$ is the unitary matrix,
the positive-semidefiniteness condition of the diffusion matrix $\bm{\mathcal{P}}^{\dagger}\bm{\mathcal{A}}^{\vec{s}}\bm{\mathcal{P}} \succeq 0$ is equivalent to the condition to perform the Hubbard-Stratonovich transformation Eq.~\eqref{eq:feasible condition of the Hubbard-Stratonovich transformation}:
Only when Eq.~\eqref{eq:feasible condition of the Hubbard-Stratonovich transformation} is satisfied, Eq.~\eqref{eq:Fokker-Planck equation} reduces to the stochastic differential equation~\eqref{eq:stochastic differential equation}.

We make some remarks on three special cases.
(i) The case of an isolated system, i.e., $\gamma_k=0$ for $\forall k$. When we choose $s_m=0$ for $\forall m$, we obtain $\bm{\mathcal{A}}^{\vec{0}}=\bm{0}$ even when the Hamiltonian includes two- or higher-body interactions, that is, the stochastic term disappears in Eq.~\eqref{eq:stochastic differential equation}.
This is the well-known result of the TWA for an isolated system.
On the other hand, when we choose the same $s_m=1$ or $-1$ for $\forall m$, we obtain $\bm{\Lambda}^{\vec{s}}=\bm{0}$.
It follows that $\bm{\mathcal{A}}^{\vec{s}}$ has a particle-hole symmetry and always has pairs of positive and negative eigenvalues with the same absolute value.
The existence of negative eigenvalues means that the Fokker-Planck equations for the Glauber-Sudarshan P and Husimi Q functions do not reduce to the stochastic differential equations.
The only exception is the free boson Hamiltonian.
(ii) The case of the Wigner function, i.e., $s_m=0$ for $\forall m$.
In this case, the matrix elements of $\bm{\mathcal{A}}^{\vec{0}}$ do not include the terms depending on the Hamiltonian.
Thus, the condition Eq.~\eqref{eq:feasible condition of the Hubbard-Stratonovich transformation} for obtaining the stochastic differential equation only depends on the details of the jump operators.
(iii) The case when the matrices $\bm{\lambda}^{\vec{s}}$ and $\bm{\Lambda}^{\vec{s}}$ are diagonal, i.e., 
\begin{align}
    \label{eq:lambda and Lambda when the jump operators do not couple different degrees of feedom}
    \lambda^{\vec{s}}_{mn}(\vec{\alpha},\vec{\alpha}^*) =
    \begin{cases}
        \lambda^{s_m}_{mm}(\alpha_m,\alpha^*_m) & (n=m) \\
        0 & (n\neq m)
    \end{cases},\quad
    \Lambda^{\vec{s}}_{mn}(\vec{\alpha},\vec{\alpha}^*) =
    \begin{cases}
        \Lambda^{s_m}_{mm}(\alpha_m,\alpha^*_m) & (n=m) \\
        0 & (n\neq m)
    \end{cases}.
\end{align}
This condition is satisfied when the jump operators do not couple different degrees of freedom and the Hamiltonian satisfies
\begin{align}
    \label{eq:condition for Hamiltonian do not couple different degrees of freedom}
    (s_m + s_n)\frac{\partial H_{\vec{s}}}{\partial\alpha_m\partial\alpha_n} = (s_m - s_n)\frac{\partial H_{\vec{s}}}{\partial\alpha_m\partial\alpha_n} = 0~\text{for}~\forall m,n\neq m.
\end{align}
Under the restriction Eq.~\eqref{eq:lambda and Lambda when the jump operators do not couple different degrees of feedom}, we can rewrite the Fokker-Planck equation~\eqref{eq:Fokker-Planck equation} as
\begin{align}
    \label{eq:Fokker-Planck equation when jumo operators do not couple different degrees of feeedom}
    i\hbar\frac{dW_{\vec{s}}(\vec{\alpha},\vec{\alpha}^*,t)}{dt} =& -\sum_{m=1}^{M}\frac{\partial}{\partial\alpha_m}\left[ \left\{\frac{\partial H_{\vec{s}}}{\partial\alpha^*_m} + \frac{i\hbar}{2}\sum_{k}\gamma_{k}\left(L^*_{k\vec{s}}\star_{\vec{s}}\frac{\partial L_{k\vec{s}}}{\partial\alpha^*_m} - \frac{\partial L^*_{k\vec{s}}}{\partial\alpha^*_m}\star_{\vec{s}}L_{k\vec{s}}\right) \right\}W_{\vec{s}}(\vec{\alpha},\vec{\alpha}^*,t)\right] \nonumber\\
    & -i\hbar\sum_{m=1}^M\frac{\partial^2}{\partial\alpha_m\partial\alpha_m}\left[\lambda^{s_m}_{mm}W_{\vec{s}}(\vec{\alpha},\vec{\alpha}^*,t)\right] + i\hbar\sum_{m=1}^M\frac{\partial^2}{\partial\alpha_m\partial\alpha^*_m}\left[\Lambda^{s_m}_{mm}W_{\vec{s}}(\vec{\alpha},\vec{\alpha}^*,t)\right] - {\rm c.c.}
\end{align}
Here, when $\lambda^{s_m}_{mm}$ and $\Lambda^{s_m}_{mm}$ satisfies the condition:
\begin{align}
    \label{eq:condition for obtaining stochastic differential equaitons when the jump operators do not couple different degrees of feedom}
    \Lambda^{s_m}_{mm} \geq |\lambda^{s_m}_{mm}|
\end{align}
for each $m=1,2,\cdots,M$, we can perform the Hubbard-Stratonovich transformation and obtain the following stochastic differential equation:
\begin{align}
    \label{eq:stochastic differential equation simplified version}
    i\hbar d\alpha_m = \left[\frac{\partial H_{\vec{s}}}{\partial\alpha^*_{m}} + \frac{i\hbar}{2}\sum_{k}\gamma_{k}\left(L^*_{k\vec{s}}\star_{\vec{s}}\frac{\partial L_{k\vec{s}}}{\partial\alpha^*_m} - \frac{\partial L^*_{k\vec{s}}}{\partial\alpha^*_m}\star_{\vec{s}}L_{k\vec{s}}\right)\right]dt + i\hbar e^{i\theta_m/2}\left(\sqrt{\Lambda^{s_m}_{mm} - |\lambda^{s_m}_{mm}|}\cdot d\mathcal{W}_{2m} + i\sqrt{\Lambda^{s_m}_{mm} + |\lambda^{s_m}_{mm}|}\cdot d\mathcal{W}_{2m+1} \right),
\end{align}
where we choose $\bm{\mathcal{Q}}$ as the identity matrix, and $\theta_m(\alpha_m,\alpha^*_m) = {\rm arg}(\lambda^{s_m}_{mm}(\alpha_m,\alpha^*_m))$.
When we choose $s_m=0$ for $\forall m$, Eqs.~\eqref{eq:Fokker-Planck equation when jumo operators do not couple different degrees of feeedom}--\eqref{eq:stochastic differential equation simplified version} are reduced to the ones obtained in Ref.~\cite{Yoneya2025}.
Solving Eq.~\eqref{eq:stochastic differential equation simplified version} needs low numerical cost rather than solving Eq.~\eqref{eq:stochastic differential equation} because we can avoid the diagonalization of $\bm{\mathcal{P}}^{\dagger}\bm{\mathcal{A}}^{\vec{s}}\bm{\mathcal{P}}$ at each time step.
We provide the derivations of Eqs.~\eqref{eq:condition for obtaining stochastic differential equaitons when the jump operators do not couple different degrees of feedom} and \eqref{eq:stochastic differential equation simplified version} in \ref{appendix:stochastic differential equation}.

\subsubsection{\label{subsubsec:Case studies: Bose-Hubbard model with on-site two- and three-body losses}Case studies: Bose-Hubbard model with on-site two- and three-body losses}
We investigate the feasibility of the second-order approximation of the experimentally realizable models, the Bose-Hubbard model with the on-site two- and three-body losses of atoms \cite{Menegatti,Tomita,Stenger,Roberts,Weber}.
Here, we consider the Bose-Hubbard model obeying the following GKSL equation:
\begin{gather}
    \frac{d\hat{\rho}(t)}{dt} = -\frac{i}{\hbar}\left[\hat{H}_{\rm BH},\hat{\rho}(t)\right]_- + \sum_m\gamma_m\left(\hat{L}_m\hat{\rho}(t)\hat{L}^{\dagger}_m - \frac{1}{2}\left[\hat{L}^{\dagger}_m\hat{L}_m,\hat{\rho}(t)\right]_+\right), \\
    \label{eq:Hamiltonian case study}
    \hat{H}_{\rm BH} = -\mu\sum_{m=1}^M\hat{a}^{\dagger}_m\hat{a}_m - J\sum_{\braket{m,n}}(\hat{a}^{\dagger}_m\hat{a}_n + {\rm h.c.}) + \frac{1}{2}\sum_{m=1}^MU_{mm}\hat{a}^{\dagger}_m\hat{a}^{\dagger}_m\hat{a}_m\hat{a}_m,
\end{gather}
where the subscript $m$ identify the lattice site, $\hat{a}_m^\dagger$ and $\hat{a}_m$ are the bosonic creation and annihilation operators of atoms at site $m$, $\mu$ is the chemical potential, $J$ is the hopping amplitude between the nearest-neighboring lattice sites $\braket{m,n}$, $U_{mm}$ is the on-site interaction energy, and $\gamma_m$ represents the strength of the on-site two- or three-body losses of atoms at site $m$, corresponding to the jump operators
\begin{align}
    \label{eq:jump operators for two- and three-body loss}
    \hat{L}_m =
    \begin{cases}
        \hat{a}_m\hat{a}_m & \text{on-site two-body loss} \\
        \hat{a}_m\hat{a}_m\hat{a}_m & \text{on-site three-body loss}
    \end{cases}.
\end{align}
Since $\hat{H}_{\rm BH}$ in Eq.~\eqref{eq:Hamiltonian case study} satisfies the condition Eq.~\eqref{eq:condition for Hamiltonian do not couple different degrees of freedom} and $\hat{L}_m$ in Eq.~\eqref{eq:jump operators for two- and three-body loss} for $\forall m$ do not couple different degrees of freedom, the matrices $\bm{\lambda}^{\vec{s}}$ and $\bm{\Lambda}^{\vec{s}}$ are diagonal as in Eq.~\eqref{eq:lambda and Lambda when the jump operators do not couple different degrees of feedom} and we can investigate the feasibility of the second-order calculation according to Eq.~\eqref{eq:condition for obtaining stochastic differential equaitons when the jump operators do not couple different degrees of feedom}.
Below, we analytically calculate $\lambda^{s_m}_{mm}$ and $\Lambda^{s_m}_{mm}$ for the two- and three-body jump operators and investigate the feasibility of the second-order calculation.
For simplicity, we restrict ourselves to consider $s_m = s = 1,0,-1$ for $\forall m$.

{\it Two-body loss} $\hat{L}_m = \hat{a}_m\hat{a}_m$--We first consider the Bose-Hubbard model with the on-site two-body losses.
From Eqs.~\eqref{eq:definitnion of lambdamn} and \eqref{eq:definitnion of Lambdamn}, we obtain $\lambda^{s_m = s}_{mm}$ and $\Lambda^{s_m = s}_{mm}$ as
\begin{gather}
    \lambda^s_{mm} = \frac{s\gamma_m\alpha_m^2}{2}\left(1 + \frac{iU_{mm}}{\hbar\gamma_m}\right), \\
    \Lambda^s_{mm} = \gamma_m(1-s)\left(|\alpha_m|^2 - \frac{1-s}{2}\right),
\end{gather}
and $\Lambda^s_{mm} - |\lambda^s_{mm}|$ as
\begin{gather}
    \Lambda^s_{mm} - |\lambda^s_{mm}| =
    \begin{cases}
        \displaystyle -\frac{\gamma_m|\alpha_m|^2}{2}\sqrt{1 + \left(\frac{U_{mm}}{\hbar\gamma_m}\right)^2} \leq 0 & \text{for}~s=1 \\
        \displaystyle\gamma_m\left(|\alpha_m|^2 - \frac{1}{2}\right) & \text{for}~s=0 \\
        \displaystyle2\gamma_m\left\{|\alpha_m|^2\left(1 - \frac{1}{4}\sqrt{1 + \left(\frac{U_{mm}}{\hbar\gamma_m}\right)^2}\right) - 1\right\} & \text{for}~s=-1
    \end{cases}.
\end{gather}
From these results, the second-order calculation is always unfeasible for $s=1$.
For the cases of $s=0$ and $-1$, the feasibility of the second-order calculation depends on the value of $\alpha_m$, i.e., there are regions in the phase space where the second-order calculation is feasible or unfeasible.
When we choose $s=0$ and $s=-1$, $\alpha_m$ need to satisfy
\begin{align}
    \begin{cases}
        \displaystyle|\alpha_m|^2 \geq \frac{1}{2} & \text{for}~s=0 \\
        \displaystyle|\alpha_m|^2\left(1 - \frac{1}{4}\sqrt{1 + \left(\frac{U_{mm}}{\hbar\gamma_m}\right)^2}\right) \geq 1 &\text{for}~s=-1
    \end{cases}
\end{align}
In both cases, these conditions set a lower bound for $|\alpha_m|^2$, and the bound for $s=-1$ becomes smaller as $U_{mm}/(\hbar\gamma_m)$ decreases.
The second-order calculation remains feasible until the two-body loss reduces $|\alpha_m|$ such that the distribution function acquires a nonzero value below the lower bound.

{\it Three-body loss} $\hat{L}_m = \hat{a}_m\hat{a}_m\hat{a}_m$--Next, we consider the Bose-Hubbard model with the on-site three-body losses of atoms.
Using Eqs.~\eqref{eq:definitnion of lambdamn} and \eqref{eq:definitnion of Lambdamn}, we obtain $\lambda^{s}_{mm}$ and $\Lambda^{s}_{mm}$ as
\begin{gather}
    \lambda^s_{mm} = \frac{3s\gamma_m\alpha^2_m}{2}\left\{|\alpha_m|^2 - \frac{3(1-s)}{2} + \frac{iU_{mm}}{3\hbar\gamma_m}\right\}, \\
    \Lambda^s_{mm} = \frac{9\gamma_m(1-s)}{4}\left\{|\alpha_m|^2 - \frac{\sqrt{2} + 1}{\sqrt{2}}(1-s)\right\}\left\{|\alpha_m|^2 - \frac{\sqrt{2} - 1}{\sqrt{2}}(1-s)\right\},
\end{gather}
and $\Lambda^s_{mm} - |\lambda^s_{mm}|$ as
\begin{gather}
\Lambda^s_{mm} - |\lambda^s_{mm}| =
    \begin{cases}
        \displaystyle -\frac{3\gamma_m|\alpha_m|^2}{2}\sqrt{|\alpha_m|^4 + \frac{1}{9}\left(\frac{U_{mm}}{\hbar\gamma_m}\right)^2}  \leq 0 & \text{for}~s=1 \\
        \displaystyle \frac{9\gamma_m}{4}\left(|\alpha_m|^2 - \frac{\sqrt{2} + 1}{\sqrt{2}}\right)\left(|\alpha_m|^2 - \frac{\sqrt{2} - 1}{\sqrt{2}}\right) & \text{for}~s=0 \\
        \displaystyle\frac{9\gamma_m}{2}\left[\left\{|\alpha_m|^2 - (2+\sqrt{2})\right\}\left\{|\alpha_m|^2 - (2-\sqrt{2})\right\} - \frac{|\alpha_m|^2}{3}\sqrt{(|\alpha_m|^2 - 3)^2 + \frac{1}{9}\left(\frac{U_{mm}}{\hbar\gamma_m}\right)^2}\right] & \text{for}~s=-1
    \end{cases}.
\end{gather}
In this model, the second-order calculation is always unfeasible for $s=1$ and is feasible depending on the value of $\alpha_m$ for $s=0$ and $s=-1$.
For the case of $s=0$, the second-order calculation is feasible when
\begin{align}
    |\alpha_m|^2\geq \frac{\sqrt{2} + 1}{\sqrt{2}}\quad\text{or}\quad|\alpha_m|^2\leq \frac{\sqrt{2} - 1}{\sqrt{2}}.
\end{align}
The former condition gives us a lower bound for $|\alpha_m|^2$.
For the case of $s=-1$, the lower bound becomes smaller as $U_{mm}/(\hbar\gamma_m)$ decreases.
In both cases, as in the case of the system the on-site two-body losses, the second-order calculation remains feasible until the three-body loss reduces $|\alpha_m|$ such that the probability takes a nonzero value below the lower bound.


\subsection{\label{subsec:Non-equal time correlation functions}Non-equal time correlation functions}
The integral expression of the GKSL equation enables us to calculate the non-equal time correlation functions within the first and second order of quantum fluctuations.
In this section, we derive formulas for calculating the non-equal two-time correlation functions.

The non-equal two-time correlation is defined by \cite{Breuer}
\begin{align}
    \label{eq:definition of two time correlation function}
    \braket{\hat{A}(t)\hat{B}(t_0)} = {\rm Tr}\left[\hat{A}\hat{\mathcal{V}}(t,t_0)\hat{B}\hat{\rho}(t_0)\right]\quad t_0 \leq t,
\end{align}
where $\hat{\mathcal{V}}(t,t_0)$ is defined by Eq.~\eqref{eq:Kraus_representation}.
In the phase space, Eq.~\eqref{eq:definition of two time correlation function} becomes
\begin{align}
    \label{eq:frist line of two time correlation function in the phase space}
    \braket{\hat{A}(t)\hat{B}(t_0)} &= \int\frac{d^2\vec{\alpha}_{\rm f}}{\pi^M}A_{\vec{s}}(\vec{\alpha}_{\rm f},\vec{\alpha}^*_{\rm f})\left[\hat{\mathcal{V}}(t,t_0)\hat{B}\hat{\rho}(t_0)\right]_{-\vec{s}}(\vec{\alpha}_{\rm f},\vec{\alpha}^*_{\rm f}) \\
    \label{eq:two time correlation function in the phase space}
    &= \int\frac{d^2\vec{\alpha}_{\rm f}d^2\vec{\alpha}_0}{\pi^{2M}}A_{\vec{s}}(\vec{\alpha}_{\rm f},\vec{\alpha}^*_{\rm f})\varUpsilon_{\vec{s}}(\vec{\alpha}_{\rm f},t;\vec{\alpha}_0,t_0)\left[B_{-\vec{s}}(\vec{\alpha}_0,\vec{\alpha}^*_0)\star_{-\vec{s}} W_{\vec{s}}(\vec{\alpha}_0,\vec{\alpha}^*_0,t_0)\right],
\end{align}
where we use Eq.~\eqref{eq:s-parametrized representation of TrAB} to transform Eq.~\eqref{eq:definition of two time correlation function} into Eq.~\eqref{eq:frist line of two time correlation function in the phase space}.
The derivation of Eq.~\eqref{eq:two time correlation function in the phase space} is provided in \ref{appendix:Non-equal two-time correlation function in the phase space}.
Eq.~\eqref{eq:two time correlation function in the phase space} is a general phase space representation of the non-equal two-time correlation function.

For the case of a Markovian open quantum system, we apply the Markov condition Eq.~\eqref{eq:Markov conditoin in the phase space} to Eq.~\eqref{eq:two time correlation function in the phase space} and obtain
\begin{align}
    \label{eq:two time correlation function in the phase space Markov}
    \braket{\hat{A}(t)\hat{B}(t_0)} = \int\frac{d^2\vec{\alpha}_{\rm f}}{\pi^M}A_{\vec{s}}(\vec{\alpha}_{\rm f},\vec{\alpha}^*_{\rm f})\lim_{\Delta t\to 0}\prod_{j=0}^{N_t - 1}\int\frac{d^2\vec{\alpha}_j}{\pi^M}\varUpsilon_{\vec{s}}(\vec{\alpha}_{j+1},t_{j+1};\vec{\alpha}_j,t_j)\left[B_{-\vec{s}}(\vec{\alpha}_0,\vec{\alpha}^*_0)\star_{-\vec{s}} W_{\vec{s}}(\vec{\alpha}_0,\vec{\alpha}^*_0,t_0)\right].
\end{align}
This is a general expression of the non-equal two-time correlation under the Markov condition.
As a specific case, we choose $\hat{B}=\hat{a}^{\dagger}_{m_c}$ with $m_c\in\{1,2,\cdots,M\}$ and assume that the initial state is given by a pure coherent state $\hat{\rho}(t_0) = \bigotimes_m\hat{\rho}_m(t_0)$ with $\hat{\rho}_m(t_0) = \ket{\alpha_{{\rm I}m}}\bra{ \alpha_{{\rm I}m}}$ where $\hat{a}_m|\alpha_{{\rm I}m}\rangle=\alpha_{{\rm I}m}|\alpha_{{\rm I}m}\rangle$.
The corresponding quasiprobability distribution function for $\hat{\rho}_m(t_0)$ is a Gaussian function $(2/(1-s_m))e^{-2|\alpha_m-\alpha_{{\rm I}m}|^2/(1-s_m)}$ for $s_m\neq 1$, and is a Dirac delta function $\pi\delta^{(2)}(\alpha_{m} - \alpha_{{\rm I}m})$ for $s_m=1$.
Under these conditions, when we choose $s_{m_c}\neq 1$, Eq.~\eqref{eq:two time correlation function in the phase space Markov} becomes
\begin{align}
    \label{eq:two time correlation function Markov B=adagger coherent state}
    \braket{\hat{A}(t)\hat{a}^{\dagger}_{m_c}(t_0)} = \int\frac{d^2\vec{\alpha}_{\rm f}}{\pi^M}A_{\vec{s}}(\vec{\alpha}_{\rm f},\vec{\alpha}^*_{\rm f})\lim_{\Delta t\to 0}\prod_{j=0}^{N_t - 1}\int\frac{d^2\vec{\alpha}_j}{\pi^M}\varUpsilon_{\vec{s}}(\vec{\alpha}_{j+1},t_{j+1};\vec{\alpha}_j,t_j)\left(\frac{2}{1-s_{m_c}}\alpha^*_{m_c,0} - \frac{1+s_{m_c}}{1-s_{m_c}}\alpha^*_{{\rm I}m_c}\right)W_{\vec{s}}(\vec{\alpha}_0,\vec{\alpha}^*_0,t_0).
\end{align}
Using the propagators \eqref{eq:first-order propagator} and \eqref{eq:second-order propagator}, we can approximately calculate Eq.~\eqref{eq:two time correlation function Markov B=adagger coherent state} using the first- and second-order approximations by the Monte Carlo simulation.
For example, when $s_{m_c}\neq1$, within the first-order [second-order] quantum fluctuations, we first solve the equation of motion of $\alpha_m$, Eq.~\eqref{eq:stochastic differential equation} [Eq.~\eqref{eq:classical equation of motion}] for $\forall m$ iteratively with different initial conditions stochastically sampled from the initial $\vec{s}$-ordered quasiprobability distribution function $W_{\vec{s}}(\vec{\alpha}_0,\vec{\alpha}^*_0,t_0)$ and calculate the value $(2\alpha^*_{m_c,0}/(1-s_{m_c}) - \alpha^*_{{\rm I}m_c}(1+s_{m_c})/(1-s_{m_c}))A_{\vec{s}}(\vec{\alpha}_{\rm f},\vec{\alpha}^*_{\rm f})$.
Then, we can calculate Eq.~\eqref{eq:two time correlation function Markov B=adagger coherent state} by taking an ensemble average over the results.
On the other hand, when $s_{m_c}=1$, Eq.~\eqref{eq:two time correlation function in the phase space Markov} becomes
\begin{align}
    \label{eq:two time correlation function Markov B=adagger coherent state s=1}
    \braket{\hat{A}(t)\hat{a}^{\dagger}_{m_c}(t_0)} = &\int\frac{d^2\vec{\alpha}_{\rm f}}{\pi^M}A_{\vec{s}}(\vec{\alpha}_{\rm f},\vec{\alpha}^*_{\rm f})\lim_{\Delta t\to 0}\prod_{j=0}^{N_t - 1}\int\frac{d^2\vec{\alpha}_j}{\pi^{M-1}}\varUpsilon_{\vec{s}}(\vec{\alpha}_{j+1},t_{j+1};\vec{\alpha}_j,t_j)\alpha^*_{m_c,0}\delta^{(2)}(\alpha_{m_c,0} - \alpha_{{\rm I}m_c})\prod_{\substack{m=1\\m\neq m_c}}^MW_{s_m}(\alpha_{m,0},\alpha^*_{m,0}) \nonumber \\
    &-\int\frac{d^2\vec{\alpha}_{\rm f}}{\pi^M}A_{\vec{s}}(\vec{\alpha}_{\rm f},\vec{\alpha}^*_{\rm f})\lim_{\Delta t\to 0}\prod_{j=0}^{N_t - 1}\int\frac{d^2\vec{\alpha}_j}{\pi^{M-1}}\varUpsilon_{\vec{s}}(\vec{\alpha}_{j+1},t_{j+1};\vec{\alpha}_j,t_j)\frac{\partial}{\partial\alpha_{m_c,0}}\delta^{(2)}(\alpha_{m_c,0} - \alpha_{{\rm I}m_c})\prod_{\substack{m=1\\m\neq m_c}}^MW_{s_m}(\alpha_{m,0},\alpha^*_{m,0}).
\end{align}
To the best of our knowledge, it is intractable to calculate Eq.~\eqref{eq:two time correlation function Markov B=adagger coherent state s=1} by using the sampling from the initial distribution function because of the presence of the derivative of the Dirac delta function.
If we know the analytical expression of $W_{\vec{s}}(\vec{\alpha}_{\rm f},\vec{\alpha}^*_{\rm f},t)$ by directly solving Eq.~\eqref{eq:generalized Liouville equation} or Eq.~\eqref{eq:Fokker-Planck equation}, we can calculate Eq.~\eqref{eq:two time correlation function Markov B=adagger coherent state s=1}.
However, it is beyond the scope of this paper.

Similarly, choosing $\hat{B} = \hat{a}_{m_c}$ and $\hat{\rho}(t_0)=\bigotimes_m\ket{\alpha_{{\rm I}m}}\bra{\alpha_{{\rm I}m}}$ in Eq.~\eqref{eq:definition of two time correlation function} leads to
\begin{align}
    \label{eq:two time correlation function Markov B=a coherent state}
    \braket{\hat{A}(t)\hat{a}_{m_c}(t_0)} = \alpha_{{\rm I}m_c}\braket{\hat{A}(t)},
\end{align}
which is more tractable than Eq.~\eqref{eq:two time correlation function Markov B=adagger coherent state} as it is factorized as a product of $\braket{\hat{a}_{m_c}(t_0)}$ and $\braket{\hat{A}(t)}$.
In particular, when we choose $\hat{A}=\hat{a}_{n_c}^\dagger$ with $n_c\in\{1,2,\cdots,M\}$, Eq.~\eqref{eq:two time correlation function Markov B=a coherent state} gives the time-normally ordered correlation function as $\braket{\hat{a}_{n_c}^\dagger(t)\hat{a}_{m_c}(t_0)} =\braket{\hat{a}^\dagger_{n_c}(t)}\braket{\hat{a}_{m_c}(t_0)}$.

Next, we consider the non-equal three-time correlation function $\braket{\hat{A}(t)\hat{B}(t_j)\hat{C}(t_0)} = {\rm Tr}[\hat{A}\hat{\mathcal{V}}(t,t_j)\hat{B}\hat{\mathcal{V}}(t_j,t_0)\hat{C}\hat{\rho}(t_0)]$ $(t_0\leq t_j \leq t)$.
The phase-space representation is
\begin{align}
    \label{eq:three_time_correlation_phase_space}
    \braket{\hat{A}(t)\hat{B}(t_j)\hat{C}(t_0)} = \int\frac{d^2\vec{\alpha}_{\rm f}d^2\vec{\alpha}_j d^2\vec{\alpha}_0}{\pi^{3M}}A_{\vec{s}}(\vec{\alpha}_{\rm f},\vec{\alpha}^*_{\rm f})\varUpsilon_{\vec{s}}(\vec{\alpha}_{\rm f},t;\vec{\alpha}_j,t_j)\left[B_{-\vec{s}}(\vec{\alpha}_j,\vec{\alpha}^*_j)\star_{-\vec{s}}\left\{\varUpsilon_{\vec{s}}(\vec{\alpha}_j,t_j;\vec{\alpha}_0,t_0)\left[C_{-\vec{s}}(\vec{\alpha}_0,\vec{\alpha}^*_0)\star_{-\vec{s}} W_{\vec{s}}(\vec{\alpha}_0,\vec{\alpha}^*_0,t_0)\right]\right\}\right],
\end{align}
which is obtained by following the same procedure to derive Eq.~\eqref{eq:two time correlation function in the phase space Markov}.
The calculation of Eq.~\eqref{eq:three_time_correlation_phase_space} is much more complicated than that of Eq.~\eqref{eq:two time correlation function in the phase space Markov} due to the presence of $\star_{-\vec{s}}$ which acts on $B_{-\vec{s}}(\vec{\alpha}_j,\vec{\alpha}^*_j)$ and subsequent terms.
However, by choosing $\hat{B} = \hat{a}^{\dagger}_{m_c}$, $\hat{C} = \hat{a}_{n_c}$, $s_{m_c} = -1$, and $s_{n_c} = 1$ in Eq.~\eqref{eq:three_time_correlation_phase_space}, we can remove $\star_{-\vec{s}}$ and rewrite Eq.~\eqref{eq:three_time_correlation_phase_space} as
\begin{align}
    \label{eq:three_time_correlation_phase_space detail}
    \braket{\hat{A}(t)\hat{a}^{\dagger}_{m_c}(t_j)\hat{a}_{n_c}(t_0)} = \int\frac{d^2\vec{\alpha}_{\rm f}d^2\vec{\alpha}_j d^2\vec{\alpha}_0}{\pi^{3M}}A_{\vec{s}}(\vec{\alpha}_{\rm f},\vec{\alpha}^*_{\rm f})\varUpsilon_{\vec{s}}(\vec{\alpha}_{\rm f},t;\vec{\alpha}_j,t_j)\alpha^*_{m_c,j}\varUpsilon_{\vec{s}}(\vec{\alpha}_j,t_j;\vec{\alpha}_0,t_0)\alpha_{n_c,0} W_{\vec{s}}(\vec{\alpha}_0,\vec{\alpha}^*_0,t_0),
\end{align}
which can be calculated by using the ensemble of the stochastic differential equations or the classical equations of motion.
Similarly, by appropriately ordering the operators and choosing the quasiprobability distribution function, we can calculate the third and higher order of non-equal time correlation function.
The general framework for the Glauber-Sudarshan P, Wigner, and Husimi Q functions is discussed in Refs.~\cite{Gardiner,Deuar2021,Agarwal3,Polkovnikov2009}.
We expect that by changing the values of $s_m$ depending on $m$, i.e., hybridizing the different quasiprobability distribution functions, we can investigate the open quantum many-body dynamics by calculating various non-equal time correlation functions.
However, it is out of the scope of this paper.

\section{\label{sec:Benchmark calculations}Benchmark calculations}
We numerically study the validity of the stochastic differential equation~\eqref{eq:stochastic differential equation} by using four models whose exact solutions are numerically obtainable.
In all the cases, we consider systems with two degrees of freedom and identify them by using the subscript $m=1, 2$.
In Secs.~\ref{subsec:Model1}--\ref{subsec:Model3}, we choose $s_m = s$ for both $m=1,2$ with $s=1,0,-1$ and consider first- and second-order approximations.
The corresponding $\vec{s}$-ordered quasiprobability distribution function is the Glauber-Sudarshan P function ($s=1$), the Wigner function ($s=0$), and the Husimi Q function ($s=-1$).
In Sec.~\ref{subsec:Model4}, we perform the approximation by hybridizing the different quasiprobability distribution functions by choosing $s_1\neq s_2$.
In Tab.~\ref{tab:quasiprobability and abbreviation},
we summarize the relation between the values of $(s_1,s_2)$ and the quasiprobability distribution functions and their abbreviations.
In the following benchmark calculations, we choose the jump operators that couple different degrees of freedom, to which our previous work \cite{Yoneya2025} cannot be applicable.
In addition, to highlight the importance of the choice of the distribution function and the second-order approximation, we selected the models for which the second-order approximation is possible for $s=1,0,-1$, $s=0,-1$, and $s=0$ in Secs.~\ref{subsec:Model1}, \ref{subsec:Model2}, and \ref{subsec:Model3}, respectively.
In Sec.~\ref{subsec:Model4}, we choose the model where the second-order calculation is not feasible for using the Glauber-Sudarshan P and Husimi Q functions, but feasible when we hybridize the different quasiprobability distribution functions to emphasize the feasibility of the second-order calculation by the hybridization.
In the following benchmark calculations, we choose $\bm{\mathcal{Q}}$ as the identity matrix in the stochastic differential equations.

\subsection{\label{sec:Common setup}Common setup}
\begin{table}[t]
    \centering
    \caption{Quasiprobability distribution functions and abbreviations}
    \label{tab:quasiprobability and abbreviation}
    \begin{tabular}{ccc}
        \hline\hline
        $(s_1,s_2)$ & Quasiprobability distribution function & Abbreviation \\ \hline
        $(1,1)$ & Glauber-Sudarshan P & P \\ 
        $(0,0)$ & Wigner & W \\ 
        $(-1,-1)$ & Husimi Q & Q \\ 
        $(0,-1)$ & Wigner and Husimi Q & W $+$ Q \\ \hline\hline
    \end{tabular}
\end{table}
\begin{table}[t]
    \centering
    \caption{Feasibility of the second-order calculation}
    \label{tab:feasibility of second-order calculation}
    \begin{tabular}{ccccc}
        \hline\hline
        \multirow{2}{*}{Model} &  \multicolumn{4}{c}{Second-order calculation}\\
        & P & W & Q & W $+$ Q \\\hline
        \multicolumn{1}{l}{Sec.~\ref{subsec:Model1}: Non-interacting bosons} &  $\checkmark$ &$\checkmark$ & $\checkmark$ & -- \\ 
        \multicolumn{1}{l}{Sec.~\ref{subsec:Model2}: Bose-Hubbard model} & $\times$ & $\checkmark$ & $\checkmark$ & -- \\ 
        \multicolumn{1}{l}{Sec.~\ref{subsec:Model3}: Bose-Einstein condensate} &  $\times$ & $\times$ &  $\checkmark$ & -- \\ 
        \multicolumn{1}{l}{Sec.~\ref{subsec:Model4}: Bose-Hubbard model} & $\times$ & $\times$ & $\checkmark$ & $\checkmark$\  \\ \hline\hline
    \end{tabular}
\end{table}
At the initial state, we prepare a pure coherent state $\hat{\rho}(0) = \ket{\alpha_{{\rm I}1},\alpha_{{\rm I}2}}\bra{\alpha_{{\rm I}1},\alpha_{{\rm I}2}}$, where $\hat{a}_m\ket{\alpha_{{\rm I}1},\alpha_{{\rm I}2}} = \alpha_{{\rm I}m}\ket{\alpha_{{\rm I}1},\alpha_{{\rm I}2}}$ for $m = 1$ and $2$, and $\alpha_{{\rm I}1} = \sqrt{N_{{\rm I}1}}e^{i\pi/8}$ and $\alpha_{{\rm I}2} = \sqrt{N_{{\rm I}2}}e^{i\pi/4}$ with the mean atomic numbers $N_{{\rm I}1}$ and $N_{{\rm I}2}$.
The corresponding initial quasiprobability distribution function becomes $W_{\vec{s}}(\vec{\alpha},\vec{\alpha}^*,t_0 = 0) = \prod_{m=1,2}W_{s_m}(\alpha_m,\alpha^*_m,0)$, where $W_{s_m}(\alpha_m,\alpha^*_m,0)$ is a Gaussian function for $s_m =0,-1$ and a Dirac delta function for $s_m=1$, i.e.,
\begin{align}
    W_{s_m}(\alpha_m,\alpha^*_m,0) = 
    \begin{cases}
    \dfrac{2}{1-s_m}e^{-2|\alpha_m-\alpha_{{\rm I}m}|^2/(1-s_m)} &\text{for}~ s_m=0,-1 \\
    \pi\delta^{(2)}(\alpha_m - \alpha_{{\rm I}m}) &\text{for}~ s_m=1.
    \end{cases}
\end{align}

In the second-order approximation, the stochastic differential equation~\eqref{eq:stochastic differential equation} is not always obtainable because the matrix $\bm{\mathcal{A}}^{\vec{s}}$ can violate the positive-semidefiniteness condition Eq.~\eqref{eq:feasible condition of the Hubbard-Stratonovich transformation}.
When the second-order calculation is unfeasible, we ignore the effects of the second order of quantum fluctuations and use the first-order approximation.
In Tab.~\ref{tab:feasibility of second-order calculation}, we summarize the feasibility of the second-order calculation for each quasiprobability distribution functions: We can (not) obtain the stochastic differential equation when $\checkmark$ ($\times$).
Below, we abbreviate the numerically exact result as ``Exact'' and the results of the first- and second-order approximations as ``Prob:$1$st'' and ``Prob:$2$nd'', respectively, where Prob $=$ P, W, Q, and W $+$ Q.

Under these setups, we investigate the time evolution of the fraction difference of the remaining atoms in each state, and those of the equal and non-equal time correlation between atoms at different states which are respectively defined by
\begin{align}
    \label{eq:physical quantities}
    n_{12} = \frac{\braket{\hat{a}_1^{\dagger}\hat{a}_1 - \hat{a}_2^{\dagger}\hat{a}_2}}{N_{\rm I}},\quad C_{12} = \frac{\braket{\hat{a}^{\dagger}_1\hat{a}_2 + \hat{a}^{\dagger}_2\hat{a}_1}}{\sqrt{2}N_{\rm I}},\quad G_{12}(t,0) = \frac{\braket{\hat{a}^{\dagger}_1(t)\hat{a}_2(0)}}{N_{\rm I}},\quad \bar{G}_{12}(t,0) = \frac{\braket{\hat{a}_1(t)\hat{a}^{\dagger}_2(0)}}{N_{\rm I}},
\end{align}
where $N_{\rm I}$ is a total mean atomic number in the initial state, $N_{\rm I} = N_{{\rm I}1} + N_{{\rm I}2}$, and $G_{12}(t,0) = G^{\rm re}_{12}(t,0) + iG^{\rm im}_{12}(t,0)\in \mathbb{C}$ and $\bar{G}_{12}(t,0) = \bar{G}^{\rm re}_{12}(t,0) + i\bar{G}^{\rm im}_{12}(t,0)\in \mathbb{C}$ take complex values with $G^{\rm re}_{12}(t,0),G^{\rm im}_{12}(t,0),\bar{G}^{\rm re}_{12}(t,0),\bar{G}^{\rm im}_{12}(t,0)\in \mathbb{R}$.
We also calculate the difference between the results of the first- and second-order approximation ($A^{\vec{s}}_{\rm approx}$) and the numerically exact one ($A_{\rm Exact}$) defined by $\delta A^{\vec{s}} = A^{\vec{s}}_{\rm approx} - A_{\rm Exact}$ with $A$ being one of the physical quantities in Eq.~\eqref{eq:physical quantities}.
Here, we note that when we use the Glauber-Sudarshan P function, although we can calculate $G_{12}(t,0)$ (Sec.~\ref{subsec:Model1}),
we cannot calculate $\bar{G}_{12}(t,0)$ (Sec.~\ref{subsec:Model2}-Sec.~\ref{subsec:Model4}) as discussed in Sec.~\ref{subsec:Non-equal time correlation functions}.

In the numerical calculations, we use the fourth-order Runge-Kutta method to solve the GKSL equation and the classical equations of motion.
For the stochastic differential equations, we use the Platen method \cite{Platen} in Sec.~\ref{subsec:Model1} and the Euler-Maruyama method \cite{Platen} in Secs.~\ref{subsec:Model2}--\ref{subsec:Model4}.
We take $N_{\rm initial} = 1000$ samples for the initial conditions and $N_{\rm stoch} = 100$ samples for the stochastic processes.
In the second-order calculations, by solving the stochastic differential equations, we obtain $\alpha_{m,ij}(t)$ for $m=1,2,\cdots,M$, where the subscripts $i$ and $j$ respectively identify the label of the initial conditions and stochastic processes.
Here, we choose the same initial state for the same $i$, but use independent stochastic processes for the same $j$ when $i$ differs.
The physical quantity $A$ is evaluated within the second-order approximation as
\begin{align}
    A^{\vec{s}}_{\rm approx}(N_{\rm initial},N_{\rm stoch}) = \frac{1}{N_{\rm initial}N_{\rm stoch}}\sum_{i=1}^{N_{\rm initial}}\sum_{j=1}^{N_{\rm stoch}}A^{\vec{s}}_{ij},
\end{align}
where $A^{\vec{s}}_{ij}$ is a $\vec{s}$-ordered phase-space representation of the physical quantity $\hat{A}$ using $\alpha_{m,ij}(t)$ for $m=1,2,\cdots,M$.
We also evaluate the standard error \cite{Barlow} defined by
\begin{align}
    \label{eq:standard error}
    \sigma^{\vec{s}}_{A} = \frac{1}{\sqrt{N_{\rm initial}}}\sqrt{\sum_{i=1}^{N_{\rm initial}}\sum_{j=1}^{N_{\rm stoch}}\frac{(A^{\vec{s}}_{ij} - \bar{A}^{\vec{s}}_{j})^2}{N_{\rm initial}N_{\rm stoch}}},
\end{align}
where $\bar{A}^{\vec{s}}_{j}$ is an averaged value of $A^{\vec{s}}_{ij}$ with respect to the sampling of the initial conditions:
\begin{align}
    \bar{A}^{\vec{s}}_j = \frac{1}{N_{\rm initial}}\sum_{i=1}^{N_{\rm initial}}A^{\vec{s}}_{ij}.
\end{align}
Note that it is not possible to define a standard error with respect to the total number of samples, $N_{\mathrm{initial}} N_{\mathrm{stoch}}$, since we perform two kinds of sampling; 
Such a definition would lead to an unphysical result, because the standard error would vanish even if only one of these numbers tends to infinity. 
Eq.~\eqref{eq:standard error} represents the standard error with respect to the sampling over the initial states; that is, it is a quantity that vanishes in the limit $N_{\mathrm{initial}} \to \infty$ for a fixed $N_{\mathrm{stoch}}$. 
In addition, in the case of the first-order calculation, the standard error can also be evaluated in the same manner by setting $N_{\mathrm{stoch}} = 1$ in Eq.~\eqref{eq:standard error}.
Below, we depict the standard error by shading the region between $A^{\vec{s}}_{\rm approx}\pm\sigma^{\vec{s}}_A$ in the figures of the numerical results.


\subsection{\label{subsec:Model1}Model 1: Non-interacting atoms}
We first consider two-site non-interacting atoms obeying the following GKSL equation:
\begin{gather}
    \label{eq:GKSL equation Model 1}
    \frac{d\hat{\rho}(t)}{dt} = -\frac{i}{\hbar}\left[\hat{H}_{\rm FB},\hat{\rho}(t)\right]_- + \gamma\left(\hat{L}\hat{\rho}(t)\hat{L}^{\dagger} - \frac{1}{2}\left[\hat{L}^{\dagger}_k\hat{L}_k,\hat{\rho}(t)\right]_+\right), \\
    \hat{H}_{\rm FB} = -\mu\sum_{m=1,2}\hat{a}^{\dagger}_{m}\hat{a}_m - J(\hat{a}^{\dagger}_2\hat{a}_1 + \hat{a}^{\dagger}_1\hat{a}_2), \\
    \label{eq: jump operator for Model1}
    \hat{L} = \hat{a}_1 + \hat{a}_2,
\end{gather}
where $\hat{a}^{\dagger}_m$ and $\hat{a}_m$ are the creation and annihilation operators, respectively, for atoms at site $m=1,2$, and $\hat{H}_{\rm FB}$ is the Hamiltonian for two-site non-interacting atoms with $\mu$ being a chemical potential and $J$ the hopping amplitude.
The jump operator \eqref{eq: jump operator for Model1} describes a non-local loss of bosons, and $\gamma$ represents its strength.
\begin{figure}[t]
	\centering 
	\includegraphics[width = \linewidth]{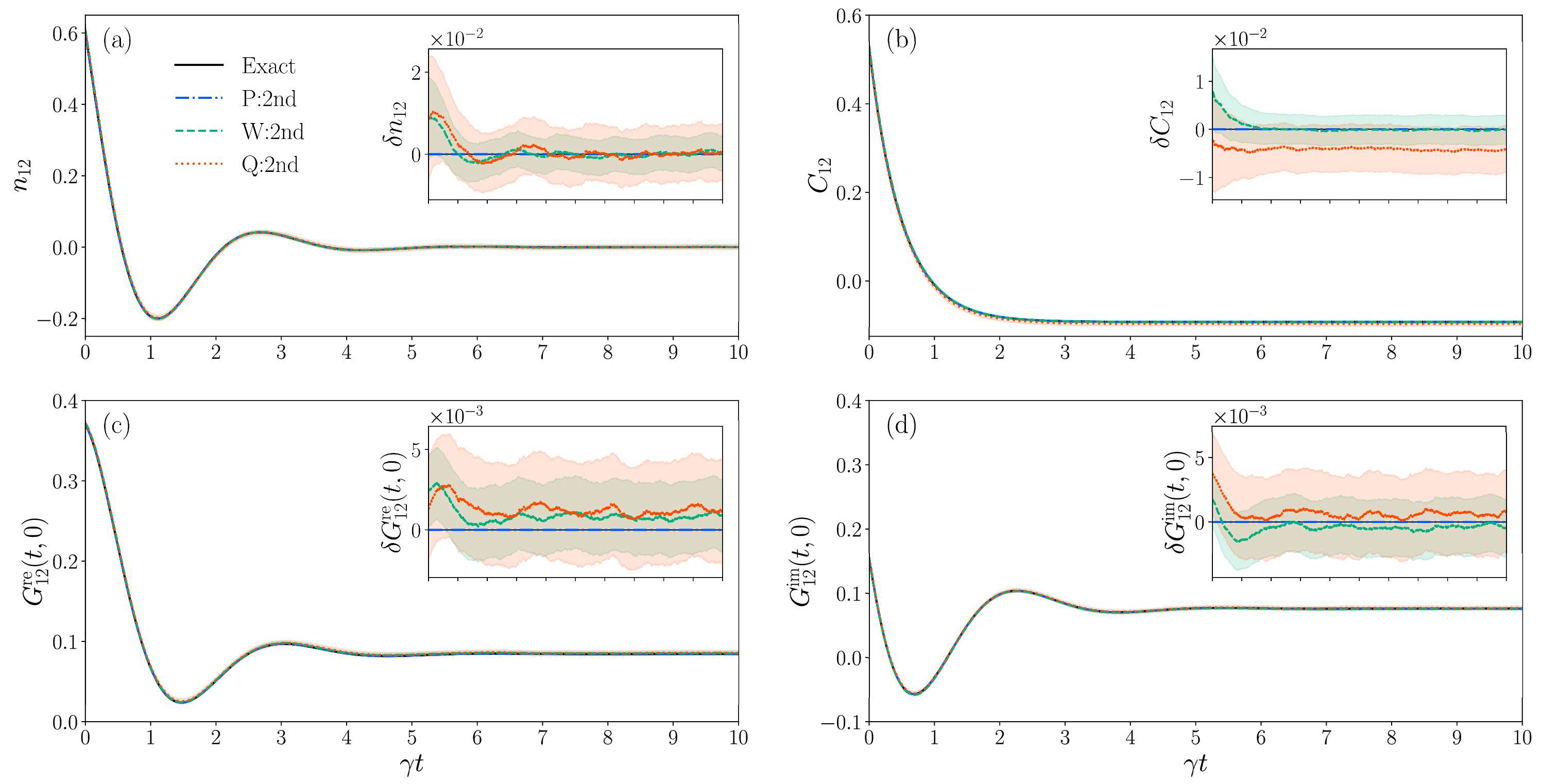}
	\caption{Relaxation dynamics of two-site non-interacting atoms obeying the GKSL equation~\eqref{eq:GKSL equation Model 1} starting from the pure coherent state $\hat{\rho}(0) = \ket{\alpha_{{\rm I}1},\alpha_{{\rm I}2}}\bra{\alpha_{{\rm I}1},\alpha_{{\rm I}2}}$, where $\alpha_{{\rm I}1} = \sqrt{N_{{\rm I}1}}e^{i\pi/8}$ and $\alpha_{{\rm I}2} = \sqrt{N_{{\rm I}2}}e^{i\pi/4}$ with $N_{{\rm I}1} = 8$ and $N_{{\rm I}2} = 2$. Shown are (a) the difference of the remaining fractions of atoms at each sites $n_{12}$, (b) the correlation of atoms at different sites $C_{12}$, and real (c) and imaginary (d) parts of the non-equal time correlation function $G_{12}(t,0)$, which are defined by Eq.~\eqref{eq:physical quantities}. In each panel, we compare the numerically exact result (Exact) obtained by directly solving the GKSL equation~\eqref{eq:GKSL equation Model 1} and the ones of the second-order approximation using the Glauber-Sudarshan P (P:$2$nd), Wigner (W:$2$nd), and Husimi Q function (Q:$2$nd). We choose $\mu/(\hbar\gamma) = 1$ and $J/(\hbar\gamma) = 1$, and take 1000 samples for the initial conditions and 100 samples for the stochastic processes in the second-order approximation. The insets depict the difference between the results of each approximation and the numerically exact one.
    The shaded regions represent the standard error with respect to the sampling over the initial states.
    }
	\label{fig:Model1}
\end{figure}

Since the GKSL equation~\eqref{eq:GKSL equation Model 1} is quadratic with respect to $\hat{a}^{\dagger}_m$ and $\hat{a}_m$, the second-order approximation becomes exact where we can always obtain the stochastic differential equations as shown in the following.
In this model, the matrices $\bm{\lambda}^{\vec{s} = (s,s)}$ and $\bm{\Lambda}^{\vec{s} = (s,s)}$ are given by
\begin{gather}
    \label{eq:lambda for Model1}
    \bm{\lambda}^{\vec{s} = (s,s)} =
    \begin{bmatrix}
        0 & 0 \\
        0 & 0
    \end{bmatrix},\\
    \label{eq:Lambda for Model1}
    \bm{\Lambda}^{\vec{s} = (s,s)} = \frac{\gamma}{4}(1-s)
    \begin{bmatrix}
        1 & 1 \\
        1 & 1
    \end{bmatrix}.
\end{gather}
Substituting Eqs.~\eqref{eq:lambda for Model1} and \eqref{eq:Lambda for Model1} into Eq.~\eqref{eq:definitnioa of the diffusion matrix A}, we obtain the matrix $\bm{\mathcal{A}}^{\vec{s}=(s,s)}$, which can be analytically diagonalized as
\begin{align}
    \label{eq:diagonalized matrix A Model 1}
    \bm{\mathcal{U}}^{\vec{s}\dagger}\bm{\mathcal{A}}^{\vec{s}}\bm{\mathcal{U}}^{\vec{s}} = \bm{\mathcal{A}}_{\rm diag}^{\vec{s}} =
    \begin{bmatrix}
        \gamma(1-s) & 0 & 0 & 0 \\
        0 & \gamma(1-s) & 0 & 0 \\
        0 & 0 & 0 & 0 \\
        0 & 0 & 0 & 0
    \end{bmatrix},
\end{align}
where the unitary matrix $\bm{\mathcal{U}}^{\vec{s}}$ is given by 
\begin{align}
    \label{eq:diagonalizing matrix U Model 1}
    \bm{\mathcal{U}}^{\vec{s}} = \bm{\mathcal{P}}\bm{\mathcal{V}}^{\vec{s}} = \bm{\mathcal{P}}\frac{1}{\sqrt{2}}
    \begin{bmatrix}
        0 & 1 & -1 & 0 \\
        0 & 1 & 1 & 0 \\
        1 & 0 & 0 & -1 \\
        1 & 0 & 0 & 1
    \end{bmatrix}=
    \frac{1}{2}
    \begin{bmatrix}
        i & 1 & -1 & -i \\
        i & 1 & 1 & i \\
        -i & 1 & -1 & i \\
        -i & 1 & 1 & -i
    \end{bmatrix}.
\end{align}
Substituting Eqs.~\eqref{eq:diagonalized matrix A Model 1} and \eqref{eq:diagonalizing matrix U Model 1} into Eq.~\eqref{eq:stochastic differential equation}, we obtain the following stochastic differential equations:
\begin{gather}
    \label{eq:SDE for site1 Model1}
    i\hbar d\alpha_{1} = \left[-\mu\alpha_{1} - J\alpha_{2} - \frac{i\hbar\gamma}{2}(\alpha_{1} + \alpha_{2})\right]dt + i\hbar\sqrt{\frac{\gamma}{4}(1-s)}\cdot(d\mathcal{W}_1 + id\mathcal{W}_2), \\
    \label{eq:SDE for site2 Model1}
    i\hbar d\alpha_{2} = \left[-\mu\alpha_{2} - J\alpha_{1} - \frac{i\hbar\gamma}{2}(\alpha_{1} + \alpha_{2})\right]dt + i\hbar\sqrt{\frac{\gamma}{4}(1-s)}\cdot(d\mathcal{W}_1 + id\mathcal{W}_2),
\end{gather}
where $\mathcal{W}_1$ and $\mathcal{W}_2$ are the Wiener processes which are independent of each other.
It is interesting that when we choose $s=1$, the stochastic terms in Eqs.~\eqref{eq:SDE for site1 Model1} and \eqref{eq:SDE for site2 Model1} become zero.
Considering that the initial Glauber-Sudarshan P function is a Dirac delta function, we can calculate the exact dynamics of the GKSL equation~\eqref{eq:GKSL equation Model 1} merely by solving the classical equations of motion with the initial conditions $\alpha_1(0) = \alpha_{{\rm I}1}$ and $\alpha_2(0) = \alpha_{{\rm I}2}$.
On the other hand, when we choose $s=0$ or $-1$, we use the Monte Carlo trajectory sampling of the stochastic differential equations~\eqref{eq:SDE for site1 Model1} and \eqref{eq:SDE for site2 Model1}.

Fig.~\ref{fig:Model1} shows the relaxation dynamics of $n_{12}$, $C_{12}$, and $G_{12}(t,0)$.
The initial mean atomic numbers are $N_{{\rm I}1} = 8$ and $N_{{\rm I}2} = 2$, and we choose the parameters in the Hamiltonian as $\mu/(\hbar\gamma) = 1$ and $J/(\hbar\gamma) = 1$.
The inset of each panel depicts the difference between the results of the second-order approximation (P:$2$nd, W:$2$nd, and Q:$2$nd) and the numerically exact one (Exact).
In all the panels, the results of the second-order approximations show good agreement with the numerically exact one.
We note that the derivations of the results using the Wigner function and the Husimi Q function shown in the insets become smaller as we increase the number of samples for the initial conditions and stochastic processes.
We also note that the results become smoother as the number of samples for the stochastic processes increases. We obtain the same dependence on the number of samples in the models in the following sections \ref{subsec:Model2}--\ref{subsec:Model4}.

From Fig.~\ref{fig:Model1}, we can conclude that the Glauber-Sudarshan P function is an appropriate and efficient choice to simulate this model, because we do not need to take an ensemble average over the results with respect to the initial conditions and stochastic processes.
The same result for the Glauber-Sudarshan P function can be obtained for a system with a non-interacting Hamiltonian $\hat{H}=\sum_{m}h^{(1)}_{m}\hat{a}_m + \sum_{m,n}h^{(2)}_{mn}\hat{a}^{\dagger}_m\hat{a}_n + {\rm c.c.}$ with $h^{(1)}_{m},h^{(2)}_{mn}\in\mathbb{C}$ and linear jump operators involving only one-body loss terms $\hat{L}_k = \sum_ml_{km}\hat{a}_m$ with $l_{km}\in\mathbb{C}$ for $\forall k$.
By substituting these Hamiltonian and jump operators into Eq.~\eqref{eq:definitnioa of the diffusion matrix A} with choosing $s_m = 1$ for $\forall m$, we can confirm that all the matrix elements of $\bm{\mathcal{A}}^{\vec{1}}$ with $\vec{1} = (1,\cdots,1) \in \mathbb{R}^M $ are zero and thus the GKSL equation is exactly mapped into the generalized Liouville equation, which greatly reduces the numerical cost as shown in the benchmark calculation.

\subsection{\label{subsec:Model2}Model 2: Bose-Hubbard model}
Next, we consider the two-site Bose-Hubbard model obeying the following GKSL equation:
\begin{gather}
    \label{eq:GKSL equation Model2}
    \frac{d\hat{\rho}(t)}{dt} = -\frac{i}{\hbar}\left[\hat{H}_{\rm BH},\hat{\rho}(t)\right]_- + \sum_{k=1,2,3,4}\gamma_k\left(\hat{L}_k\hat{\rho}(t)\hat{L}^{\dagger}_k - \frac{1}{2}\left[\hat{L}^{\dagger}_k\hat{L}_k,\hat{\rho}(t)\right]_+\right), \\
    \label{eq:Hamiltonian Model2}
    \hat{H}_{\rm BH} = -\mu\sum_{m=1,2}\hat{a}^{\dagger}_m\hat{a}_m - J(\hat{a}^{\dagger}_2\hat{a}_1 + \hat{a}_1^{\dagger}\hat{a}_2) + \frac{1}{2}\sum_{m=1,2}U_{mm}\hat{a}^{\dagger}_m\hat{a}^{\dagger}_m\hat{a}_m\hat{a}_m, \\
    \label{eq:Jump operators Model2}
    \hat{L}_1 = \hat{a}_1,~\hat{L}_2 = \hat{a}_2,~\hat{L}_3 = (\hat{a}^{\dagger}_1 + \hat{a}^{\dagger}_2)(\hat{a}_1 + \hat{a}_2),
\end{gather}
where $\hat{H}_{\rm BH}$ is the Bose-Hubbard Hamiltonian with $U_{11}$ and $U_{22}$ being on-site interaction energies for atoms at site $1$ and $2$, respectively.
We consider one-body loss jump operators at site 1 ($\hat{L}_1$) and site 2 ($\hat{L}_2$) and a non-local two-body jump operator $\hat{L}_3$, where $\gamma_1$, $\gamma_2$, and $\gamma_3$ respectively represent their strengths.
\begin{figure}[t]
	\centering 
	\includegraphics[width = \linewidth]{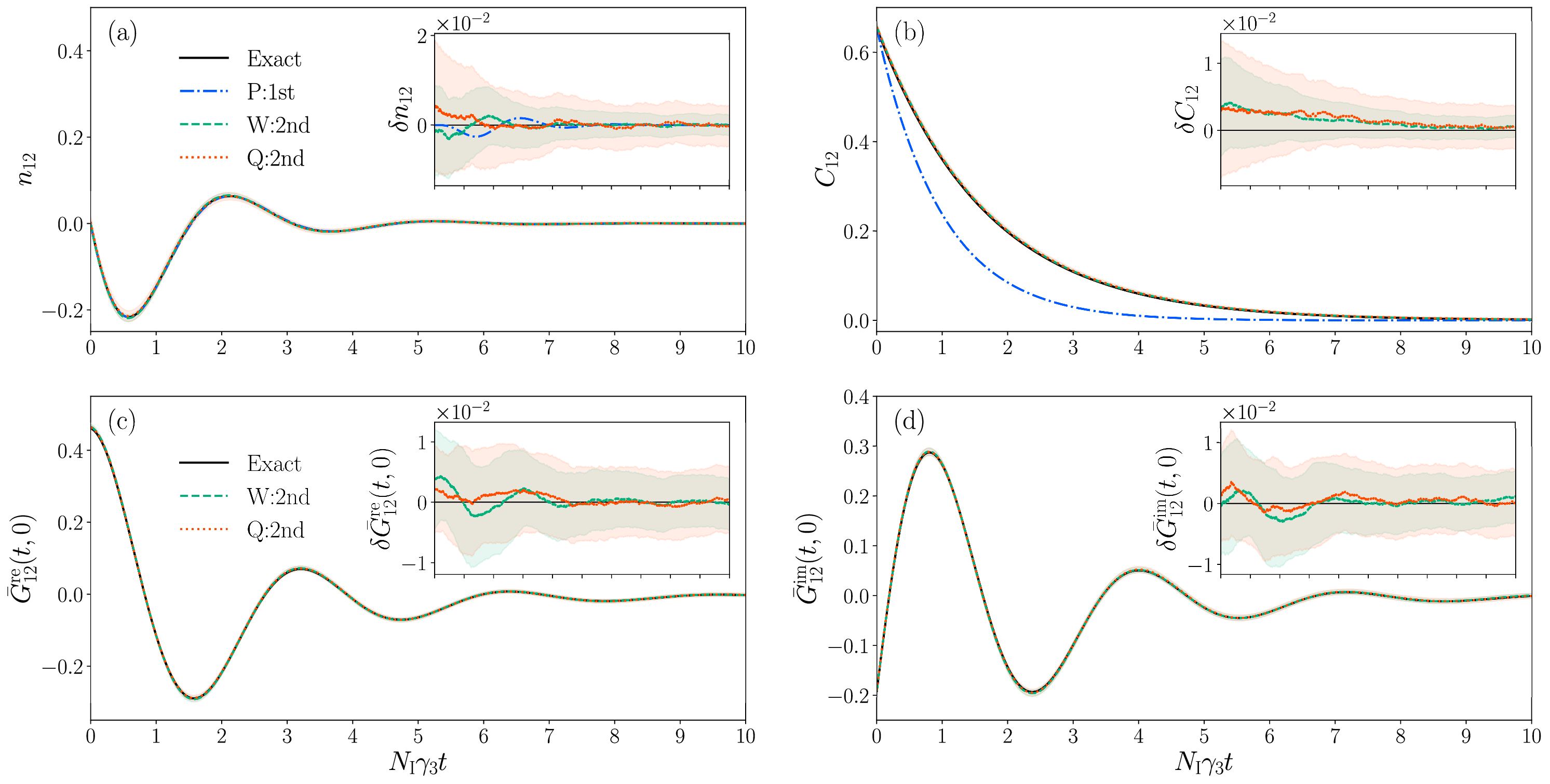}
	\caption{Relaxation dynamics of two-site Bose-Hubbard model obeying the GKSL equation~\eqref{eq:GKSL equation Model2} starting from the pure coherent state $\hat{\rho}(0) = \ket{\alpha_{{\rm I}1},\alpha_{{\rm I}2}}\bra{\alpha_{{\rm I}1},\alpha_{{\rm I}2}}$, where $\alpha_{{\rm I}1} = \sqrt{N_{{\rm I}1}}e^{i\pi/8}$ and $\alpha_{{\rm I}2} = \sqrt{N_{{\rm I}2}}e^{i\pi/4}$ with $N_{{\rm I}1} = N_{{\rm I}2} = 5$. 
    The quantities and notations are the same as those in Fig.~\ref{fig:Model1} except that we calculate the non-equal time correlation function $\bar{G}_{12}(t,0)$ instead of $G_{12}(t,0)$, and use the first-order approximation for the Glauber-Sudarshan P function.
    The other parameters are $\mu/(\hbar N_{\rm I}\gamma) = 1$, $J/(\hbar N_{\rm I}\gamma) = 1$, $U_{11}/(\hbar\gamma_3) = U_{22}/(\hbar\gamma_3) = 0.2$, and $\gamma_1/(N_{\rm I}\gamma_3) = \gamma_2/(N_{\rm I}\gamma_3) = 0.6$, where $N_{\rm I} = N_{{\rm I}1} + N_{{\rm I}2} = 10$, and we take 1000 samples for the initial conditions and 100 samples for the stochastic processes (W:$2$nd and Q:$2$nd). The insets depict the difference between the results of the approximations (P:$1$st, W:$2$nd, and Q:$2$nd) and the numerically exact one (Exact).}
	\label{fig:Model2}
\end{figure}

In the second-order approximation, we cannot always obtain the stochastic differential equations depending on the value of $s$ as shown in the following.
The matrices $\bm{\lambda}^{\vec{s} = (s,s)}$ and $\bm{\Lambda}^{\vec{s} = (s,s)}$ are respectively given by
\begin{gather}
    \label{eq:lambda for Model2}
    \bm{\lambda}^{\vec{s} = (s,s)} = \frac{1}{2}
    \begin{bmatrix}
        \gamma_3(\alpha_1 + \alpha_2)^2 + i\dfrac{sU_{11}}{\hbar}\alpha_1^2 & \gamma_3(\alpha_1 + \alpha_2)^2 \\
        \gamma_3(\alpha_1 + \alpha_2)^2 & \gamma_3(\alpha_1 + \alpha_2)^2 + i\dfrac{sU_{22}}{\hbar}\alpha_2^2
    \end{bmatrix},\\
    \label{eq:Lambda for Model2}
    \bm{\Lambda}^{\vec{s} = (s,s)} = \frac{1}{4}
    \begin{bmatrix}
        \gamma_1(1-s) + 2\gamma_3|\alpha_1 + \alpha_2|^2& 2\gamma_3|\alpha_1 + \alpha_2|^2 \\
        2\gamma_3|\alpha_1 + \alpha_2|^2 & \gamma_2(1-s) + 2\gamma_3|\alpha_1 + \alpha_2|^2
    \end{bmatrix}.
\end{gather}
When the matrix $\bm{\mathcal{A}}^{\vec{s}}$ is positive-semidefinite, we can obtain the following stochastic differential equations:
\begin{gather}
    \label{eq:SDE for site1 Model2}
    i\hbar d\alpha_1 = \left[-\mu\alpha_1 - J\alpha_2 + U_{11}\alpha_1(|\alpha_1|^2 - 1 + s) - \frac{i\hbar}{2}\left\{(\gamma_1 + 2\gamma_3)\alpha_1 + 2\gamma_3\alpha_2\right\}\right]dt + i\hbar\left[i\bm{\mathcal{U}}^{\vec{s}}\sqrt{\bm{\mathcal{A}}^{\vec{s}}_{\rm diag}}\cdot d\overrightarrow{\mathcal{W}}\right]_1, \\
    \label{eq:SDE for site2 Model2}
    i\hbar d\alpha_2 = \left[-\mu\alpha_2 - J\alpha_1 + U_{22}\alpha_2(|\alpha_2|^2 - 1 + s) - \frac{i\hbar}{2}\left\{2\gamma_3\alpha_1 + (\gamma_2 + 2\gamma_3)\alpha_2\right\}\right]dt + i\hbar\left[i\bm{\mathcal{U}}^{\vec{s}}\sqrt{\bm{\mathcal{A}}^{\vec{s}}_{\rm diag}}\cdot d\overrightarrow{\mathcal{W}}\right]_2.
\end{gather}
When we choose $s=0$, $\bm{\mathcal{A}}^{\vec{s}=(0,0)}$ is always positive-semidefinite.
For the case of $s=-1$, $\bm{\mathcal{A}}^{\vec{s}=(-1,-1)}$ can have negative eigenvalues depending on the values of $\vec{\alpha}$.
However, we numerically confirm that $\bm{\mathcal{A}}^{\vec{s}=(-1,-1)}$ is always positive-semidefinite during the simulation at least under our parameter setting, which we will show below, and we can perform the second-order calculation.
In the second-order calculations, we numerically diagonalize the matrix $\bm{\mathcal{P}}^{\dagger}\bm{\mathcal{A}}^{\vec{s}}\bm{\mathcal{P}}$ in each time step to obtain $\bm{\mathcal{U}}^{\vec{s}}$ and $\bm{\mathcal{A}}^{\vec{s}}_{\rm diag}$, and we use the same procedure in Secs.~\ref{subsec:Model3} and \ref{subsec:Model4}.
On the other hand, when we choose $s=1$, we can analytically show that $\bm{\mathcal{A}}^{\vec{s}=(1,1)}$ always involves negative eigenvalues independently of parameters.
Hence, for the case of $s=1$, we ignore the second-order terms and solve the classical equations of motion, which is given by Eqs.~\eqref{eq:SDE for site1 Model2} and \eqref{eq:SDE for site2 Model2} with neglecting the stochastic terms (last terms).

Fig.~\ref{fig:Model2} shows the relaxation dynamics of $n_{12}$, $C_{12}$, and $\bar{G}_{12}(t,0)$, where the insets depict the difference between the result of the first- and second-order approximations and the numerically exact one.
We prepare the initial mean atomic numbers as $N_{{\rm I}1} = N_{{\rm I}2} = 5$ and choose the other parameters as $\mu/(\hbar N_{\rm I}\gamma_3) = 1$, $J/(\hbar N_{\rm I}\gamma_3) = 1$, $U_{11}/(\hbar\gamma_3) = U_{22}/(\hbar\gamma_3) = 0.2$, and $\gamma_1/(N_{\rm I}\gamma_3) = \gamma_2/(N_{\rm I}\gamma_3) = 0.6$.
Here, when we use the Glauber-Sudarshan P function, we cannot calculate the non-equal time correlation function $\bar{G}_{12}(t,0)$ as discussed in Sec.~\ref{subsec:Non-equal time correlation functions}.
The situation is the same for the models in Secs.~\ref{subsec:Model3} and \ref{subsec:Model4}.
In all the panels, there are good agreements between the results of the second-order approximations (W:$2$nd and Q:$2$nd) and the numerically exact one (Exact).
Although we ignore the effect of the second order of quantum fluctuations in the first-order approximation (P:$1$st), it well reproduces the exact dynamics of $n_{12}$ as shown in Figs.~\ref{fig:Model2}(a).
On the other hand, one can see a significant deviation of the first-order approximation from the exact result in Fig.~\ref{fig:Model2}(b).
This result suggests that the effect of the second order of quantum fluctuations strongly affects the dynamics of $C_{12}$, and the use of the second-order approximation using the Wigner function or the Husimi Q function is necessary to reproduce the exact dynamics.


\subsection{\label{subsec:Model3}Model 3: Two-component Bose-Einstein condensate}
As a third model, we consider a two-component Bose-Einstein condensate (BEC) obeying the following GKSL equation:
\begin{gather}
    \label{eq:GKSL equation Model3}
    \frac{d\hat{\rho}(t)}{dt} = -\frac{i}{\hbar}\left[\hat{H}_{\rm BEC},\hat{\rho}(t)\right]_- + \sum_{k=1,2,3}\gamma_k\left(\hat{L}_k\hat{\rho}(t)\hat{L}^{\dagger}_k - \frac{1}{2}\left[\hat{L}^{\dagger}_k\hat{L}_k,\hat{\rho}(t)\right]_+\right), \\
    \label{eq:Hamiltonian Model3}
    \hat{H}_{\rm BEC} = -\mu\sum_{m=1,2}\hat{a}^{\dagger}_m\hat{a}_m + \frac{1}{2}\sum_{m=1,2}U_{mn}\hat{a}^{\dagger}_m\hat{a}^{\dagger}_n\hat{a}_m\hat{a}_n, \\
    \label{eq:Jump operators Model3}
    \hat{L}_1 = \hat{a}_1,~\hat{L}_2 = \hat{a}_2,~\hat{L}_3 = \hat{a}^{\dagger}_2\hat{a}_1.
\end{gather}
This Hamiltonian describes Bose atoms with two internal degrees of freedom, denoted by $m=1$ and $2$, trapped in a strongly confined potential such that the spatial degrees of freedom of atoms are frozen \cite{Cirac,Gordon}.
$U_{mn}$ denotes the interaction energy between atoms in states $m$ and $n$.
We consider three jump operators: two are one-body losses from each internal state ($\hat{L}_1$ and $\hat{L}_2$), and the third is the incoherent transfer between the internal states ($\hat{L}_3$).
\begin{figure}[t]
	\centering 
	\includegraphics[width = \linewidth]{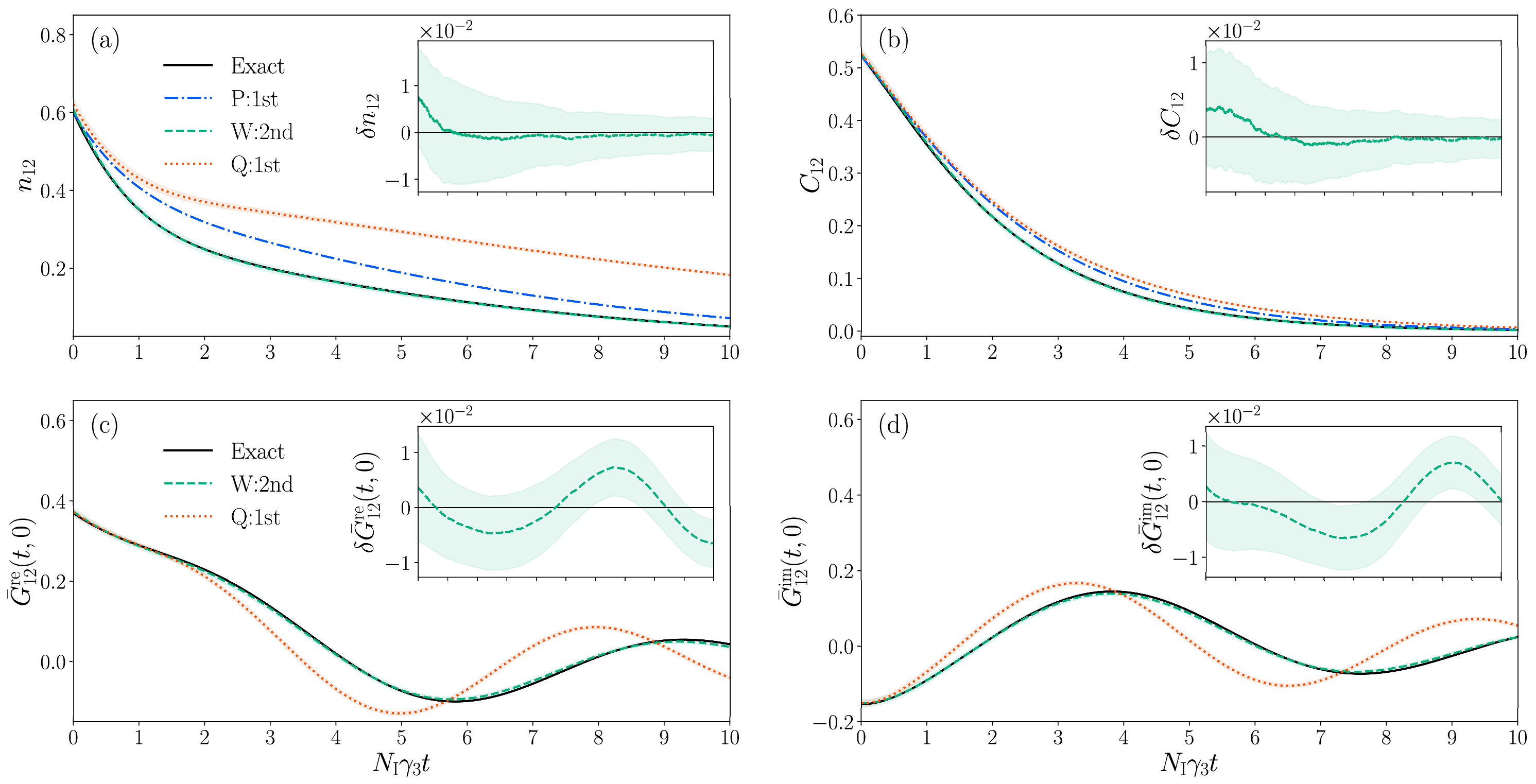}
	\caption{Relaxation dynamics of two-component BEC obeying the GKSL equation~\eqref{eq:GKSL equation Model3} starting from the pure coherent state $\hat{\rho}(0) = \ket{\alpha_{{\rm I}1},\alpha_{{\rm I}2}}\bra{\alpha_{{\rm I}1},\alpha_{{\rm I}2}}$, where $\alpha_{{\rm I}1} = \sqrt{N_{{\rm I}1}}e^{i\pi/8}$ and $\alpha_{{\rm I}2} = \sqrt{N_{{\rm I}2}}e^{i\pi/4}$ with $N_{{\rm I}1} = 8$ and $N_{{\rm I}2} = 2$ at $\mu/(\hbar N_{\rm I}\gamma_3) = 1$, $U_{11}/(\hbar\gamma_3) = U_{22}/(\hbar\gamma_3) = U_{12}/(\hbar\gamma_3) = 1$, $\gamma_1/(N_{\rm I}\gamma_3) = 0.1$, and $\gamma_2/(N_{\rm I}\gamma_3) = 1$, where $N_{\rm I} = N_{{\rm I}1} + N_{{\rm I}2} = 10$.
    The quantities and notations are the same as those in Fig.~\ref{fig:Model2} except that we use the first-order approximation using the Husimi Q function.
    The insets depict the difference between the results of the second-order approximation using the Wigner function (W:$2$nd) and the numerically exact one (Exact).}
	\label{fig:Model3}
\end{figure}

In this system, the second-order calculation is feasible when we use the Wigner function.
The matrices $\bm{\lambda}^{\vec{s} = (s,s)}$ and $\bm{\Lambda}^{\vec{s} = (s,s)}$ are given by
\begin{gather}
    \label{eq:lambda for Model3}
    \bm{\lambda}^{\vec{s} = (s,s)} = \frac{1}{2}
    \begin{bmatrix}
        i\dfrac{sU_{11}}{\hbar}\alpha_1^2 & \left(\dfrac{\gamma_3}{2} + i\dfrac{sU_{12}}{\hbar}\right)\alpha_1\alpha_2 \\
        \left(\dfrac{\gamma_3}{2} + i\dfrac{sU_{12}}{\hbar}\right)\alpha_1\alpha_2 & i\dfrac{sU_{22}}{\hbar}\alpha_2^2
    \end{bmatrix},\\
    \label{eq:Lambda for Model3}
    \bm{\Lambda}^{\vec{s} = (s,s)} = \frac{1}{4}
    \begin{bmatrix}
        \gamma_1(1-s) + \gamma_3(1-s)\left(|\alpha_2|^2 + \dfrac{1+s}{2}\right) & 0 \\
        0 & \gamma_2(1-s) + \gamma_3(1+s)\left(|\alpha_1|^2 - \dfrac{1-s}{2}\right)
    \end{bmatrix}.
\end{gather}
Substituting Eqs.~\eqref{eq:lambda for Model3} and \eqref{eq:Lambda for Model3} into Eq.~\eqref{eq:definitnioa of the diffusion matrix A}, we obtain the matrix $\bm{\mathcal{A}}^{\vec{s}=(s,s)}$.
When $\bm{\mathcal{A}}^{\vec{s}}$ is positive-semidefinite, we obtain the following stochastic differential equations:
\begin{gather}
    \label{eq:SDE for site1 Model3}
    i\hbar d\alpha_1 = \left[-\mu\alpha_1 + U_{11}\alpha_1(|\alpha_1|^2 - 1 + s) + U_{12}\alpha_1\left(|\alpha_2|^2 - \frac{1-s}{2}\right) - \frac{i\hbar}{2}\alpha_1\left\{\gamma_3\left(|\alpha_2|^2 + \frac{1+s}{2}\right) + \gamma_1\right\}\right]dt + i\hbar\left[i\bm{\mathcal{U}}^{\vec{s}}\sqrt{\bm{\mathcal{A}}^{\vec{s}}_{\rm diag}}\cdot d\overrightarrow{\mathcal{W}}\right]_1, \\
    \label{eq:SDE for site2 Model3}
    i\hbar d\alpha_2 = \left[-\mu\alpha_2 + U_{22}\alpha_2(|\alpha_2|^2 - 1 + s) + U_{12}\alpha_2\left(|\alpha_1|^2 - \frac{1-s}{2}\right) + \frac{i\hbar}{2}\alpha_2\left\{\gamma_3\left(|\alpha_1|^2 - \frac{1+s}{2}\right) - \gamma_2\right\}\right]dt + i\hbar\left[i\bm{\mathcal{U}}^{\vec{s}}\sqrt{\bm{\mathcal{A}}^{\vec{s}}_{\rm diag}}\cdot d\overrightarrow{\mathcal{W}}\right]_2.
\end{gather}
When we choose $s=0$ and analytically diagonalize $\bm{\mathcal{A}}^{\vec{s}=(0,0)}$, we find that if $\gamma_2$ and $\gamma_3$ satisfy the condition:
\begin{align}
    \label{eq:conditnion for FPE to SDE Model3}
    \frac{\gamma_2}{\gamma_3} \geq \frac{1}{2},
\end{align}
$\bm{\mathcal{A}}^{\vec{s}=(0,0)}$ becomes positive-semidefinite.
Interestingly, the condition Eq.~\eqref{eq:conditnion for FPE to SDE Model3} do not impose any restrictions on $\gamma_1$.
We choose the parameters for the numerical simulation such that Eq.~\eqref{eq:conditnion for FPE to SDE Model3} is satisfied.
On the other hand, for the case of $s=1$, we can analytically show that $\bm{\mathcal{A}}^{\vec{s}=(1,1)}$ always involves negative eigenvalues independently of parameters.
When we choose $s=-1$, we numerically confirm that $\bm{\mathcal{A}}^{\vec{s}=(-1,-1)}$ involves at least one negative eigenvalue at almost all initial points sampled from the initial Husimi Q function.
Hence, for the cases of $s=1$ and $s=-1$, we use the first-order approximation and solve the classical equations of motion, which are given by Eqs.~\eqref{eq:SDE for site1 Model3} and \eqref{eq:SDE for site2 Model3} without the stochastic terms.

Fig.~\ref{fig:Model3} shows the relaxation dynamics of $n_{12}$, $C_{12}$, and $\bar{G}_{12}(t,0)$.
Initially, $N_{{\rm I}1} = 8$ and $N_{{\rm I}2} = 2$ atoms are condensed into a pure coherent state.
We choose the other parameters as $\mu/(\hbar N_{\rm I}\gamma_3) = 1$, $J/(\hbar N_{\rm I}\gamma_3) = 1$, $U_{11}/(\hbar\gamma_3) = U_{22}/(\hbar\gamma_3) = U_{12}/(\hbar\gamma_3) = 1$, $\gamma_1/(N_{\rm I}\gamma_3) = 0.1$, and $\gamma_2/(N_{\rm I}\gamma_3) = 1$.
The inset shows the difference between the result of the second-order approximation using the Wigner function (W:$2$nd) and the numerically exact one (Exact).
In all the panels, although the results of the second-order approximation (W:$2$nd) well reproduces the exact dynamics, there are large discrepancies between the results of the first-order approximations (P:$1$st and Q:$1$st) and the numerically exact one.
This result suggests that the effects of the second order of quantum fluctuations strongly affect the relaxation dynamics, and the use of the second-order approximation using the Wigner function is an appropriate choice for simulating this model in the phase space.


\subsection{\label{subsec:Model4}Model 4: Bose-Hubbard model with a hybrid of quasiprobability distribution functions}
Below, we perform the benchmark calculation by hybridizing the different quasiprobability distribution functions, i.e.,
we choose $s_m\neq s_n$ for $m\neq n$.
We consider the Bose-Hubbard model obeying the following GKSL equation:
\begin{gather}
    \label{eq:GKSL equation Model4}
    \frac{d\hat{\rho}(t)}{dt} = -\frac{i}{\hbar}\left[\hat{H}_{\rm BH},\hat{\rho}(t)\right]_- + \sum_{k=1,2,3}\gamma_k\left(\hat{L}_k\hat{\rho}(t)\hat{L}_k^{\dagger} - \frac{1}{2}\left[\hat{L}_k^{\dagger}\hat{L}_k,\hat{\rho}(t)\right]_+\right), \\
    \label{eq:Jump operators Model4}
    \hat{L}_1 = \hat{a}_1 + \hat{a}_2,~\hat{L}_2 = \hat{a}_2,~\hat{L}_3 = \hat{a}^{\dagger}_1\hat{a}_1,
\end{gather}
where $\hat{H}_{\rm BH}$ is the Hamiltonian of the two-site Bose-Hubbard model given by Eq.~\eqref{eq:Hamiltonian Model2}, $\hat{L}_1$, $\hat{L}_2$, and $\hat{L}_3$ respectively describe the non-local one-body loss of atoms at site $1$ and $2$, one-body loss of atoms at site $2$, and the dephasing of atoms at site $1$, and $\gamma_1$, $\gamma_2$, and $\gamma_3$ represent their strength, respectively.
\begin{figure}[t]
	\centering 
	\includegraphics[width = \linewidth]{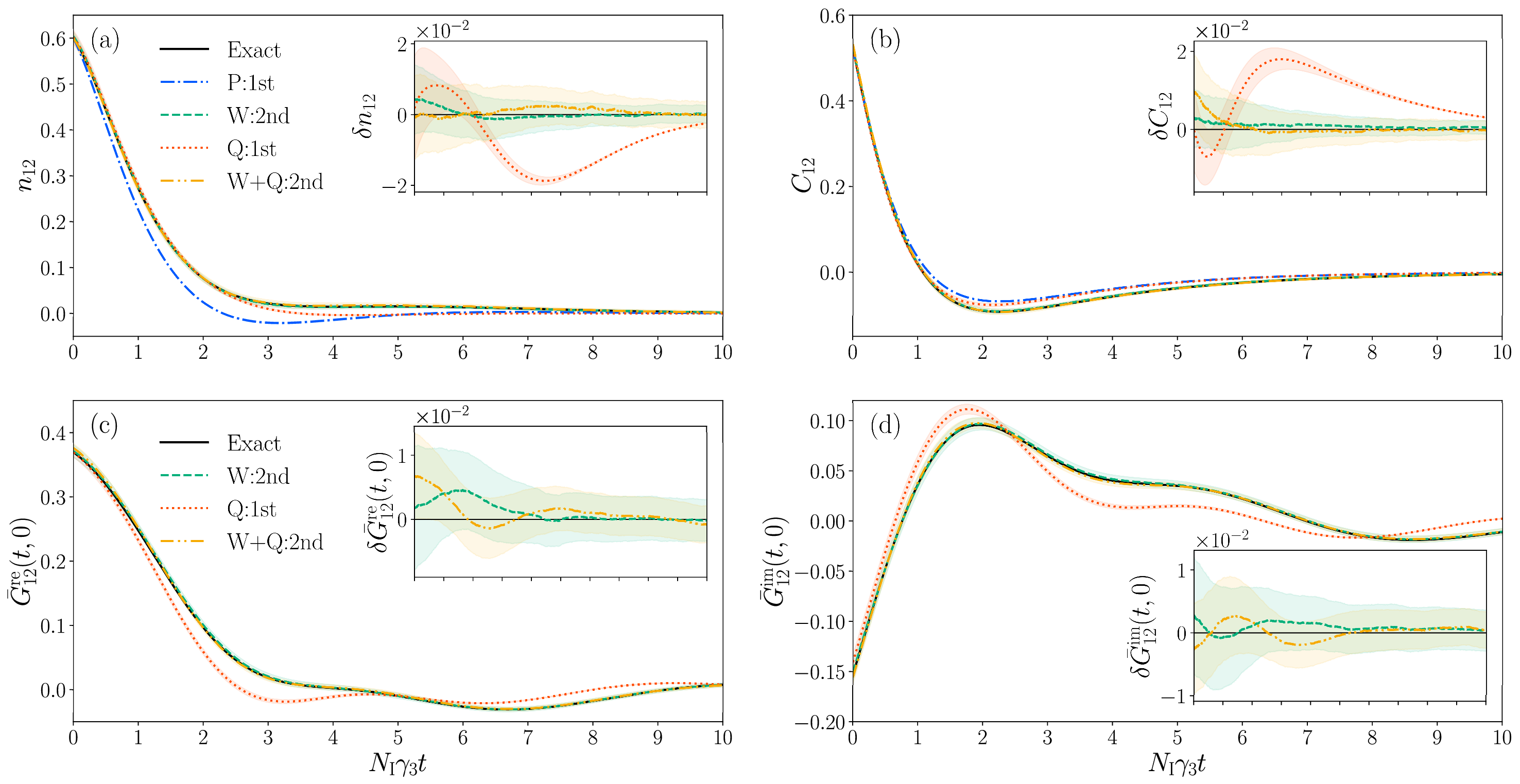}
	\caption{Relaxation dynamics of two-site Bose-Hubbard model obeying the GKSL equation~\eqref{eq:GKSL equation Model4} starting from the pure coherent state $\hat{\rho}(0) = \ket{\alpha_{{\rm I}1},\alpha_{{\rm I}2}}\bra{\alpha_{{\rm I}1},\alpha_{{\rm I}2}}$, where $\alpha_{{\rm I}1} = \sqrt{N_{{\rm I}1}}e^{i\pi/8}$ and $\alpha_{{\rm I}2} = \sqrt{N_{{\rm I}2}}e^{i\pi/4}$ with $N_{{\rm I}1} = 8$ and $N_{{\rm I}2} = 2$ The other parameters are $\mu/(\hbar N_{\rm I}\gamma_3) = 1$, $J/(\hbar N_{\rm I}\gamma_3) = 0.4$, $U_{11}/(\hbar\gamma_3) = 1$, $U_{22}/(\hbar\gamma_3) = 0.2$, $\gamma_1/(N_{\rm I}\gamma_3) = 0.25$, and $\gamma_2/(N_{\rm I}\gamma_3) = 1.0$, where $N_{\rm I} = N_{{\rm I}1} + N_{{\rm I}2} = 10$. 
    The notations in this figure are the same as those in Figs.~\ref{fig:Model3} except that we use the second-order approximation for a hybrid of the Wigner and Husimi Q functions.
    The insets depict the difference between the results of the second-order approximations (W:$2$nd and W+Q:$2$nd) and the numerically exact one (Exact).}
	\label{fig:Model4}
\end{figure}

In this model, we can perform the second-order calculation when we use the Wigner function and a hybrid of the Wigner and Husimi Q functions.
The matrices $\bm{\lambda}^{\vec{s}}$ and $\bm{\Lambda}^{\vec{s}}$ are given by
\begin{gather}
    \label{eq:lambda for Model4}
    \bm{\lambda}^{\vec{s}} = \frac{1}{2}
    \begin{bmatrix}
        \gamma_3\alpha_1^2 + i\dfrac{s_1U_{11}}{\hbar}\alpha^2_1 & 0 \\
        0 & i\dfrac{s_2U_{22}}{\hbar}\alpha^2_2
    \end{bmatrix},\\
    \label{eq:Lambda for Model4}
    \bm{\Lambda}^{\vec{s}} = \frac{1}{4}
    \begin{bmatrix}
        \gamma_1 (1-s_1) + 2\gamma_3|\alpha_1|^2& \gamma_1\left(1 - \dfrac{s_1 + s_2}{2}\right) - i\dfrac{s_1-s_2}{\hbar}J \\
        \gamma_1\left(1 - \dfrac{s_1+s_2}{2}\right) + i\dfrac{s_1-s_2}{\hbar}J & (\gamma_1+\gamma_2)(1-s_2)
    \end{bmatrix},
\end{gather}
where $\vec{s} = (s_1,s_2)$.
Substituting Eqs.~\eqref{eq:lambda for Model4} and \eqref{eq:Lambda for Model4} into Eq.~\eqref{eq:definitnioa of the diffusion matrix A}, we obtain the matrix $\bm{\mathcal{A}}^{\vec{s}}$.
When $\bm{\mathcal{A}}^{\vec{s}}$ is positive-semidefinite, we obtain the following stochastic differential equations:
\begin{align}
    \label{eq:SDE for site1 Model4}
    i\hbar d\alpha_1 &= \left[-\mu\alpha_1 - J\alpha_2 + U_{11}\alpha_1(|\alpha_1|^2 - 1 + s_1) - \frac{i\hbar}{2}\left\{(\gamma_1 + \gamma_3)\alpha_1 + \gamma_1\alpha_2\right\}\right]dt + i\hbar\left[i\bm{\mathcal{U}}^{\vec{s}}\sqrt{\bm{\mathcal{A}}^{\vec{s}}_{\rm diag}}\cdot d\overrightarrow{\mathcal{W}}\right]_1, \\
    \label{eq:SDE for site2 Model4}
    i\hbar d\alpha_2 &= \left[-\mu\alpha_2 - J\alpha_1 + U_{22}\alpha_2(|\alpha_2|^2 - 1 + s_2) - \frac{i\hbar}{2}\left\{\gamma_1\alpha_1 + (\gamma_1 + \gamma_2)\alpha_2\right\} \right]dt + i\hbar\left[i\bm{\mathcal{U}}^{\vec{s}}\sqrt{\bm{\mathcal{A}}^{\vec{s}}_{\rm diag}}\cdot d\overrightarrow{\mathcal{W}}\right]_2.
\end{align}
When we choose $(s_1,s_2) = (0,0)$, by analytically diagonalizing the the matrix $\bm{\mathcal{A}}^{\vec{s}=(0,0)}$, we can confirm that $\bm{\mathcal{A}}^{\vec{s}=(0,0)}$ is always positive-semidefinite.
When we choose $(s_1,s_2) = (0,-1)$, although $\bm{\mathcal{A}}^{\vec{s}=(0,-1)}$ is not always positive-definite depending on the values of $\vec{\alpha}$, we numerically confirm that $\bm{\mathcal{A}}^{\vec{s}=(0,-1)}$ is always positive semidefinite at least under our parameter setting, which we will show below, and we can simulate the second-order calculation.
On the other hand, for the case of $(s_1,s_2) = (1,1)$, we can show that $\bm{\mathcal{A}}^{\vec{s}=(1,1)}$ always involves negative eigenvalues independently of parameters.
When we choose $(s_1,s_2) = (-1,-1)$, we numerically confirm $\bm{\mathcal{A}}^{\vec{s}=(-1,-1)}$ involves at least one negative eigenvalue at almost all initial points sampled from the initial Husimi Q function.
Hence, for the cases of $(s_1,s_2) = (1,1)$ and $(-1,-1)$, we use the first-order approximation and solve the classical equations of motion, which are given by Eqs~\eqref{eq:SDE for site1 Model4} and \eqref{eq:SDE for site2 Model4} with neglecting the stochastic terms.

Fig.~\ref{fig:Model4} shows the relaxation dynamics of $n_{12}$, $C_{12}$, and $\bar{G}_{12}(t,0)$.
The initial mean atomic numbers are $N_{{\rm I}1} = 8$ and $N_{{\rm I}2} = 2$, and the parameters are $\mu/(\hbar N_{\rm I}\gamma_3) = 1$, $J/(\hbar N_{\rm I}\gamma_3) = 0.4$, $U_{11}/(\hbar\gamma_3) = 1$, $U_{22}/(\hbar\gamma_3) = 0.2$, $\gamma_1/(N_{\rm I}\gamma_3) = 0.25$, and $\gamma_2/(N_{\rm I}\gamma_3) = 1.0$.
In all the panels, there are large discrepancies between the results of the first-order approximation (P:$1$st and Q:$1$st) and the numerically exact one (Exact),
whereas the results of the second-order approximation (W:$2$nd and W+Q:$2$nd) well reproduce the exact dynamics.

\section{\label{sec:Summary and conclusions}Summary and conclusions}
The phase-space formalism of a quantum state enables us to investigate the bosonic quantum many-body dynamics while taking into account the effects of quantum fluctuations, where bosonic operators are mapped into $c$-number functions and the density operator is represented as a quasiprobability distribution function, such as the Glauber-Sudarshan P, Wigner, and Husimi Q function.
In the phase space, the GKSL equation is approximated into the Fokker-Planck equation for the quasiprobability distribution function.
To investigate the dynamics following the Fokker-Planck equation, we usually derive the corresponding stochastic differential equations and perform the Monte Carlo simulation.
However, the Fokker-Planck equation does not always reduce to the stochastic differential equation because the diffusion matrix is not necessarily positive-semidefinite and may have negative eigenvalue depending on the details of the Hamiltonian and jump operators and the choice of the quasiprobability distribution function.
In this work, we have analytically derived the diffusion matrix and stochastic differential equation [Eq.~\eqref{eq:stochastic differential equation}] for arbitrary Hamiltonian and jump operators without using the Fokker-Planck equation, obtaining the conditions for describing quantum systems with stochastic differential equations.

Our derivation is based on the path-integral formalism.
In the course of the derivation, we use the $s$-ordered quasiprobability distribution function, which reduces to the Glauber-Sudarshan, Wigner, and Husimi Q function by choosing $s=1$, $s=0$, and $s=-1$, respectively.
For a system with multiple degrees of freedom, we hybridize the quasiprobability distribution functions by choosing different values of $s$ for different degrees of freedom.
Based on the $s$-ordered phase-space mapping, we formulate the path-integral representation for the GKSL equation [Eqs.~\eqref{eq:path-integral representaiton discrete} and \eqref{eq:path-integral representaiton continuous}],
where the action includes classical and quantum fields.
Expanding the action with respect to the quantum fields up to second order leads to the stochastic differential equation.
Here, in order to integrate out the quantum fields,
we perform the Hubbard-Stratonovich transformation, which is feasible when the Hamiltonian and jump operators satisfy the condition [Eq.~\eqref{eq:feasible condition of the Hubbard-Stratonovich transformation}].
This condition is identical to the positive-semidefiniteness condition for the diffusion matrix of the Fokker-Planck equation.

In the benchmark calculations, we investigate the relaxation dynamics of physical quantities including non-equal time correlation functions.
In all the models we calculated, the second-order approximation, when available,  reproduces the exact dynamics regardless of the values of $s$. However, whether we can use the second-order approximation, i.e., whether the stochastic differential equation is available, strongly depends on the choice of the quasiprobability distribution function and the details of the Hamiltonian and jump operators. When it is unavailable, we used the first-order approximation and found a non-negligible deviation from the exact result in some physical quantities.

The condition [Eq.~\eqref{eq:feasible condition of the Hubbard-Stratonovich transformation}] for obtaining the stochastic differential equations is well satisfied when we use the Wigner function, which corresponds to the truncated Wigner approximation.
Empirically, the use of the Glauber-Sudarshan P function tends to violate the condition rather than using the Husimi Q function as shown in the benchmark calculation in Sec.~\ref{subsec:Model2}.
However, if we consider a non-interacting Hamiltonian and linear jump operators involving only loss terms, the use of the Glauber-Sudarshan P function is most efficient because of the absence of the effects of second order of quantum fluctuations, which enables us to avoid handling the stochastic terms and reduce the numerical cost as shown in Sec.~\ref{subsec:Model1}.

As shown in Sec.~\ref{subsec:Non-equal time correlation functions}, the hybridized use of the quasiprobability distribution functions is required for the calculation of the higher-order non-equal time correlation functions of different degrees of freedom.
However, in this case, the second-order approximation is often unfeasible.
In Ref.~\cite{Huber}, the authors found that in some cases, the diffusion matrix can be proved to be positive semidefinite if they removed several terms in the diffusion matrix. They thus proposed to ignore the terms, the positive diffusion approximation.
We expect that by using the hybridized quasiprobability distribution functions with the positive diffusion approximation, we can calculate various non-equal time correlation functions while considering the effects of the second order of quantum fluctuations as much as possible.
It is also interesting to investigate the properties of the Hamiltonian and jump operators where the diffusion matrix is always positive-semidefinite, which will be published elsewhere.
In isolated systems, the path-integral representation enables us to take into account a part of the effects of third order of quantum fluctuations and go beyond the truncated Wigner approximation, which is referred to as the quantum jump method \cite{Polkovnikov2003}.
We expect that by using the quantum jump method with our formulation, we can investigate the open quantum many-body dynamics by taking into account the effects of higher order of quantum fluctuations.

As argued in Sec.~\ref{subsec:Second order of quantum fluctuations}, when the matrix $\bm{\mathcal{A}}^{\vec{s}}$ is positive-semidefinite, we can obtain the stochastic differential equation [Eq.~\eqref{eq:stochastic differential equation}] and perform the second-order calculation.
However, when $\bm{\mathcal{A}}^{\vec{s}}$ violates the positive-semidefiniteness condition, our formulation cannot provide any computable scheme and give it to any physical interpretation.
In order to study this point, we need to directly solve the Fokker-Planck equation [Eq.~\eqref{eq:Fokker-Planck equation}] and investigate the details of the quasiprobability distribution function, which needs high numerical cost.
This is out of the scope of this work and is a remaining issue.

\section*{Acknowledgements}
This work was supported by JSPS, Japan KAKENHI (Grant Numbers JP24K00557 and JP23K13029) and Grant-in-Aid for JSPS Fellows (Grant Number JP25KJ1414).
\appendix


\section{Kraus representation in the phase space}
We derive the propagator of the $\vec{s}$-ordered quasiprobability distribution function \eqref{eq: phase space representation of non Markov propagator} and show that it satisfies the Markov condition Eq.~\eqref{eq:Markov conditoin in the phase space} when the Kraus operator satisfies the Markov condition.

\subsection{\label{appendix:The propagator in the Kraus representation}The propagator in the Kraus representation{\rm :} Derivation of Eq.~\texorpdfstring{\eqref{eq: phase space representation of non Markov propagator}}{TEXT}}
Using Eqs.~\eqref{eq:definition of the s-ordered quasiprobability distribution function} and \eqref{eq:definition of the characteristic function of the s-ordered quasiprobability distribution function}, we obtain the ($-\vec{s}$)-ordered phase-space representation of the left- and right-hand sides of Eq.~\eqref{eq:Kraus_representation} as follows:
\begin{align}
    \label{eq:phase space representation of the Kraus representation 1}
    W_{\vec{s}}(\vec{\alpha}_{\rm f},\vec{\alpha}^*_{\rm f},t) &=  \left[\sum_k\hat{M}_k(t,t_0)\hat{\rho}(t_0)\hat{M}^{\dagger}_k(t,t_0)\right]_{-\vec{s}}(\vec{\alpha}_{\rm f},\vec{\alpha}^*_{\rm f}) \\
    &= \int\frac{d^2\vec{\xi} }{\pi^M}\sum_k{\rm Tr}\left[\hat{D}^{\dagger}(\vec{\xi},\vec{s})\hat{M}_k(t,t_0)\hat{\rho}(t_0)\hat{M}^{\dagger}_k(t,t_0)\right]e^{\vec{\alpha}_{\rm f}^*\cdot\vec{\xi} - \vec{\alpha}_{\rm f}\vec{\xi}^*} \\
    \label{eq:phase space representation of the Kraus representation 2}
    &= \int\frac{d^2\vec{\alpha}_0d^2\vec{\xi} }{\pi^{2M}}\sum_k\left[\hat{M}^{\dagger}_k(t,t_0)\hat{D}^{\dagger}(\vec{\xi},\vec{s})\hat{M}_k(t,t_0)\right]_{\vec{s}}(\vec{\alpha}_0,\vec{\alpha}^*_0)e^{\vec{\alpha}_{\rm f}^*\cdot\vec{\xi}-\vec{\alpha}_{\rm f}\cdot\vec{\xi}^* }W_{\vec{s}}(\vec{\alpha}_0,\vec{\alpha}^*_0,t_0),
\end{align}
where we explicitly denote the characteristic function \eqref{eq:definition of the characteristic function of the s-ordered quasiprobability distribution function} of the Kraus representation for the second equality and obtain the last line by using the cyclic property of the trace and Eq.~\eqref{eq:s-parametrized representation of TrAB}.
Here, we express $[\hat{M}^{\dagger}_k(t,t_0)\hat{D}^{\dagger}(\vec{\xi},\vec{s})\hat{M}_k(t,t_0)]_{\vec{s}}$ on the right-hand side of Eq.~\eqref{eq:phase space representation of the Kraus representation 2} in the integral form by using Eqs.~\eqref{eq:s-parametrized mapping of A} and \eqref{eq:definition of the characteristic function} as
\begin{align}
    \label{eq:phase space representation of MDM}
    \left[\hat{M}^{\dagger}_k(t,t_0)\hat{D}^{\dagger}(\vec{\xi},\vec{s} )\hat{M}_k(t,t_0)\right]_{\vec{s}}(\vec{\alpha}_0,\vec{\alpha}^*_0) = \int\frac{d^2\vec{\eta}}{\pi^M}{\rm Tr}\left[\hat{D}^{\dagger}(\vec{\eta},-\vec{s})\hat{M}^{\dagger}_k(t,t_0)\hat{D}^{\dagger}(\vec{\xi},\vec{s} )\hat{M}_k(t,t_0)\right]e^{\vec{\alpha}^*_0\cdot\vec{\eta} - \vec{\alpha}_0\cdot\vec{\eta}^*}.
\end{align}
Substituting this expression into Eq.~\eqref{eq:phase space representation of the Kraus representation 2}, we obtain
\begin{gather}
    W_{\vec{s}}(\vec{\alpha}_{\rm f},\vec{\alpha}^*_{\rm f},t) = \int\frac{d^2\vec{\alpha}_0}{\pi^M}\varUpsilon_{\vec{s}}(\vec{\alpha}_{\rm f},t;\vec{\alpha}_0,t_0)W_{\vec{s}}(\vec{\alpha}_0,\vec{\alpha}^*_0,t_0), \\
    \label{eq: phase space representation of non Markov propagator appendix}
    \varUpsilon_{\vec{s}}(\vec{\alpha}_{\rm f},t;\vec{\alpha}_0,t_0) = \int\frac{d^2\vec{\xi} d^2\vec{\eta} }{\pi^{2M}}\sum_k{\rm Tr}\left[\hat{D}^{\dagger}(\vec{\xi},\vec{s})\hat{M}_k(t,t_0)\hat{D}^{\dagger}(\vec{\eta},-\vec{s})\hat{M}^{\dagger}_k(t,t_0)\right]e^{\vec{\alpha}_{\rm f}^*\cdot\vec{\xi} - \vec{\alpha}_{\rm f}\cdot\vec{\xi}^*}e^{\vec{\alpha}^*_0\cdot\vec{\eta} - \vec{\alpha}_0\cdot\vec{\eta}^*},
\end{gather}
which are Eqs.~\eqref{eq:Kraus representation in the phase space} and \eqref{eq: phase space representation of non Markov propagator}, respectively.


\subsection{\label{appendix:Markov condition for the propagator}Markov condition for the propagator{\rm :} Derivation of Eq.~\texorpdfstring{\eqref{eq:Markov conditoin in the phase space}}{TEXT}}
We can write the Markov dynamics of the system as $\hat{\rho}(t) = \hat{\mathcal{V}}(t,t_0)[\hat{\rho}(t_0)] = \hat{\mathcal{V}}(t,t_j)[\hat{\mathcal{V}}(t_j,t_0)[\hat{\rho}(t_0)]]$ for $t \geq t_j \geq t_0$, which is equivalent to
\begin{align}
    \label{eq:Markov_conditoin_kraus_representation}
    \sum_k\hat{M}_k(t,t_0)\hat{\rho}(t_0)\hat{M}^{\dagger}_k(t,t_0) = \sum_{k,k'}\hat{M}_{k'}(t,t_j)\hat{M}_{k}(t_j,t_0)\hat{\rho}(t_0)\hat{M}^{\dagger}_{k}(t_j,t_0)\hat{M}^{\dagger}_{k'}(t,t_j)
\end{align}
in the Kraus representation.
Applying the same procedure from Eq.~\eqref{eq:phase space representation of the Kraus representation 1} to Eq.~\eqref{eq: phase space representation of non Markov propagator appendix} to the right-hand side of Eq.~\eqref{eq:Markov_conditoin_kraus_representation}, we obtain
\begin{align}
    \varUpsilon_{\vec{s}}(\vec{\alpha}_{\rm f},t;\vec{\alpha}_0,t_0) &= \int\frac{d^2\vec{\xi}' d^2\vec{\eta}}{\pi^{2M}}\sum_{k,k'}{\rm Tr}\left[\hat{D}^{\dagger}(\xi',\vec{s})\hat{M}_{k'}(t,t_j)\hat{M}_k(t_j,t_0)\hat{D}^{\dagger}(\vec{\eta},-\vec{s})\hat{M}^{\dagger}_k(t_j,t_0)\hat{M}^{\dagger}_{k'}(t,t_j)\right]e^{\vec{\alpha}_{\rm f}^*\cdot\vec{\xi}' - \vec{\alpha}_{\rm f}\cdot\vec{\xi}'^* }e^{\vec{\alpha}^*_0\cdot\vec{\eta} - \vec{\alpha}_0\cdot\vec{\eta}^*} \\
    &= \int\frac{d^2\vec{\alpha}_j}{\pi^M}\int\frac{d^2\vec{\xi}'}{\pi^M}\sum_{k'}\left[\hat{M}^{\dagger}_{k'}(t,t_j)\hat{D}^{\dagger}(\vec{\xi}',\vec{s} )\hat{M}_{k'}(t,t_j)\right]_{\vec{s}}(\vec{\alpha}_j,\vec{\alpha}^*_j)e^{\vec{\alpha}_{\rm f}^*\cdot\vec{\xi}' - \vec{\alpha}_{\rm f}\cdot\vec{\xi}'^*}\nonumber \\
    \label{eq:Markov condition in the phase space 1}
    &\hphantom{\int\frac{d^2\vec{\alpha}_j}{\pi^M}}\times\int\frac{d^2\vec{\eta}}{\pi^M}\sum_k\left[\hat{M}_k(t_j,t_0)\hat{D}^{\dagger}(\vec{\eta},-\vec{s})\hat{M}^{\dagger}_k(t_j,t_0)\right]_{-\vec{s}}(\vec{\alpha}_j,\vec{\alpha}^*_j)e^{\vec{\alpha}^*_0\cdot\vec{\eta} - \vec{\alpha}_0\cdot\vec{\eta}^*},
\end{align}
where we use the cyclic property of the trace and Eq.~\eqref{eq:s-parametrized representation of TrAB} for the second equality.
By using Eqs.~\eqref{eq:s-parametrized mapping of A} and \eqref{eq:definition of the characteristic function}, we can rewrite 
$[\hat{M}^{\dagger}_{k'}(t,t_j)\hat{D}^{\dagger}(\vec{\xi}',\vec{s})\hat{M}_{k'}(t,t_j)]_{\vec{s}}$ and $[\hat{M}_k(t_j,t_0)\hat{D}^{\dagger}(\vec{\eta},-\vec{s})\hat{M}^{\dagger}_k(t_j,t_0)]_{-\vec{s}}$
in the integral forms as follows:
\begin{gather}
    \left[\hat{M}^{\dagger}_{k'}(t,t_j)\hat{D}^{\dagger}(\vec{\xi}',\vec{s})\hat{M}_{k'}(t,t_j)\right]_{\vec{s}}(\vec{\alpha}_j,\vec{\alpha}^*_j) = \int\frac{d^2\vec{\eta}'}{\pi^M}{\rm Tr}\left[\hat{D}^{\dagger}(\vec{\xi}',\vec{s})\hat{M}_{k'}(t,t_j)\hat{D}^{\dagger}(\vec{\eta}',-\vec{s})\hat{M}^{\dagger}_{k'}(t,t_j)\right]e^{\vec{\alpha}^*_j\cdot\vec{\eta}' - \vec{\alpha}_j\cdot\vec{\eta}'^*},\\
    \left[\hat{M}_k(t_j,t_0)\hat{D}^{\dagger}(\vec{\eta},-\vec{s})\hat{M}^{\dagger}_k(t_j,t_0)\right]_{-\vec{s}}(\vec{\alpha}_j,\vec{\alpha}^*_j) = \int\frac{d^2\vec{\xi}}{\pi^M}{\rm Tr}\left[\hat{D}^{\dagger}(\vec{\xi},\vec{s})\hat{M}_k(t_j,t_0)\hat{D}^{\dagger}(\vec{\eta},-\vec{s})\hat{M}^{\dagger}_k(t_j,t_0)\right]e^{\vec{\alpha}^*_j\cdot\vec{\xi} - \vec{\alpha}_j\cdot\vec{\xi}^*}.
\end{gather}
Substituting these expressions into Eq.~\eqref{eq:Markov condition in the phase space 1}, we obtain
\begin{align}
    \label{eq:Markov condition in the phase space 2}
    \varUpsilon_{\vec{s}}(\vec{\alpha}_{\rm f},t;\vec{\alpha}_0,t_0) &= \int\frac{d^2\vec{\alpha}_j}{\pi^M}\int\frac{d^2\vec{\xi}'d^2\vec{\eta}'}{\pi^{2M}}\sum_{k'}{\rm Tr}\left[\hat{D}^{\dagger}(\vec{\xi}',\vec{s})\hat{M}_{k'}(t,t_j)\hat{D}^{\dagger}(\vec{\eta}',-\vec{s})\hat{M}^{\dagger}_{k'}(t,t_j)\right]e^{\vec{\alpha}_{\rm f}^*\cdot\vec{\xi}' - \vec{\alpha}_{\rm f}\cdot\vec{\xi}'^*}e^{\vec{\alpha}^*_j\cdot\vec{\eta}' - \vec{\alpha}_j\cdot\vec{\eta}'^*}\nonumber \\
    &\hphantom{\int\frac{d^2\vec{\alpha}_j}{\pi^M}}\times\int\frac{d^2\vec{\xi} d^2\vec{\eta}}{\pi^{2M}}\sum_k{\rm Tr}\left[\hat{D}^{\dagger}(\vec{\xi},\vec{s})\hat{M}_k(t_j,t_0)\hat{D}^{\dagger}(\vec{\eta},-\vec{s})\hat{M}^{\dagger}_k(t_j,t_0)\right]e^{\vec{\alpha}^*_j\cdot\vec{\xi} - \vec{\alpha}_j\cdot\vec{\xi}^*}e^{\vec{\alpha}^*_0\cdot\vec{\eta} - \vec{\alpha}_0\cdot\vec{\eta}^*}.
\end{align}
Finally, using the expressions
\begin{gather}
    \varUpsilon_{\vec{s}}(\vec{\alpha}_{\rm f},t;\vec{\alpha}_j,t_j) = \int\frac{d^2\vec{\xi}'d^2\vec{\eta}'}{\pi^{2M}}\sum_{k'}{\rm Tr}\left[\hat{D}^{\dagger}(\vec{\xi}',\vec{s})\hat{M}_{k'}(t,t_j)\hat{D}^{\dagger}(\vec{\eta}',-\vec{s})\hat{M}^{\dagger}_{k'}(t,t_j)\right]e^{\vec{\alpha}_{\rm f}^*\cdot\vec{\xi}' - \vec{\alpha}_{\rm f}\cdot\vec{\xi}'^*}e^{\vec{\alpha}^*_j\cdot\vec{\eta}' - \vec{\alpha}_j\cdot\vec{\eta}'^*},\\
    \varUpsilon_{\vec{s}}(\vec{\alpha}_j,t_j;\vec{\alpha}_0,t_0) = \int\frac{d^2\vec{\xi} d^2\vec{\eta}}{\pi^{2M}}\sum_k{\rm Tr}\left[\hat{D}^{\dagger}(\vec{\xi},\vec{s})\hat{M}_k(t_j,t_0)\hat{D}^{\dagger}(\vec{\eta},-\vec{s})\hat{M}^{\dagger}_k(t_j,t_0)\right]e^{\vec{\alpha}^*_j\cdot\vec{\xi} - \vec{\alpha}_j\cdot\vec{\xi}^*}e^{\vec{\alpha}^*_0\cdot\vec{\eta} - \vec{\alpha}_0\cdot\vec{\eta}^*},
\end{gather}
we can rewrite Eq.~\eqref{eq:Markov condition in the phase space 2} as
\begin{align}
    \varUpsilon_{\vec{s}}(\vec{\alpha}_{\rm f},t;\vec{\alpha}_0,t_0) = \int\frac{d^2\vec{\alpha}_j}{\pi^M}\varUpsilon_{\vec{s}}(\vec{\alpha}_{\rm f},t;\vec{\alpha}_j,t_j)\varUpsilon_{\vec{s}}(\vec{\alpha}_j,t_j;\vec{\alpha}_0,t_0).
\end{align}
This is the Markov condition for the propagator of the $\vec{s}$-ordered quasiprobability distribution function.


\section{\label{appendix:Infinitesimal time propagator}Infinitesimal time propagator{\rm :} Derivation of Eq.~\texorpdfstring{\eqref{eq:s-parametrized phase space representation of Markov propagator}}{TEXT}}


\subsection{Preliminary calculations}
Before deriving Eq.~\eqref{eq:s-parametrized phase space representation of Markov propagator}, we introduce three useful relations for the $\vec{s}$-ordered phase-space representation.
The first one is the relation between $A_{\vec{s}}(\vec{\alpha},\vec{\alpha}^*)$ and $A_{\vec{0}}(\vec{\alpha},\vec{\alpha}^*)$:
\begin{gather}
    \label{eq:relation between A0 and As}
    {\rm exp}\left(\sum_{m=1}^{M}\frac{s_m}{2}\frac{\partial^2}{\partial\alpha_m\partial\alpha_m^*}\right)A_{\vec{0}}(\vec{\alpha},\vec{\alpha}^*) = A_{\vec{s}}(\vec{\alpha},\vec{\alpha}^*), \\
    \label{eq:relation between A stare B and A stars B}
    {\rm exp}\left(\sum_{m=1}^{M}\frac{s_m}{2}\frac{\partial^2}{\partial\alpha_m\partial\alpha_m^*}\right)\left[A_{\vec{0}}(\vec{\alpha} + \vec{\zeta},\vec{\alpha}^* + \vec{\zeta}^*)\star^e B_{\vec{0}}(\vec{\alpha} + \vec{\xi},\vec{\alpha}^* + \vec{\xi}^*)\right] = A_{\vec{s}}(\vec{\alpha} + \vec{\zeta},\vec{\alpha}^* + \vec{\zeta}^*)\star_{\vec{s}} B_{\vec{s}}(\vec{\alpha} + \vec{\xi},\vec{\alpha}^* + \vec{\xi}^*), \\
    \label{eq:relation between A star B and Ase stars Bse}
    {\rm exp}\left\{\sum_{m=1}^M\left(\zeta_{m}\frac{\partial}{\partial\alpha_{m}} + \xi^*_{m}\frac{\partial}{\partial\alpha^*_{m}}\right)\right\}\left[A_{\vec{s}}(\vec{\alpha},\vec{\alpha}^*)\star_{\vec{s}}B_{\vec{s}}(\vec{\alpha},\vec{\alpha}^*)\right] = A^e_{\vec{s}}(\vec{\alpha} + \vec{\zeta},\vec{\alpha}^* + \vec{\xi}^*)\star_{\vec{s}}B^e_{\vec{s}}(\vec{\alpha} + \vec{\zeta},\vec{\alpha}^* + \vec{\xi}^*),
\end{gather}
where the extended Moyal product $\star^e$ and differential operator $\star_{\vec{s}}$ are defined by Eqs.~\eqref{eq:definition of the extended Moyal product} and \eqref{eq:definition of the s-ordered Moyal product}, respectively, and $A^e_{\vec{s}}(\vec{\alpha} + \vec{\zeta},\vec{\alpha}^* + \vec{\xi}^*)$ and $B^e_{\vec{s}}(\vec{\alpha} + \vec{\zeta},\vec{\alpha}^* + \vec{\xi}^*)$ are the extended $\vec{s}$-ordered phase-space representations defined by Eq.~\eqref{eq:definition of extended s-ordered phase-space representation of A}.
When we choose $\vec{\xi} = \vec{\zeta}$ in Eq.~\eqref{eq:relation between A stare B and A stars B}, it reduces to the one with replacing $\star^e$ as the Moyal product $\star$.
Below, we respectively derive Eqs.~\eqref{eq:relation between A0 and As}--\eqref{eq:relation between A star B and Ase stars Bse}.

{\it Eq.~\eqref{eq:relation between A0 and As}}--Substituting Eqs.~\eqref{eq:s-parametrized mapping of A} and \eqref{eq:definition of the characteristic function} with $s_m=0$ for $\forall m$ into the left-hand side of Eq.~\eqref{eq:relation between A0 and As}, we obtain
\begin{align}
    \text{LHS of Eq.~\eqref{eq:relation between A0 and As}} &= {\rm exp}\left(\sum_{m=1}^{M}\frac{s_m}{2}\frac{\partial^2}{\partial\alpha_m\partial\alpha_m^*}\right)\int\frac{d^2\vec{\eta}}{\pi^M}{\rm Tr}\left[\hat{A}\hat{D}^{\dagger}(\vec{\eta},\vec{0})\right]e^{\vec{\alpha}^*\cdot\vec{\eta} - \vec{\alpha}\cdot\vec{\eta}^*} \\
    &= \int\frac{d^2\vec{\eta}}{\pi^M}{\rm Tr}\left[\hat{A}\hat{D}^{\dagger}(\vec{\eta},\vec{0})\right]e^{\vec{\alpha}^*\cdot\vec{\eta} - \vec{\alpha}\cdot\vec{\eta}^* - \sum_ms_m|\eta_m|^2/2} \\
    &= \int\frac{d^2\vec{\eta}}{\pi^M} {\rm Tr}[AD^\dagger(\vec{\eta},-\vec{s})]e^{\vec{\alpha}^*\cdot\vec{\eta}-\vec{\alpha}\cdot\vec{\eta}^*} \\
    &= A_{\vec{s}}(\vec{\alpha},\vec{\alpha}^*) \\
    &= \text{RHS of Eq.~\eqref{eq:relation between A0 and As}},
\end{align}
where we have used $\hat{D}^{\dagger}(\vec{\eta},\vec{0})e^{- \sum_ms_m|\eta_m|^2/2} = \hat{D}^{\dagger}(\vec{\eta},-\vec{s})$ for the third equality.

{\it Eq.~\eqref{eq:relation between A stare B and A stars B}}--Using Eq.~\eqref{eq:definition of the extended Moyal product}, we can rewrite the left-hand side of Eq.~\eqref{eq:relation between A stare B and A stars B} as
\begin{align}
    \label{eq:derivation of B2}
    \text{LHS of Eq.~\eqref{eq:relation between A stare B and A stars B}} &= \left.
    {\rm exp}\left(\sum_{m=1}^{M}\frac{s_m}{2}\frac{\partial^2}{\partial\alpha_m\partial\alpha_m^*}\right)\left[{\rm exp}\left\{\sum_{m=1}^M\left(\frac{1}{2}\frac{\partial^2}{\partial\psi_m\partial\phi^*_m} - \frac{1}{2}\frac{\partial^2}{\partial\psi^*_m\partial\phi_m}\right)\right\}A_{\vec{0}}(\vec{\psi},\vec{\psi}^*)B_{\vec{0}}(\vec{\phi},\vec{\phi}^*)\right|_{\substack{\vec{\psi} = \vec{\alpha} + \vec{\zeta},\vec{\phi} = \vec{\alpha} + \vec{\xi}}}\right].
\end{align}
Below, we first operate ${\rm exp}[\sum_m(s_m/2)\partial^2/(\partial\alpha_m\partial\alpha^*_m)]$ on $A_{\vec{0}}(\vec{\alpha} + \vec{\zeta},\vec{\alpha}^* + \vec{\zeta}^*)B_{\vec{0}}(\vec{\alpha} + \vec{\xi},\vec{\alpha}^* + \vec{\xi}^*)$.
Here, we introduce $C(\vec{\alpha},\vec{\zeta},\vec{\xi},\vec{\alpha}^*,\vec{\zeta}^*,\vec{\xi}^*)$ as
\begin{align}
    \label{eq:definitnion of the function C appendix}
    C(\vec{\alpha},\vec{\zeta},\vec{\xi},\vec{\alpha}^*,\vec{\zeta}^*,\vec{\xi}^*) =
    {\rm exp}\left(\sum_{m=1}^{M}\frac{s_m}{2}\frac{\partial^2}{\partial\alpha_m\partial\alpha_m^*}\right)\left[A_{\vec{0}}(\vec{\alpha} + \vec{\zeta},\vec{\alpha}^* + \vec{\zeta}^*)B_{\vec{0}}(\vec{\alpha} + \vec{\xi},\vec{\alpha}^* + \vec{\xi}^*)\right].
\end{align}
In order to perform the differential calculation in the right-hand side of Eq.~\eqref{eq:definitnion of the function C appendix}, we use the formula \cite{plimak2009}:
\begin{align}
    \label{eq:formula for the differential calculation}
    D\left(\frac{\partial}{\partial \vec{\alpha}},\frac{\partial}{\partial \vec{\alpha}^*}\right)\left[F(\vec{\alpha},\vec{\alpha}^*)G(\vec{\alpha},\vec{\alpha}^*)\right] = \left.D\left(\frac{\partial}{\partial \vec{\alpha}} + \frac{\partial}{\partial\vec{\beta}},\frac{\partial}{\partial \vec{\alpha}^*} + \frac{\partial}{\partial\vec{\beta}^*}\right)\left[F(\vec{\alpha},\vec{\alpha}^*)G(\vec{\beta},\vec{\beta}^*)\right]\right|_{\vec{\beta}=\vec{\alpha}},
\end{align}
where $D(\partial/\partial\vec{\alpha},\partial/\partial\vec{\alpha}^*)$ is an arbitrary polynomial function of differential operators $\partial/\partial \alpha_j$ and $\partial/\partial \alpha_j$ ($j=1, 2, \cdots, M$), and $F(\vec{\alpha},\vec{\alpha}^*)$ and $G(\vec{\alpha},\vec{\alpha}^*)$ are arbitrary $c$-number functions.
Applying Eq.~\eqref{eq:formula for the differential calculation} to the left-hand side of Eq.~\eqref{eq:definitnion of the function C appendix} with $D(\partial/\partial\vec{\alpha},\partial/\partial\vec{\alpha}^*) = {\rm exp}[\sum_m(s_m/2)\partial^2/(\partial\alpha_m\partial\alpha^*_m)]$, $F(\vec{\alpha},\vec{\zeta},\vec{\alpha}^*,\vec{\zeta}^*) = A_{\vec{0}}(\vec{\alpha} + \vec{\zeta},\vec{\alpha}^* + \vec{\zeta}^*)$ and $G(\vec{\alpha},\vec{\xi},\vec{\alpha}^*,\vec{\xi}^*) = B_{\vec{0}}(\vec{\alpha} + \vec{\xi},\vec{\alpha}^* + \vec{\xi}^*)$, we obtain
\begin{align}
    \label{eq:calculaiton of the function C appendix 1}
    C(\vec{\alpha},\vec{\zeta},\vec{\xi},\vec{\alpha}^*,\vec{\zeta}^*,\vec{\xi}^*)  = \left.{\rm exp}\left\{\sum_{m=1}^{M}\frac{s_m}{2}
    \left(\frac{\partial}{\partial\alpha_{m}} + \frac{\partial}{\partial\beta_{m}}\right)
    \left(\frac{\partial}{\partial\alpha^*_{m}} + \frac{\partial}{\partial\beta^*_{m}}\right)
    \right\}A_{\vec{0}}(\vec{\alpha} + \vec{\zeta},\vec{\alpha}^* + \vec{\zeta}^*)B_{\vec{0}}(\vec{\beta} + \vec{\xi},\vec{\beta}^* + \vec{\xi}^*)\right|_{\vec{\beta} = \vec{\alpha}}.
\end{align}
Using Eq.~\eqref{eq:relation between A0 and As}, we can rewrite the right-hand side of Eq.~\eqref{eq:calculaiton of the function C appendix 1} as
\begin{align}
    \label{eq:calculaiton of the function C appendix 2}
    C(\vec{\alpha},\vec{\zeta},\vec{\xi},\vec{\alpha}^*,\vec{\zeta}^*,\vec{\xi}^*) &= \left.{\rm exp}\left\{\sum_{m=1}^M\frac{s_m}{2}\left(\frac{\partial^2}{\partial\alpha_{m}\partial\beta^*_{m}} + \frac{\partial^2}{\partial\alpha^*_{m}\partial\beta_{m}}\right)\right\}A_{\vec{s}}(\vec{\alpha} + \vec{\zeta},\vec{\alpha}^* + \vec{\zeta}^*)B_{\vec{s}}(\vec{\beta} + \vec{\xi},\vec{\beta}^* + \vec{\xi}^*)\right|_{\vec{\beta} = \vec{\alpha}} \\
    &= \left.{\rm exp}\left\{\sum_{m=1}^M\frac{s_m}{2}\left(\frac{\partial^2}{\partial\psi_{m}\partial\phi^*_{m}} + \frac{\partial^2}{\partial\psi^*_{m}\partial\phi_{m}}\right)\right\}A_{\vec{s}}(\vec{\psi},\vec{\psi}^*)B_{\vec{s}}(\vec{\phi},\vec{\phi}^*)\right|_{\vec{\psi} = \vec{\alpha} + \vec{\zeta},\vec{\phi} = \vec{\alpha} + \vec{\xi}}.
\end{align}
Finally, by using Eqs.~\eqref{eq:derivation of B2} and \eqref{eq:calculaiton of the function C appendix 2}, we obtain
\begin{align}
    \text{LHS of Eq.~\eqref{eq:relation between A stare B and A stars B}} &= \left.{\rm exp}\left\{\sum_{m=1}^M\left(\frac{1+s_m}{2}\frac{\partial^2}{\partial\psi_m\partial\phi^*_m} - \frac{1-s_m}{2}\frac{\partial^2}{\partial\psi^*_m\partial\phi_m}\right)\right\}A_{\vec{s}}(\vec{\psi},\vec{\psi}^*)B_{\vec{s}}(\vec{\phi},\vec{\phi}^*)\right|_{\vec{\psi} = \vec{\alpha} + \vec{\zeta},\vec{\phi} = \vec{\alpha} + \vec{\xi}} \\
    &= A_{\vec{s}}(\vec{\alpha} + \vec{\zeta},\vec{\alpha}^* + \vec{\zeta}^*)\star_{\vec{s}} B_{\vec{s}}(\vec{\alpha} + \vec{\xi},\vec{\alpha}^* + \vec{\xi}^*).
\end{align}
This completes the derivation of Eq.~\eqref{eq:relation between A stare B and A stars B}.

{\it Eq.~\eqref{eq:relation between A star B and Ase stars Bse}}--We can derive Eq.~\eqref{eq:relation between A star B and Ase stars Bse} by applying Eq.~\eqref{eq:formula for the differential calculation} to the left-hand side and using Eq.~\eqref{eq:definition of extended s-ordered phase-space representation of A}.


\subsection{\label{appendix:derivation of infinitesimal time propagator}Derivation of the infinitesimal time propagator of Eq.~\texorpdfstring{\eqref{eq:s-parametrized phase space representation of Markov propagator}}{TEXT}}

Substituting Eq.~\eqref{eq:propagato for the Wigner function} into Eq.~\eqref{eq:relation between s-ordered propagator and s=0 propagator} and performing the derivative with respect to $\alpha_{m,j+1}$ and $\alpha^*_{m,j+1}$  as 
\begin{align}
{\rm exp}\left(\sum_{m=1}^{M}\frac{s_m}{2}\frac{\partial^2}{\partial\alpha_{m,j+1}\partial\alpha^*_{m,j+1}}\right)
e^{\vec{\eta}_{j+1}^*\cdot(\vec{\alpha}_{j+1} - \vec{\alpha}_j) - \vec{\eta}_{j+1}\cdot(\vec{\alpha}_{j+1}^* - \vec{\alpha}_j^*)} = e^{\vec{\eta}_{j+1}^*\cdot(\vec{\alpha}_{j+1} - \vec{\alpha}_j) - \vec{\eta}_{j+1}\cdot(\vec{\alpha}_{j+1}^* - \vec{\alpha}_j^*) + \sum_ms_m|\vec{\eta}_{m,j+1}|^2/2},
\end{align}
we obtain
\begin{align}
    \label{eq:calculation of the propagator appendix 2}
    \varUpsilon_{\vec{s}}&(\vec{\alpha}_{j+1},t_{j+1};\vec{\alpha}_j,t_j) =  {\rm exp}\left(\sum_{m=1}^{M}\frac{s_m}{2}\frac{\partial^2}{\partial\alpha_{m,j}\partial\alpha^*_{m,j}}\right)\int\frac{d^2\vec{\eta}_{j+1}}{\pi^M}e^{\vec{\eta}_{j+1}^*\cdot(\vec{\alpha}_{j+1} - \vec{\alpha}_j) - \vec{\eta}_{j+1}\cdot(\vec{\alpha}_{j+1}^* - \vec{\alpha}_j^*) + \sum_ms_m|\vec{\eta}_{m,j+1}|^2/2}\nonumber \\
    &\times\left[1 + \frac{i\Delta t}{\hbar}\left\{\sum_{n=0,1}(-1)^nH_{\vec{0}}\left(\vec{\alpha}_j + \frac{(-1)^n}{2}\vec{\eta}_{j+1},\vec{\alpha}^*_j + \frac{(-1)^n}{2}\vec{\eta}^*_{j+1}\right) - i\hbar\mathcal{D}_{\vec{0}}\left(\vec{\alpha}_j + \frac{1}{2}\vec{\eta}_{j+1},\vec{\alpha}^*_j + \frac{1}{2}\vec{\eta}^*_{j+1},\vec{\alpha}_j - \frac{1}{2}\vec{\eta}_{j+1},\vec{\alpha}^*_j - \frac{1}{2}\vec{\eta}^*_{j+1}\right)\right\}\right].
\end{align}
In order to perform the derivative with respect to $\alpha_{m,j}$ and $\alpha^*_{m,j}$, we apply Eq.~\eqref{eq:formula for the differential calculation} to Eq.~\eqref{eq:calculation of the propagator appendix 2} with $D = {\rm exp}[\sum_m(s_m/2)\partial^2/(\partial\alpha_{m,j}\partial\alpha^*_{m,j})]$, $F = e^{\vec{\eta}_{j+1}^*\cdot(\vec{\alpha}_{j+1} - \vec{\alpha}_j) - \vec{\eta}_{j+1}\cdot(\vec{\alpha}_{j+1}^* - \vec{\alpha}_j^*) + \sum_ms_m|\vec{\eta}_{m,j+1}|^2/2}$ and $G$ being the remaining integrand of the right-hand side of Eq.~\eqref{eq:calculation of the propagator appendix 2}, obtaining
\begin{align}
    \label{eq:calculation of the propagator appendix 3}
    \varUpsilon_{\vec{s}}&(\vec{\alpha}_{j+1},t_{j+1};\vec{\alpha}_j,t_j) =  {\rm exp}\left\{\sum_{m=1}^{M}\frac{s_m}{2}
    \left(\frac{\partial}{\partial\alpha_{m,j}} + \frac{\partial}{\partial\beta_{m,j}}\right)
    \left(\frac{\partial}{\partial\alpha^*_{m,j}} + \frac{\partial}{\partial\beta^*_{m,j}}\right)
    \right\}\int\frac{d^2\vec{\eta}_{j+1}}{\pi^M}e^{\vec{\eta}_{j+1}^*\cdot(\vec{\alpha}_{j+1} - \vec{\alpha}_j) - \vec{\eta}_{j+1}\cdot(\vec{\alpha}_{j+1}^* - \vec{\alpha}_j^*) + \sum_ms_m|\vec{\eta}_{m,j+1}|^2/2}\nonumber \\
    &\left.\times\left[1 + \frac{i\Delta t}{\hbar}\left\{\sum_{n=0,1}(-1)^nH_{\vec{0}}\left(\vec{\beta}_j + \frac{(-1)^n}{2}\vec{\eta}_{j+1},\vec{\beta}^*_j + \frac{(-1)^n}{2}\vec{\eta}^*_{j+1}\right) - i\hbar\mathcal{D}_{\vec{0}}\left(\vec{\beta}_j + \frac{1}{2}\vec{\eta}_{j+1},\vec{\beta}^*_j + \frac{1}{2}\vec{\eta}^*_{j+1},\vec{\beta}_j - \frac{1}{2}\vec{\eta}_{j+1},\vec{\beta}^*_j - \frac{1}{2}\vec{\eta}^*_{j+1}\right)\right\}\right]\right|_{\vec{\beta}_j = \vec{\alpha}_j}.
\end{align}
We first perform the $\alpha_{m,j}$- and $\alpha^*_{m,j}$-derivatives in Eq.~\eqref{eq:calculation of the propagator appendix 3}.
By factorizing the exponential function of the differential operators and calculating ${\rm exp}[\sum_m(s_m/2)\partial^2/(\partial\alpha_{m,j}\partial\alpha^*_{m,j})]e^{\vec{\eta}_{j+1}^*\cdot(\vec{\alpha}_{j+1} - \vec{\alpha}_j) - \vec{\eta}_{j+1}\cdot(\vec{\alpha}_{j+1}^* - \vec{\alpha}_j^*)+ \sum_ms_m|\vec{\eta}_{m,j+1}|^2/2} = e^{\vec{\eta}_{j+1}^*\cdot(\vec{\alpha}_{j+1} - \vec{\alpha}_j) - \vec{\eta}_{j+1}\cdot(\vec{\alpha}_{j+1}^* - \vec{\alpha}_j^*)}$,
we can rewrite Eq.~\eqref{eq:calculation of the propagator appendix 3} as
\begin{align}
    \label{eq:calculation of the propagator appendix 3.5}
    \varUpsilon_{\vec{s}}&(\vec{\alpha}_{j+1},t_{j+1};\vec{\alpha}_j,t_j) =
    {\rm exp}\left\{\sum_{m=1}^M\frac{s_m}{2}\left(\frac{\partial^2}{\partial\alpha_{m,j}\partial\beta^*_{m,j}} + \frac{\partial^2}{\partial\alpha^*_{m,j}\partial\beta_{m,j}}\right)\right\}
    \int\frac{d^2\vec{\eta}_{j+1}}{\pi^M}
    e^{\vec{\eta}_{j+1}^*\cdot(\vec{\alpha}_{j+1} - \vec{\alpha}_j) - \vec{\eta}_{j+1}\cdot(\vec{\alpha}_{j+1}^* - \vec{\alpha}_j^*)}{\rm exp}\left(\sum_{m=1}^{M}\frac{s_m}{2}\frac{\partial^2}{\partial\beta_{m,j}\partial\beta^*_{m,j}}\right)\nonumber \\
    &\times \left[1 + \frac{i\Delta t}{\hbar}\left\{\sum_{n=0,1}(-1)^n H_{\vec{0}}\left(\vec{\beta}_j + \frac{(-1)^n}{2}\vec{\eta}_{j+1},\vec{\beta}^*_j + \frac{(-1)^n}{2}\vec{\eta}^*_{j+1}\right)\left.- i\hbar\mathcal{D}_{\vec{0}}\left(\vec{\beta}_j + \frac{1}{2}\vec{\eta}_{j+1},\vec{\beta}^*_j + \frac{1}{2}\vec{\eta}^*_{j+1},\vec{\beta}_j - \frac{1}{2}\vec{\eta}_{j+1},\vec{\beta}^*_j - \frac{1}{2}\vec{\eta}^*_{j+1}\right)\right\}\right]\right|_{\vec{\beta}_j = \vec{\alpha}_j}.
\end{align}
The calculation of the remaining $\alpha_{m,j}$- and $\alpha_{m,j}^*$-derivatives leads
\begin{align}
        {\rm exp}\left\{\sum_{m=1}^M\frac{s_m}{2}\left(\frac{\partial^2}{\partial\alpha_{m,j}\partial\beta^*_{m,j}} + \frac{\partial^2}{\partial\alpha^*_{m,j}\partial\beta_{m,j}}\right)\right\}e^{\vec{\eta}_{j+1}^*\cdot(\vec{\alpha}_{j+1} - \vec{\alpha}_j) - \vec{\eta}_{j+1}\cdot(\vec{\alpha}_{j+1}^* - \vec{\alpha}_j^*)}
        = e^{\vec{\eta}_{j+1}^*\cdot(\vec{\alpha}_{j+1} - \vec{\alpha}_j) - \vec{\eta}_{j+1}\cdot(\vec{\alpha}_{j+1}^* - \vec{\alpha}_j^*)}{\rm exp}\left\{\sum_{m=1}^M\frac{s_m}{2}\left(\eta_{m,j+1}\frac{\partial}{\partial\beta_{m,j}} - \eta^*_{m,j+1}\frac{\partial}{\partial\beta^*_{m,j}}\right)\right\},
\end{align}
from which we can rewrite Eq.~\eqref{eq:calculation of the propagator appendix 3.5} with the replacement of $\vec{\beta}$ to $\vec{\alpha}$ as
\begin{align}
    \label{eq:calculation of the propagator appendix 4}
    \varUpsilon_{\vec{s}}&(\vec{\alpha}_{j+1},t_{j+1};\vec{\alpha}_j,t_j) =
    \int\frac{d^2\vec{\eta}_{j+1}}{\pi^M}
    e^{\vec{\eta}_{j+1}^*\cdot(\vec{\alpha}_{j+1} - \vec{\alpha}_j) - \vec{\eta}_{j+1}\cdot(\vec{\alpha}_{j+1}^* - \vec{\alpha}_j^*)}{\rm exp}\left\{\sum_{m=1}^M\frac{s_m}{2}\left(\eta_{m,j+1}\frac{\partial}{\partial\alpha_{m,j}} - \eta^*_{m,j+1}\frac{\partial}{\partial\alpha^*_{m,j}}\right)\right\}{\rm exp}\left(\sum_{m=1}^{M}\frac{s_m}{2}\frac{\partial^2}{\partial\alpha_{m,j}\partial\alpha^*_{m,j}}\right)\nonumber \\
    &\times\left[1 + \frac{i\Delta t}{\hbar}\left\{\sum_{n=0,1}(-1)^n 
     H_{\vec{0}}\left(\vec{\alpha}_j + \frac{(-1)^n}{2}\vec{\eta}_{j+1},\vec{\alpha}^*_j + \frac{(-1)^n}{2}\vec{\eta}^*_{j+1}\right) - i\hbar\mathcal{D}_{\vec{0}}\left(\vec{\alpha}_j + \frac{1}{2}\vec{\eta}_{j+1},\vec{\alpha}^*_j + \frac{1}{2}\vec{\eta}^*_{j+1},\vec{\alpha}_j - \frac{1}{2}\vec{\eta}_{j+1},\vec{\alpha}^*_j - \frac{1}{2}\vec{\eta}^*_{j+1}\right)\right\}\right].
\end{align}
From Eqs.~\eqref{eq:relation between A0 and As} and \eqref{eq:relation between A stare B and A stars B}, the operation of ${\rm exp}[\sum_m(s_m/2)\partial^2/(\partial\alpha_{m,j}\partial\alpha_{m,j}^*)]$ to $H_{\vec{0}}$ and $\mathcal{D}_{\vec{0}}$ leads to $H_{\vec{s}}$ and $\mathcal{D}_{\vec{s}}$, respectively, without changing the arguments:
\begin{align}
    \label{eq:calculation of the propagator appendix 5}
    \varUpsilon_{\vec{s}}&(\vec{\alpha}_{j+1},t_{j+1};\vec{\alpha}_j,t_j) =
    \int\frac{d^2\vec{\eta}_{j+1}}{\pi^M}
    e^{\vec{\eta}_{j+1}^*\cdot(\vec{\alpha}_{j+1} - \vec{\alpha}_j) - \vec{\eta}_{j+1}\cdot(\vec{\alpha}_{j+1}^* - \vec{\alpha}_j^*)}{\rm exp}\left\{\sum_{m=1}^M\frac{s_m}{2}\left(\eta_{m,j+1}\frac{\partial}{\partial\alpha_{m,j}} - \eta^*_{m,j+1}\frac{\partial}{\partial\alpha^*_{m,j}}\right)\right\}\nonumber \\
    &\times\left[1 + \frac{i\Delta t}{\hbar}\left\{\sum_{n=0,1}(-1)^n 
    H_{\vec{s}}\left(\vec{\alpha}_j + \frac{(-1)^n}{2}\vec{\eta}_{j+1},\vec{\alpha}^*_j + \frac{(-1)^n}{2}\vec{\eta}^*_{j+1}\right) - i\hbar \mathcal{D}_{\vec{s}}\left(\vec{\alpha}_j + \frac{1}{2}\vec{\eta}_{j+1},\vec{\alpha}^*_j + \frac{1}{2}\vec{\eta}^*_{j+1},\vec{\alpha}_j - \frac{1}{2}\vec{\eta}_{j+1},\vec{\alpha}^*_j - \frac{1}{2}\vec{\eta}^*_{j+1}\right)\right\}\right].
\end{align}
Here, $\mathcal{D}_{\vec{s}}(\vec{\alpha},\vec{\alpha}^*,\vec{\gamma},\vec{\gamma}^*)$ is given by Eq.~\eqref{eq:definition of the non-unitary term of the propagator of the s-ordered quasiprobability distributino function} with substituting $\vec{\beta}=\vec{\alpha}^*$ and $\vec{\delta}=\vec{\gamma}^*$ and using $L^{e}_{k\vec{s}}(\vec{\alpha},\vec{\alpha}^*)=L_{k\vec{s}}(\vec{\alpha},\vec{\alpha}^*)$ and $\bar{L}^e_{k\vec{s}}(\vec{\alpha},\vec{\alpha}^*)=L^*_{k\vec{s}}(\vec{\alpha},\vec{\alpha}^*)$.
Finally, by using Eqs.~\eqref{eq:definition of extended s-ordered phase-space representation of A}, \eqref{eq:relation between A star B and Ase stars Bse} and \eqref{eq:definition of the non-unitary term of the propagator of the s-ordered quasiprobability distributino function}, we can rewrite Eq.~\eqref{eq:calculation of the propagator appendix 5} as
\begin{align}
    \varUpsilon_{\vec{s}}(\vec{\alpha}_{j+1},t_{j+1};\vec{\alpha}_j,t_j) = \int\frac{d^2\vec{\eta}_{j+1}}{\pi^M}e^{\vec{\eta}_{j+1}^*\cdot(\vec{\alpha}_{j+1} - \vec{\alpha}_j) - \vec{\eta}_{j+1}\cdot(\vec{\alpha}_{j+1}^* - \vec{\alpha}_j^*)}
    \left[1 + \frac{i\Delta t}{\hbar}\left\{H^e_{\vec{s}}(\vec{\psi}^+_{\vec{s},j},\vec{\psi}^{+*}_{-\vec{s},j}) - H^e_{\vec{s}}(\vec{\psi}^-_{-\vec{s},j},\vec{\psi}^{-*}_{\vec{s},j}) - i\hbar\mathcal{D}_{\vec{s}}(\vec{\psi}^+_{\vec{s},j},\vec{\psi}^{+*}_{-\vec{s},j},\vec{\psi}^-_{-\vec{s},j},\vec{\psi}^{-*}_{\vec{s},j})\right\}\right],
\end{align}
where the vectors $\psi^+_{\vec{s},j}$ and $\psi^-_{\vec{s},j}$ are defined by Eqs.~\eqref{eq:+ vectors contains alpha and eta discrete} and \eqref{eq:- vectors contains alpha and eta discrete}, respectively.
This completes the derivation of Eq.~\eqref{eq:s-parametrized phase space representation of Markov propagator}.

\section{\label{appendix:Equations of motion in the phase space}Equations of motion in the phase space}
We derive the generalized Liouville equation~\eqref{eq:generalized Liouville equation}, the Fokker-Planck equation~\eqref{eq:Fokker-Planck equation}, and the stochastic differential equation~\eqref{eq:stochastic differential equation simplified version} for a system satisfying the condition Eq.~\eqref{eq:lambda and Lambda when the jump operators do not couple different degrees of feedom}.


\subsection{\label{appendix:Generalized Liouville equation}Generalized Liouville equation{\rm :} Derivation of Eq.~\texorpdfstring{\eqref{eq:generalized Liouville equation}}{TEXT}}
Expanding $H^e_{\vec{s}}$ and $\mathcal{D}_{\vec{s}}$ in Eq.~\eqref{eq:s-parametrized phase space representation of Markov propagator} with respect to the quantum fields $\eta_{m,j+1}$ up to first order, we obtain
\begin{align}
    W_{\vec{s}}(\vec{\alpha}_{j+1},\vec{\alpha}^*_{j+1},t_{j+1}) =& \int\frac{d^2\vec{\alpha}_{j}d^2\vec{\eta}_{j+1}}{\pi^{2M}}e^{\vec{\eta}^*_{j+1}\cdot(\vec{\alpha}_{j+1} - \vec{\alpha}_{m,j}) - {\rm c.c.}}W_{\vec{s}}(\vec{\alpha}_j,\vec{\alpha}^*_j,t_j)\nonumber\\
    &- \frac{\Delta t}{i\hbar}\sum_{m=1}^{M}\int \frac{d^2\vec{\alpha}_jd^2\vec{\eta}_j}{\pi^{2M}}e^{\vec{\eta}^*_{j+1}\cdot(\vec{\alpha}_{j+1} - \vec{\alpha}_{m,j}) - {\rm c.c.}} \left[\eta^*_{m,j+1}\left\{\frac{\partial H_{\vec{s}}(\vec{\alpha}_j,\vec{\alpha}^*_j)}{\partial\alpha^*_{m,j}} - i\hbar \mathcal{K}^{\vec{s}}_m(\vec{\alpha}_j,\vec{\alpha}^*_j)\right\} + {\rm c.c.}\right]W_{\vec{s}}(\vec{\alpha}_j,\vec{\alpha}^*_j,t_j)\\
    =& \int\frac{d^2\vec{\alpha}_{j}d^2\vec{\eta}_{j+1}}{\pi^{2M}}e^{\vec{\eta}^*_{j+1}\cdot(\vec{\alpha}_{j+1} - \vec{\alpha}_{m,j}) - {\rm c.c.}}W_{\vec{s}}(\vec{\alpha}_j,\vec{\alpha}^*_j,t_j)\nonumber\\
    &+ \frac{\Delta t}{i\hbar}\sum_{m=1}^{M}\int\frac{d^2\vec{\alpha}_jd^2\vec{\eta}_j}{\pi^{2M}}\left[\left\{\frac{\partial H_{\vec{s}}(\vec{\alpha}_j,\vec{\alpha}^*_j)}{\partial\alpha^*_{m,j}} - i\hbar \mathcal{K}^{\vec{s}}_m(\vec{\alpha}_j,\vec{\alpha}^*_j)\right\}\frac{\partial}{\partial\alpha_{m,j}}e^{\vec{\eta}^*_{j+1}\cdot(\vec{\alpha_{j+1}} - \vec{\alpha_{j}}) - {\rm c.c.}} - {\rm c.c.}\right]W_{\vec{s}}(\vec{\alpha}_j,\vec{\alpha}^*_j,t_j),
\end{align}
where $\mathcal{K}^{\vec{s}}_m$ is given by Eq.~\eqref{eq:definition of Ksm}.
Performing the integration by part, we obtain
\begin{align}
    W_{\vec{s}}(\vec{\alpha}_{j+1},\vec{\alpha}^*_{j+1},t_{j+1})
    =& \int\frac{d^2\vec{\alpha}_{j}d^2\vec{\eta}_{j+1}}{\pi^{2M}}e^{\vec{\eta}^*_{j+1}\cdot(\vec{\alpha}_{j+1} - \vec{\alpha}_{m,j}) - {\rm c.c.}}W_{\vec{s}}(\vec{\alpha}_j,\vec{\alpha}^*_j,t_j)\nonumber\\
    \label{eq:calculation GLE appendix 1}
    &- \frac{\Delta t}{i\hbar}\sum_{m=1}^{M}\int\frac{d^2\vec{\alpha}_{j}d^2\vec{\eta}_{j+1}}{\pi^{2M}}e^{\vec{\eta}^*_{j+1}\cdot(\vec{\alpha}_{j+1} - \vec{\alpha}_{m,j}) - {\rm c.c.}}\frac{\partial}{\partial\alpha_{m,j}}\left[\left\{\frac{\partial H_{\vec{s}}(\vec{\alpha}_j,\vec{\alpha}^*_j)}{\partial\alpha^*_{m,j}} - i\hbar \mathcal{K}^{\vec{s}}_m(\vec{\alpha}_j,\vec{\alpha}^*_j)\right\}W_{\vec{s}}(\vec{\alpha}_j,\vec{\alpha}^*_j,t_j) - {\rm c.c.}\right].
\end{align}
Integrating out the quantum fields by using Eq.~\eqref{eq:definitnio of the Dirac delta function}, we can rewrite Eq.~\eqref{eq:calculation GLE appendix 1} as
\begin{align}
    W_{\vec{s}}(\vec{\alpha}_{j+1},\vec{\alpha}^*_{j+1},t_{j+1})
    =& \prod_{m=1}^{M}\int d^2\alpha_{m,j}\delta^{(2)}(\alpha_{m,j+1} - \alpha_{m,j})W_{\vec{s}}(\vec{\alpha}_j,\vec{\alpha}^*_j,t_j)\nonumber\\
    &- \frac{\Delta t}{i\hbar}\sum_{m=1}^{M}\prod_{p=1}^M\int d^2\alpha_{p,j}\delta^{(2)}(\alpha_{p,j+1} - \alpha_{p,j})\frac{\partial}{\partial\alpha_{m,j}}\left[\left\{\frac{\partial H_{\vec{s}}(\vec{\alpha}_j,\vec{\alpha}^*_j)}{\partial\alpha^*_{m,j}} - i\hbar \mathcal{K}^{\vec{s}}_m(\vec{\alpha}_j,\vec{\alpha}^*_j)\right\}W_{\vec{s}}(\vec{\alpha}_j,\vec{\alpha}^*_j,t_j) - {\rm c.c.}\right],
\end{align}
and integrating out the classical fields reads
\begin{align}
    \label{eq:calculation GLE appendix 2}
    W_{\vec{s}}(\vec{\alpha}_{j+1},\vec{\alpha}^*_{j+1},t_{j+1})- W_{\vec{s}}(\vec{\alpha}_{j+1},\vec{\alpha}^*_{j+1},t_j) = - \frac{\Delta t}{i\hbar}\sum_{m=1}^{M}\frac{\partial}{\partial\alpha_{m,j+1}}\left[ \left\{\frac{\partial H_{\vec{s}}(\vec{\alpha}_{j+1},\vec{\alpha}^*_{j+1})}{\partial\alpha_{m,j+1}} - i\hbar \mathcal{K}^{\vec{s}}_m(\vec{\alpha}_{j+1},\vec{\alpha}^*_{j+1})\right\}W_{\vec{s}}(\vec{\alpha}_{j+1},\vec{\alpha}^*_{j+1},t_j)\right] + {\rm c.c.}
\end{align}
Taking the continuous limit of Eq.~\eqref{eq:calculation GLE appendix 2} and substituting the detailed form of $\mathcal{K}^{\vec{s}}_m$ [Eq.~\eqref{eq:definition of Ksm}], we obtain the generalized Liouville equation:
\begin{align}
    i\hbar\frac{dW_{\vec{s}}(\vec{\alpha},\vec{\alpha}^*,t)}{dt} = -\sum_{m=1}^{M}\frac{\partial}{\partial\alpha_m}\left[ \left\{\frac{\partial H_{\vec{s}}}{\partial\alpha^*_m} + \frac{i\hbar}{2}\sum_k\gamma_k\left(L^*_{k\vec{s}}\star_{\vec{s}}\frac{\partial L_{k\vec{s}}}{\partial\alpha^*_m} - \frac{\partial L^*_{k\vec{s}}}{\partial\alpha^*_m}\star_{\vec{s}}L_{k\vec{s}}\right) \right\}W_{\vec{s}}(\vec{\alpha},\vec{\alpha}^*,t)\right] - {\rm c.c.}
\end{align}


\subsection{\label{appendix:Fokker-Planck equation}Fokker-Planck equation{\rm :} Derivation of Eq.~\texorpdfstring{\eqref{eq:Fokker-Planck equation}}{TEXT}}
Expanding $H^e_{\vec{s}}$ and $\mathcal{D}_{\vec{s}}$ in Eq.~\eqref{eq:s-parametrized phase space representation of Markov propagator} with respect to the quantum fields $\eta_{m,j+1}$ up to second order and using Eq.~\eqref{eq:calculation GLE appendix 2}, we obtain
\begin{align}
    W_{\vec{s}}(\vec{\alpha}_{j+1},\vec{\alpha}^*_{j+1},t_{j+1}) =& W_{\vec{s}}(\vec{\alpha}_{j+1},\vec{\alpha}^*_{j+1},t_j) - \frac{\Delta t}{i\hbar}\sum_{m=1}^{M}\frac{\partial}{\partial\alpha_{m,j+1}} \left[\left\{\frac{\partial H_{\vec{s}}(\vec{\alpha}_{j+1},\vec{\alpha}^*_{j+1})}{\partial\alpha_{m,j+1}} - i\hbar \mathcal{K}^{\vec{s}}_m(\vec{\alpha}_{j+1},\vec{\alpha}^*_{j+1})\right\}W_{\vec{s}}(\vec{\alpha}_{j+1},\vec{\alpha}^*_{j+1},t_j)\right] + {\rm c.c.} \nonumber\\
    &- \Delta t\sum_{m,n=1}^{M}\int\frac{d^2\vec{\alpha}_jd^2\vec{\eta}_{j+1}}{\pi^{2M}}e^{\vec{\eta}^*_{j+1}\cdot(\vec{\alpha}_{j+1} - \vec{\alpha}_{j}) - {\rm c.c.}}\left\{\lambda^{\vec{s}}_{mn}(\vec{\alpha}_j,\vec{\alpha}^*_j)\eta^*_{m,j+1}\eta^*_{n,j+1} + \Lambda^{\vec{s}}_{mn}(\vec{\alpha}_j,\vec{\alpha}^*_j)\eta^*_{m,j+1}\eta_{n,j+1} + {\rm c.c.}\right\}W_{\vec{s}}(\vec{\alpha}_j,\vec{\alpha}^*_j,t_j)\\
     \label{eq:calculaiton of FPE appendix 1}
    =& W_{\vec{s}}(\vec{\alpha}_{j+1},\vec{\alpha}^*_{j+1},t_j) - \frac{\Delta t}{i\hbar}\sum_{m=1}^{M}\frac{\partial}{\partial\alpha_{m,j+1}}\left[ \left\{\frac{\partial H_{\vec{s}}(\vec{\alpha}_{j+1},\vec{\alpha}^*_{j+1})}{\partial\alpha_{m,j+1}} - i\hbar \mathcal{K}^{\vec{s}}_m(\vec{\alpha}_{j+1},\vec{\alpha}^*_{j+1})\right\}W_{\vec{s}}(\vec{\alpha}_{j+1},\vec{\alpha}^*_{j+1},t_j)\right] + {\rm c.c.}\nonumber\\
    - &\Delta t\sum_{m,n=1}^{M}\int\frac{d^2\vec{\alpha}_jd^2\vec{\eta}_{j+1}}{\pi^{2M}}\left[\left\{\lambda^{\vec{s}}_{mn}(\vec{\alpha}_j,\vec{\alpha}^*_j)\frac{\partial^2}{\partial\alpha_{m,j}\partial\alpha_{n,j}} - \Lambda^{\vec{s}}_{mn}(\vec{\alpha}_j,\vec{\alpha}^*_j)\frac{\partial^2}{\partial\alpha_{m,j}\partial\alpha^*_{n,j}} + {\rm c.c.}\right\}e^{\vec{\eta}^*_{j+1}\cdot(\vec{\alpha}_{j+1} - \vec{\alpha}_{j}) - {\rm c.c.}}\right]W_{\vec{s}}(\vec{\alpha}_j,\vec{\alpha}^*_j,t_j),
\end{align}
where $\lambda^{\vec{s}}_{mn}$ and $\Lambda^{\vec{s}}_{mn}$ are given by Eqs.~\eqref{eq:definitnion of lambdamn} and \eqref{eq:definitnion of Lambdamn}, respectively.
Performing the integration by part, we can rewrite Eq.~\eqref{eq:calculaiton of FPE appendix 1} as
\begin{align}
    W_{\vec{s}}(\vec{\alpha}_{j+1},\vec{\alpha}^*_{j+1},t_{j+1}) =& W_{\vec{s}}(\vec{\alpha}_{j+1},\vec{\alpha}^*_{j+1},t_j) - \frac{\Delta t}{i\hbar}\sum_{m=1}^{M}\frac{\partial}{\partial\alpha_{m,j+1}}\left[ \left\{\frac{\partial H_{\vec{s}}(\vec{\alpha}_{j+1},\vec{\alpha}^*_{j+1})}{\partial\alpha_{m,j+1}} - i\hbar \mathcal{K}^{\vec{s}}_m(\vec{\alpha}_{j+1},\vec{\alpha}^*_{j+1})\right\}W_{\vec{s}}(\vec{\alpha}_{j+1},\vec{\alpha}^*_{j+1},t_j)\right]\nonumber\\
    &- \Delta t\sum_{m,n=1}^{M}\int\frac{d^2\vec{\alpha}_jd^2\vec{\eta}_{j+1}}{\pi^{2M}}e^{\vec{\eta}^*_{j+1}\cdot(\vec{\alpha}_{j+1} - \vec{\alpha}_{j}) - {\rm c.c.}}\frac{\partial^2}{\partial\alpha_{m,j}\partial\alpha_{n,j}}\left\{\lambda^{\vec{s}}_{mn}(\vec{\alpha}_j,\vec{\alpha}^*_j)W_{\vec{s}}(\vec{\alpha}_j,\vec{\alpha}^*_j,t_j)\right\} \nonumber\\
    &+\Delta t\sum_{m,n=1}^{M}\int\frac{d^2\vec{\alpha}_jd^2\vec{\eta}_{j+1}}{\pi^{2M}}e^{\vec{\eta}^*_{j+1}\cdot(\vec{\alpha}_{j+1} - \vec{\alpha}_{j}) - {\rm c.c.}}\frac{\partial^2}{\partial\alpha_{m,j}\partial\alpha^*_{n,j}}\left\{\Lambda^{\vec{s}}_{mn}(\vec{\alpha}_j,\vec{\alpha}^*_j)W_{\vec{s}}(\vec{\alpha}_j,\vec{\alpha}^*_j,t_j)\right\} + {\rm c.c.}
\end{align}
Integrating out the quantum fields by using Eq.~\eqref{eq:definitnio of the Dirac delta function}, we obtain
\begin{align}
    \label{eq:calculaiton of FPE appendix 1.5}
    W_{\vec{s}}(\vec{\alpha}_{j+1},\vec{\alpha}^*_{j+1},t_{j+1}) =& W_{\vec{s}}(\vec{\alpha}_{j+1},\vec{\alpha}^*_{j+1},t_j) - \frac{\Delta t}{i\hbar}\sum_{m=1}^{M}\frac{\partial}{\partial\alpha_{m,j+1}}\left[ \left\{\frac{\partial H_{\vec{s}}(\vec{\alpha}_{j+1},\vec{\alpha}^*_{j+1})}{\partial\alpha_{m,j+1}} - i\hbar \mathcal{K}^{\vec{s}}_m(\vec{\alpha}_{j+1},\vec{\alpha}^*_{j+1})\right\}W_{\vec{s}}(\vec{\alpha}_{j+1},\vec{\alpha}^*_{j+1},t_j)\right]\nonumber\\
    &- \Delta t\sum_{m,n=1}^{M}\prod_{p=1}^M\int d^2\alpha_{p,j}\delta^{(2)}(\alpha_{p,j+1} - \alpha_{p,j}) \frac{\partial^2}{\partial\alpha_{m,j}\partial\alpha_{n,j}}\left\{\lambda^{\vec{s}}_{mn}(\vec{\alpha}_j,\vec{\alpha}^*_j)W_{\vec{s}}(\vec{\alpha}_j,\vec{\alpha}^*_j,t_j)\right\} \nonumber\\
    &+\Delta t\sum_{m,n=1}^{M}\prod_{p=1}^M\int d^2\alpha_{p,j}\delta^{(2)}(\alpha_{p,j+1} - \alpha_{p,j})\frac{\partial^2}{\partial\alpha_{m,j}\partial\alpha^*_{n,j}}\left\{\Lambda^{\vec{s}}_{mn}(\vec{\alpha}_j,\vec{\alpha}^*_j)W_{\vec{s}}(\vec{\alpha}_j,\vec{\alpha}^*_j,t_j)\right\} + {\rm c.c.},
\end{align}
and integrating out the classical fields leads us to rewrite Eq.~\eqref{eq:calculaiton of FPE appendix 1.5} as
\begin{align}
    \label{eq:calculaiton of FPE appendix 2}
    W_{\vec{s}}(\vec{\alpha}_{j+1},\vec{\alpha}^*_{j+1},t_{j+1})- W_{\vec{s}}(\vec{\alpha}_{j+1},\vec{\alpha}^*_{j+1},t_j) =& - \frac{\Delta t}{i\hbar}\sum_{m=1}^{M}\frac{\partial}{\partial\alpha_{m,j+1}}\left[ \left\{\frac{\partial H_{\vec{s}}(\vec{\alpha}_{j+1},\vec{\alpha}^*_{j+1})}{\partial\alpha_{m,j+1}} - i\hbar \mathcal{K}^{\vec{s}}_m(\vec{\alpha}_{j+1},\vec{\alpha}^*_{j+1})\right\}W_{\vec{s}}(\vec{\alpha}_{j+1},\vec{\alpha}^*_{j+1},t_j)\right] \nonumber \\
    &-\Delta t\sum_{m,n = 1}^M\left[\frac{\partial^2}{\partial\alpha_{m,j+1}\partial\alpha_{n,j+1}}\left\{\lambda^{\vec{s}}_{mn}(\vec{\alpha}_{j+1},\vec{\alpha}^*_{j+1})W_{\vec{s}}(\vec{\alpha}_{j+1},\vec{\alpha}^*_{j+1},t_j)\right\}\right] \nonumber \\
    &+\Delta t\sum_{m,n = 1}^M\left[\frac{\partial^2}{\partial\alpha_{m,j+1}\partial\alpha^*_{n,j+1}}\left\{\Lambda^{\vec{s}}_{mn}(\vec{\alpha}_{j+1},\vec{\alpha}^*_{j+1})W_{\vec{s}}(\vec{\alpha}_{j+1},\vec{\alpha}^*_{j+1},t_j)\right\}\right] + {\rm c.c.}
\end{align}
Taking the continuous limit of Eq.~\eqref{eq:calculaiton of FPE appendix 2}, we obtain the Fokker-Planck equation:
\begin{align}
    i\hbar\frac{dW_{\vec{s}}(\vec{\alpha},\vec{\alpha}^*,t)}{dt} =& -\sum_{m=1}^{M}\frac{\partial}{\partial\alpha_m}\left[ \left\{\frac{\partial H_{\vec{s}}}{\partial\alpha^*_m} + \frac{i\hbar}{2}\sum_k\gamma_k\left(L^*_{k\vec{s}}\star_{\vec{s}}\frac{\partial L_{k\vec{s}}}{\partial\alpha^*_m} - \frac{\partial L^*_{k\vec{s}}}{\partial\alpha^*_m}\star_{\vec{s}}L_{k\vec{s}}\right) \right\}W_{\vec{s}}(\vec{\alpha},\vec{\alpha}^*,t)\right] \nonumber\\
    & -i\hbar\sum_{m,n=1}^M\frac{\partial^2}{\partial\alpha_m\partial\alpha_n}\left[\lambda^{\vec{s}}_{mn}W_{\vec{s}}(\vec{\alpha},\vec{\alpha}^*,t)\right] + i\hbar\sum_{m,n=1}^M\frac{\partial^2}{\partial\alpha_m\partial\alpha^*_n}\left[\Lambda^{\vec{s}}_{mn}W_{\vec{s}}(\vec{\alpha},\vec{\alpha}^*,t)\right] - {\rm c.c.},
\end{align}
where we have used Eq.~\eqref{eq:definition of Ksm}.


\subsection{\label{appendix:stochastic differential equation}Stochastic differential equation{\rm :} Derivations of Eqs.~\texorpdfstring{\eqref{eq:condition for obtaining stochastic differential equaitons when the jump operators do not couple different degrees of feedom} and \eqref{eq:stochastic differential equation simplified version}}{TEXT}}
When the the matrix elements of $\bm{\lambda}^{\vec{s}}$ and $\bm{\Lambda}^{\vec{s}}$ satisfy the condition Eq.~\eqref{eq:lambda and Lambda when the jump operators do not couple different degrees of feedom}, i.e., when they are diagonal, we can analytically diagonalize the matrix $\bm{\mathcal{A}}^{\vec{s}}$ as
\begin{align}
    \label{eq:diagonalizing matrix A when jumo operators do not couple different degrees of feeedom}
    \bm{\mathcal{U}}^{\vec{s}\dagger}\bm{\mathcal{A}}^{\vec{s}} \bm{\mathcal{U}}^{\vec{s}} = \bm{\mathcal{A}}^{\vec{s}}_{\rm diag} = 
    \left[\!\!\!\!\!
    \begin{array}{ccc} 
          {\begin{array}{cc} 2(\Lambda^{s_1}_{11} - |\lambda^{s_1}_{11}|) &  \\  & 2(\Lambda^{s_1}_{11} + |\lambda^{s_1}_{11}|) \end{array}} & &  \text{\huge 0} \\ 
          & \ddots & \\
          \text{\huge 0} & & {\begin{array}{cc} 2(\Lambda^{s_M}_{MM} - |\lambda^{s_M}_{MM}|) &  \\  & 2(\Lambda^{s_M}_{MM} + |\lambda^{s_M}_{MM}|) \end{array}} 
    \end{array}
    \!\!\!\!\!\right],
\end{align}
and the diagonalizing matrix $\bm{\mathcal{U}}^{\vec{s}}$ takes the form:
\begin{align}
    \label{eq:diagonalizing matrix U when jumo operators do not couple different degrees of feeedom}
    \bm{\mathcal{U}}^{\vec{s}} = \frac{1}{\sqrt{2}}
    \begin{bmatrix}
        \bm{U}^{\vec{s}} \\
        \bm{U}^{\vec{s}*}
    \end{bmatrix},
\end{align}
with $\bm{U}^{\vec{s}}$ being a $M\times 2M$ matrix given by
\begin{align}
    \label{eq:a patr of the matrix U}
    \bm{U}^{\vec{s}} =
    \left[\!\!\!\!\!
    \begin{array}{ccc} 
          {\begin{array}{cccc} -ie^{i\theta_1/2} & e^{i\theta_1/2} & 0 & 0 \\ 0 & 0 & -ie^{i\theta_2/2} & e^{i\theta_2/2} \end{array}} & &  \text{\huge 0} \\ 
          & \ddots & \\
          \text{\huge 0} & & {\begin{array}{cccc} -ie^{i\theta_{M-1}/2} & e^{i\theta_{M-1}/2} & 0 & 0 \\ 0 & 0 & -ie^{i\theta_M/2} & e^{i\theta_M/2} \end{array}} 
    \end{array}
    \!\!\!\!\!\right],
\end{align}
where $\theta_m(\alpha_m,\alpha^*_m) = {\rm arg}(\lambda^{s_m}_{mm}(\alpha_m,\alpha^*_m))$.
Substituting Eq.~\eqref{eq:diagonalizing matrix A when jumo operators do not couple different degrees of feeedom} into Eq.~\eqref{eq:feasible condition of the Hubbard-Stratonovich transformation}, we can rewrite the condition Eq.~\eqref{eq:feasible condition of the Hubbard-Stratonovich transformation} as Eq.~\eqref{eq:condition for obtaining stochastic differential equaitons when the jump operators do not couple different degrees of feedom},
and substituting Eqs.~\eqref{eq:diagonalizing matrix A when jumo operators do not couple different degrees of feeedom}-\eqref{eq:a patr of the matrix U} into Eq.~\eqref{eq:stochastic differential equation} and choosing $\bm{\mathcal{Q}}$ as the identity matrix, we obtain the stochastic differential equation~\eqref{eq:stochastic differential equation simplified version}.


\section{\label{appendix:Hubbard-Stratonovich transformation}Hubbard-Stratonovich transformation{\rm :} Derivation of Eq.~\texorpdfstring{\eqref{eq:Hubbard-Stratonovich transformation}}{TEXT}}
In \ref{appendix:Phase-space Gaussian integral}, we first introduce the phase-space Gaussian integral which is necessary for deriving the Hubbard-Stratonovich transformation.
In the subsequent sections \ref{appendix:Hubbard-Stratonovich transformation A>0} and \ref{appendix:Hubbard-Stratonovich transformation A contains zeros}, we derive the Hubbard-Stratonovich transformation for the cases of $\bm{\mathcal{A}}^{\vec{s}}\succ 0$ and $\bm{\mathcal{A}}^{\vec{s}}\succeq 0$, respectively.


\subsection{\label{appendix:Phase-space Gaussian integral}Phase-space Gaussian integral}
We introduce the Gaussian integral in the phase space:
\begin{align}
    \label{eq:Gaussian integral phase space}
    {\rm exp}\left(-
    \begin{bmatrix}
        \vec{\eta}^{*{\rm T}}, \vec{\eta}^{\rm T}
    \end{bmatrix}
    \bm{\mathcal{G}}
    \begin{bmatrix}
        \vec{\eta} \\
        \vec{\eta}^*
    \end{bmatrix}
    \right)
    =
    \frac{1}{\sqrt{\text{det}\bm{\mathcal{G}}}}\int\frac{d^2\vec{\xi}}{\pi^M}{\rm exp}\left(-\frac{1}{2}
    \begin{bmatrix}
        \vec{\xi}^{*{\rm T}}, \vec{\xi}^{\rm T}
    \end{bmatrix}
    \bm{\mathcal{G}}^{-1}
    \begin{bmatrix}
        \vec{\xi} \\
        \vec{\xi}^*
    \end{bmatrix}
    + \sqrt{2}i
    \begin{bmatrix}
        \vec{\eta}^{*{\rm T}}, \vec{\eta}^{\rm T}
    \end{bmatrix}
    \begin{bmatrix}
        \vec{\xi} \\
        \vec{\xi}^*
    \end{bmatrix}
    \right),
\end{align}
where $\bm{\mathcal{G}} \succ 0$ is a $2M\times 2M$ Hermitian and positive-definite matrix, 
$\vec{\eta}$ and $\vec{\xi}$ are complex vectors of dimension $M$, which are given by $\vec{\eta} = \vec{\eta}^{\rm re} + i\vec{\eta}^{\rm im}$ with $\vec{\eta}^{\rm re/{\rm im}} = (\eta_1^{\rm re/{\rm im}},\cdots,\eta_M^{\rm re/{\rm im}})^{\rm T}\in \mathbb{R}^M$, and $\vec{\xi} = \vec{\xi}^{\rm re} + i\vec{\xi}^{\rm im}$ with $\vec{\xi}^{\rm re/{\rm im}} = (\xi_1^{\rm re/{\rm im}},\cdots,\xi_M^{\rm re/{\rm im}})^{\rm T}\in \mathbb{R}^M$, respectively, and $\int d^2\vec{\xi} = \prod_{m=1}^M\int d^2\xi_m = \prod_{m}\int_{-\infty}^\infty d\xi_m^{\rm re} \int_{-\infty}^\infty d\xi_m^{\rm im}$.
Below, we calculate the right-hand side of Eq.~\eqref{eq:Gaussian integral phase space} and show it agrees with the left-hand side.
For this purpose, we introduce the following $2M\times 2M$ unitary matrix $\bm{\mathcal{P}}$:
\begin{gather}
    \bm{\mathcal{P}} = \frac{1}{\sqrt{2}}
    \begin{bmatrix}
        \bm{1} & i\bm{1} \\
        \bm{1} & -i\bm{1}
    \end{bmatrix},\\
    \bm{\mathcal{P}}^{-1} = \bm{\mathcal{P}}^{\dagger} = \frac{1}{\sqrt{2}}
    \begin{bmatrix}
        \bm{1} & \bm{1} \\
        -i\bm{1} & i\bm{1}
    \end{bmatrix},
\end{gather}
where $\bm{1}$ is the $M\times M$ identity matrix.
The matrix $\bm{\mathcal{P}}$ acts on the vector $[\vec{\xi}^{\rm T},\vec{\xi}^{*{\rm T}}]^{\rm T}$ as
\begin{gather}
    \bm{\mathcal{P}}^{\dagger}
    \begin{bmatrix}
        \vec{\xi} \\
        \vec{\xi}^*
    \end{bmatrix} = \sqrt{2}
    \begin{bmatrix}
        \vec{\xi}^{\rm re} \\
        \vec{\xi}^{\rm im}
    \end{bmatrix},\\
    \begin{bmatrix}
        \vec{\xi}^{*{\rm T}},\vec{\xi}^{\rm T}
    \end{bmatrix}\bm{\mathcal{P}} = \sqrt{2}
    \begin{bmatrix}
        \vec{\xi}^{\rm re {\rm T}}, \vec{\xi}^{\rm im {\rm T}}
    \end{bmatrix}.
\end{gather}
Substituting the identity matrix $\bm{\mathcal{P}}\bm{\mathcal{P}}^{\dagger} = \bm{1}$ into both side of $\bm{\mathcal{G}}^{-1}$ on the right-hand side of Eq.~\eqref{eq:Gaussian integral phase space} and using the equality $[\vec{\eta}^{* {\rm T}},\vec{\eta}^{\rm T}][\xi^{\rm T},\xi^{* {\rm T}}]^{\rm T} = 2[\vec{\eta}^{\rm re {\rm T}},\vec{\eta}^{\rm im {\rm T}}][\xi^{\rm re {\rm T}},\xi^{\rm im {\rm T}}]^{\rm T}$, we obtain
\begin{align}
    \label{eq:phase_space_Gaussian_integral_tochu_appendix}
    \text{RHS of Eq.~\eqref{eq:Gaussian integral phase space}}
    &=
    \frac{2^M}{\sqrt{\text{det}\bm{\mathcal{G}}}}\int\frac{d^2\vec{\xi}}{(2\pi)^M}{\rm exp}\left(-\frac{1}{2}
    \begin{bmatrix}
        \vec{\xi}^{\rm re {\rm T}}, \vec{\xi}^{\rm im {\rm T}}
    \end{bmatrix}
    2\bm{\mathcal{P}}^{\dagger}\bm{\mathcal{G}}^{-1}\bm{\mathcal{P}}
    \begin{bmatrix}
        \vec{\xi}^{\rm re} \\
        \vec{\xi}^{\rm im}
    \end{bmatrix}
    + 2\sqrt{2}i
    \begin{bmatrix}
        \vec{\eta}^{\rm re {\rm T}},\vec{\eta}^{\rm im {\rm T}}
    \end{bmatrix}
    \begin{bmatrix}
        \vec{\xi}^{\rm re} \\
        \vec{\xi}^{\rm im}
    \end{bmatrix}
    \right).
\end{align}
To implement the integration in Eq.~\eqref{eq:phase_space_Gaussian_integral_tochu_appendix}, we use the multiple-variables Gaussian integral formula:
\begin{align}
    \label{eq:formular of the Gaussian integral for multiple variables appendix}
    \int \frac{d^2\vec{\xi}}{(2\pi)^M}{\rm exp}\left(-\frac{1}{2}
    \begin{bmatrix}
        \vec{\xi}^{\rm re {\rm T}},\vec{\xi}^{\rm im {\rm T}}
    \end{bmatrix}
    \bm{\mathcal{F}}
    \begin{bmatrix}
        \vec{\xi}^{\rm re} \\
        \vec{\xi}^{\rm im}
    \end{bmatrix}
    + 
    \begin{bmatrix}
        \vec{u}^{\rm T},\vec{v}^{\rm T}
    \end{bmatrix}
    \begin{bmatrix}
        \vec{\xi}^{\rm re} \\
        \vec{\xi}^{\rm im}
    \end{bmatrix}
    \right)
    =
    \frac{1}{\sqrt{\text{det}\bm{\mathcal{F}}}}
    {\rm exp}\left(\frac{1}{2}
    \begin{bmatrix}
        \vec{u}^{\rm T},\vec{v}^{\rm T}
    \end{bmatrix}
    \bm{\mathcal{F}}^{-1}
    \begin{bmatrix}
        \vec{u} \\
        \vec{v}
    \end{bmatrix}
    \right),
\end{align}
where $\bm{\mathcal{F}} \succ 0$ is a $2M\times 2M$ symmetric and positive-definite matrix, and $\vec{u}$ and $\vec{v}$ are complex vectors of dimension $M$.
Noting that $\bm{\mathcal{P}}^{\dagger}\bm{\mathcal{G}}^{-1}\bm{\mathcal{P}}$ is a symmetric and positive-definite matrix and
substituting Eq.~\eqref{eq:formular of the Gaussian integral for multiple variables appendix} with $\bm{\mathcal{F}} = 2\bm{\mathcal{P}}^{\dagger}\bm{\mathcal{G}}^{-1}\bm{\mathcal{P}}$, $\vec{u} = 2\sqrt{2}i\vec{\eta}^{\rm re}$, and $\vec{v} = 2\sqrt{2}i\vec{\eta}^{\rm im}$ into the right-hand side of Eq.~\eqref{eq:phase_space_Gaussian_integral_tochu_appendix}, we obtain the left-hand side of Eq.~\eqref{eq:Gaussian integral phase space}.


\subsection{\label{appendix:Hubbard-Stratonovich transformation A>0}Hubbard-Stratonovich transformation for \texorpdfstring{$\bm{\mathcal{A}}^{\vec{s}} \succ 0$}{TEXT}}
We first consider the case of a positive-definite $\bm{\mathcal{A}}^{\vec{s}}$.
Substituting $\vec{\eta} = \vec{\eta}_{j+1}$ and $\bm{\mathcal{G}} = \Delta t\bm{\mathcal{A}}^{\vec{s}}/2$ into Eq.~\eqref{eq:Gaussian integral phase space}, we obtain
\begin{align}
    \label{eq:calculation of HS transformation appendix 0}
    {\rm exp}\left(-\frac{\Delta t}{2} 
    \begin{bmatrix}
        \vec{\eta}_{j+1}^{*{\rm T}},\vec{\eta}^{\rm T}_{j+1}
    \end{bmatrix}
    \bm{\mathcal{A}}^{\vec{s}}
    \begin{bmatrix}
        \vec{\eta}_{j+1}^* \\
        \vec{\eta}_{j+1}
    \end{bmatrix}
    \right) = \frac{2^M}{\sqrt{\Delta t^{2M}{\rm det}\bm{\mathcal{A}}^{\vec{s}}}}\int\frac{d^2\vec{\xi}}{\pi^M}{\rm exp}\left(-\frac{1}{\Delta t}
    \begin{bmatrix}
        \vec{\xi}^{*{\rm T}}, \vec{\xi}^{\rm T}
    \end{bmatrix}
    [\bm{\mathcal{A}}^{s}]^{-1}
    \begin{bmatrix}
        \vec{\xi} \\
        \vec{\xi}^*
    \end{bmatrix}
    + \sqrt{2}i
    \begin{bmatrix}
        \vec{\eta}_{j+1}^{*{\rm T}}, \vec{\eta}_{j+1}^{\rm T}
    \end{bmatrix}
    \begin{bmatrix}
        \vec{\xi} \\
        \vec{\xi}^*
    \end{bmatrix}
    \right),
\end{align}
where we have used ${\rm det}(\Delta t\bm{\mathcal{A}}^{\vec{s}}/2) = \Delta t^{2M}{\rm det}\bm{\mathcal{A}}^{\vec{s}}/2^{2M}$.
Here, $\bm{\mathcal{A}}^{\vec{s}}$ is defined by Eq.~\eqref{eq:definitnioa of the diffusion matrix A}, i.e.,
\begin{align}
    \label{eq:calculation of HS transformation appendix 1}
    \bm{\mathcal{A}}^{\vec{s}} = 2
    \begin{bmatrix}
        \bm{\Lambda}^{\vec{s}} & \bm{\lambda}^{\vec{s}} \\
        \bm{\lambda}^{\vec{s}*} & \bm{\Lambda}^{\vec{s}*}
    \end{bmatrix},
\end{align}
where the matrix elements of $\bm{\Lambda}^{\vec{s}}$ and $\bm{\lambda}^{\vec{s}}$ are given by Eqs.~\eqref{eq:definitnion of Lambdamn} and \eqref{eq:definitnion of lambdamn}, respectively.
Multiplying the identity matrix $\bm{\mathcal{P}}\bm{\mathcal{P}}^{\dagger} = \bm{1}$ from both sides, we rewrite $\bm{\mathcal{A}}^{\vec{s}}$ in Eq.~\eqref{eq:calculation of HS transformation appendix 1} as
\begin{align}
    \bm{\mathcal{A}}^{\vec{s}} = \bm{\mathcal{P}}\bm{\mathcal{P}}^{\dagger}\bm{\mathcal{A}}^{\vec{s}}\bm{\mathcal{P}}\bm{\mathcal{P}}^{\dagger} = 2\bm{\mathcal{P}}
    \begin{bmatrix}
        [\bm{\Lambda}^{\vec{s}}]^{\rm re} + [\bm{\lambda}^{\vec{s}}]^{\rm re} & -[\bm{\Lambda}^{\vec{s}}]^{\rm im} + [\bm{\lambda}^{\vec{s}}]^{\rm im} \\ 
        -[\bm{\Lambda}^{\vec{s}}]^{\rm im\rm T} + [\bm{\lambda}^{\vec{s}}]^{\rm im\rm T} & [\bm{\Lambda}^{\vec{s}}]^{\rm re} - [\bm{\lambda}^{\vec{s}}]^{\rm re}
    \end{bmatrix}
    \bm{\mathcal{P}}^{\dagger},
\end{align}
where $[\bm{\Lambda}^{\vec{s}}]^{\rm re}$ and $[\bm{\Lambda}^{\vec{s}}]^{\rm im}$ ($[\bm{\lambda}^{\vec{s}}]^{\rm re}$ and $[\bm{\lambda}^{\vec{s}}]^{\rm im}$) are the real and imaginary parts of the matrix  $\bm{\Lambda}^{\vec{s}}$ ($\bm{\lambda}^{\vec{s}}$), respectively, and we have used the symmetricity of $\bm{\lambda}^{\vec{s}}$ and the Hermiticity of $\bm{\Lambda}^{\vec{s}}$.
Since $\bm{\mathcal{P}}^{\dagger}\bm{\mathcal{A}}^{\vec{s}}\bm{\mathcal{P}}$ is a real symmetric matrix, we can diagonalize it by using an orthogonal matrix $\bm{\mathcal{V}}^{\vec{s}}$ as
\begin{align}
    \label{eq:calculation of HS transformation appendix 2}
    \bm{\mathcal{V}}^{\vec{s}\rm T}\bm{\mathcal{P}}^{\dagger}\bm{\mathcal{A}}^{\vec{s}}\bm{\mathcal{P}}\bm{\mathcal{V}}^{\vec{s}} = \bm{\mathcal{A}}^{\vec{s}}_{\rm diag} =
    \begin{bmatrix}
        \sigma_1 & &\text{\huge 0} \\
        & \ddots & \\
        \text{\huge 0}& & \sigma_{2M}
    \end{bmatrix},
\end{align}
where $\bm{\mathcal{A}}^{\vec{s}}_{\rm diag}$ is the diagonal matrix having the eigenvalues $\sigma_{\mu} \in \mathbb{R}_{>0}$ for $\forall \mu$ of $\bm{\mathcal{A}}^{\vec{s}}$ as diagonal entries.
Here, from the assumption $\bm{\mathcal{A}}^{\vec{s}} \succ 0$, $\sigma_{\mu}$ for $\forall \mu$ take positive values.
Eq.~\eqref{eq:calculation of HS transformation appendix 2} also shows that $\bm{\mathcal{A}}^{\vec{s}}$ is diagonalized with the unitary matrix $\bm{\mathcal{U}}^{\vec{s}}$ defined by
\begin{align}
    \label{eq:calculation of HS transformation appendix 4}
    \bm{\mathcal{U}}^{\vec{s}} = \bm{\mathcal{P}}\bm{\mathcal{V}}^{\vec{s}}.
\end{align}
Taking the inverse of both side of Eq.~\eqref{eq:calculation of HS transformation appendix 2}, we obtain
\begin{align}
    \label{eq:calculation of HS transformation appendix 3}
    [\bm{\mathcal{A}}^{\vec{s}}]^{-1} = \bm{\mathcal{U}}^{\vec{s}}[\bm{\mathcal{A}}^{\vec{s}}_{\rm diag}]^{-1}\bm{\mathcal{U}}^{\vec{s}\dagger} = \bm{\mathcal{U}}^{\vec{s}}
    \begin{bmatrix}
        \dfrac{1}{\sigma_1} & &\text{\huge 0} \\
        & \ddots & \\
       \text{\huge 0} & & \dfrac{1}{\sigma_{2M}}
    \end{bmatrix}
    \bm{\mathcal{U}}^{\vec{s}\dagger}.
\end{align}
Substituting Eq.~\eqref{eq:calculation of HS transformation appendix 3} into Eq.~\eqref{eq:calculation of HS transformation appendix 0}, we can rewrite Eq.~\eqref{eq:calculation of HS transformation appendix 0} as
\begin{align}
    \label{eq:calculation of HS transformation appendix 7}
    {\rm exp}\left(-\frac{\Delta t}{2} 
    \begin{bmatrix}
        \vec{\eta}_{j+1}^{*{\rm T}},\vec{\eta}^{\rm T}_{j+1}
    \end{bmatrix}
    \bm{\mathcal{A}}^{\vec{s}}
    \begin{bmatrix}
        \vec{\eta}_{j+1}^* \\
        \vec{\eta}_{j+1}
    \end{bmatrix}
    \right) = \frac{2^M}{\sqrt{\Delta t^{2M}{\rm det}\bm{\mathcal{A}}^{\vec{s}}}}\int\frac{d^2\vec{\xi}}{\pi^M}{\rm exp}\left(-\frac{1}{\Delta t}
    \begin{bmatrix}
        \vec{\xi}^{*{\rm T}}, \vec{\xi}^{\rm T}
    \end{bmatrix}
    \bm{\mathcal{U}}^{\vec{s}}[\bm{\mathcal{A}}^{\vec{s}}_{\rm diag}]^{-1}\bm{\mathcal{U}}^{\vec{s}\dagger}
    \begin{bmatrix}
        \vec{\xi} \\
        \vec{\xi}^*
    \end{bmatrix}
    + \sqrt{2}i
    \begin{bmatrix}
        \vec{\eta}_{j+1}^{*{\rm T}}, \vec{\eta}_{j+1}^{\rm T}
    \end{bmatrix}
    \begin{bmatrix}
        \vec{\xi} \\
        \vec{\xi}^*
    \end{bmatrix}
    \right).
\end{align}

Here, we define 
\begin{align}
\Delta\overrightarrow{\Xi}=\sqrt{2}\bm{\mathcal{U}}^{\vec{s}\dagger}
    \begin{bmatrix}
        \vec{\xi} \\
        \vec{\xi}^*
    \end{bmatrix},
\end{align}
which is real as shown below:
Dividing the real $2M\times 2M$ matrix $\bm{\mathcal{V}}^{\vec{s}}$ into four blocks as
\begin{align}
    \label{eq:calculation of HS transformation appendix 5}
    \bm{\mathcal{V}}^{\vec{s}} =
    \begin{bmatrix}
        \bm{V}^{\vec{s}}_{11} & \bm{V}^{\vec{s}}_{12} \\
        \bm{V}^{\vec{s}}_{21} & \bm{V}^{\vec{s}}_{22}
    \end{bmatrix},
\end{align}
where $\bm{V}^{\vec{s}}_{11}$, $\bm{V}^{\vec{s}}_{12}$, $\bm{V}^{\vec{s}}_{21}$, and $\bm{V}^{\vec{s}}_{22}$ are $M\times M$ real matrices, we can express $\bm{\mathcal{U}}^{\vec{s}}$ as
\begin{gather}
    \bm{\mathcal{U}}^{\vec{s}} = \bm{\mathcal{P}}\bm{\mathcal{V}}^{\vec{s}} =  \frac{1}{\sqrt{2}}
    \begin{bmatrix}
        \bm{V}^{\vec{s}}_{11} + i\bm{V}^{\vec{s}}_{21} & \bm{V}^{\vec{s}}_{12} + i\bm{V}^{\vec{s}}_{22} \\
        \bm{V}^{\vec{s}}_{11} - i\bm{V}^{\vec{s}}_{21} & \bm{V}^{\vec{s}}_{12} - i\bm{V}^{\vec{s}}_{22}
    \end{bmatrix}.
\end{gather}
It follows that all the components of $\Delta\overrightarrow{\Xi}$ are real:
\begin{align}
    \label{eq:calculation of HS transformation appendix 6}
    \Delta\overrightarrow{\Xi} = \sqrt{2}\bm{\mathcal{U}}^{\vec{s}\dagger}
    \begin{bmatrix}
        \vec{\xi} \\
        \vec{\xi}^*
    \end{bmatrix}
    = \frac{1}{\sqrt{2}}
    \begin{bmatrix}
        (\bm{V}^{\vec{s}\rm T}_{11} - i\bm{V}^{\vec{s}\rm T}_{21})\vec{\xi} + {\rm c.c.} \\
        (\bm{V}^{\vec{s}\rm T}_{12} - i\bm{V}^{\vec{s}\rm T}_{22})\vec{\xi} + {\rm c.c.}
    \end{bmatrix} \in\mathbb{R}^{2M}.
\end{align}
Taking the Hermitian conjugate of the above equation, we also have
\begin{align}
    \Delta\overrightarrow{\Xi}^{\rm T}=\sqrt{2}
    \begin{bmatrix}
        \vec{\xi}^{*{\rm T}},\vec{\xi}^{\rm T}
    \end{bmatrix}
    \bm{\mathcal{U}}^{\vec{s}}.
\end{align}

Performing the variable transformation according to Eq.~\eqref{eq:calculation of HS transformation appendix 6}, we can rewrite Eq.~\eqref{eq:calculation of HS transformation appendix 7} as follows:
\begin{align}
    {\rm exp}\left(-\frac{\Delta t}{2} 
    \begin{bmatrix}
        \vec{\eta}_{j+1}^{*{\rm T}},\vec{\eta}^{\rm T}_{j+1}
    \end{bmatrix}
    \bm{\mathcal{A}}^{\vec{s}}
    \begin{bmatrix}
        \vec{\eta}_{j+1}^* \\
        \vec{\eta}_{j+1}
    \end{bmatrix}
    \right) &= \frac{1}{\sqrt{(2\pi\Delta t)^{2M}{\rm det}\bm{\mathcal{A}}^{\vec{s}}}}\prod_{\mu = 1}^{2M}\int^{\infty}_{-\infty}d\Delta\Xi_{\mu}{\rm exp}\left(-\frac{1}{2\Delta t}
    \Delta\overrightarrow{\Xi}^{\rm T}[\bm{\mathcal{A}}^{\vec{s}}_{\rm diag}]^{-1}\Delta\overrightarrow{\Xi}
    + i
    \begin{bmatrix}
        \vec{\eta}_{j+1}^{*{\rm T}}, \vec{\eta}_{j+1}^{\rm T}
    \end{bmatrix}
    \bm{\mathcal{U}}^{\vec{s}}\Delta\overrightarrow{\Xi}
    \right) \\
    \label{eq:calculation of HS transformation appendix 8}
    &= \prod_{\mu = 1}^{2M}\int^{\infty}_{-\infty}d\Delta\Xi_{\mu}\frac{e^{-\Delta\Xi^2_{\mu}/(2\sigma_{\mu}\Delta t)}}{\sqrt{2\pi\sigma_{\mu}\Delta t}}
    {\rm exp}\left\{i
    \sum_{m=1}^M\left(
    \eta^*_{m,j+1}\left[\bm{\mathcal{U}}^{\vec{s}}\Delta\overrightarrow{\Xi}\right]_m + \eta_{m,j+1}\left[\bm{\mathcal{U}}^{\vec{s}}\Delta\overrightarrow{\Xi}\right]_{m+M}
    \right)
    \right\},
\end{align}
where we have used the Jacobian $2^{-2M}$ for the variable transformation $[\vec{\xi}^{\rm re},\vec{\xi}^{\rm im}]^{\rm T}=\frac{1}{2}\bm{\mathcal{P}}^{\dagger}\bm{\mathcal{U}}^{\vec{s}}\Delta\overrightarrow{\Xi}$ and ${\rm det}\bm{\mathcal{A}}^{\vec{s}}=\prod_{\mu = 1}^{2M} \sigma_{\mu}$.
Noting the relation
\begin{align}
    \left[\bm{\mathcal{U}}^{\vec{s}}\Delta\overrightarrow{\Xi}\right]_{m + M} = \left[\bm{\mathcal{U}}^{\vec{s}}\Delta\overrightarrow{\Xi}\right]^*_{m} =\sqrt{2}\xi_m^*~(m=1,2,\cdots, M)
\end{align}
derived from Eq.~\eqref{eq:calculation of HS transformation appendix 6}, we can rewrite Eq.~\eqref{eq:calculation of HS transformation appendix 8} as
\begin{align}
    {\rm exp}\left(-\frac{\Delta t}{2} 
    \begin{bmatrix}
        \vec{\eta}_{j+1}^{*{\rm T}},\vec{\eta}^{\rm T}_{j+1}
    \end{bmatrix}
    \bm{\mathcal{A}}^{\vec{s}}
    \begin{bmatrix}
        \vec{\eta}_{j+1}^* \\
        \vec{\eta}_{j+1}
    \end{bmatrix}
    \right)
    =
    \prod_{\mu = 1}^{2M}\int^{\infty}_{-\infty}d\Delta\Xi_{\mu}\frac{e^{-\Delta\Xi^2_{\mu}/(2\sigma_{\mu}\Delta t)}}{\sqrt{2\pi\sigma_{\mu}\Delta t}}
    \prod_{m=1}^M{\rm exp}\left(\eta^*_{m,j+1}\left[i\bm{\mathcal{U}}^{\vec{s}}\Delta\overrightarrow{\Xi}\right]_m - {\rm c.c.}\right).
\end{align}
We further transform the integration variable so that the Gaussian in the integrand has the same width for all variables. The resulting Gaussian is invariant under an orthogonal transformation of the integration variables. Thus, we define the new integration variable $\Delta\overrightarrow{\mathcal{W}}$ as
\begin{align}
    \Delta\overrightarrow{\Xi} = \sqrt{\bm{\mathcal{A}}^{\vec{s}}_{\rm diag}}\bm{\mathcal{Q}}\Delta\overrightarrow{\mathcal{W}},
\end{align}
where $\bm{\mathcal{Q}}$ is an arbitrarily chosen $2M\times 2M$ orthogonal matrix.
Then, we finally obtain
\begin{align}
    \label{eq:calculation of HS transformation appendix 14}
    {\rm exp}\left(-\frac{\Delta t}{2} 
    \begin{bmatrix}
        \vec{\eta}_{j+1}^{*{\rm T}},\vec{\eta}^{\rm T}_{j+1}
    \end{bmatrix}
    \bm{\mathcal{A}}^{\vec{s}}
    \begin{bmatrix}
        \vec{\eta}_{j+1}^* \\
        \vec{\eta}_{j+1}
    \end{bmatrix}
    \right)
    =
    \prod_{\mu = 1}^{2M}\int^{\infty}_{-\infty}d\Delta\mathcal{W}_{\mu}\frac{e^{-\Delta\mathcal{W}^2_{\mu}/(2\Delta t)}}{\sqrt{2\pi\Delta t}}
    \prod_{m=1}^M{\rm exp}\left(\eta^*_{m,j+1}\left[i\bm{\mathcal{U}}^{\vec{s}}\sqrt{\bm{\mathcal{A}}^{\vec{s}}_{\rm diag}}\bm{\mathcal{Q}}\Delta\overrightarrow{\mathcal{W}}\right]_m - {\rm c.c.}\right),
\end{align}
which is the Hubbard-Stratonovich transformation Eq.~\eqref{eq:Hubbard-Stratonovich transformation} for $\bm{\mathcal{A}}^{\vec{s}} \succ 0$.

\subsection{\label{appendix:Hubbard-Stratonovich transformation A contains zeros}Hubbard-Stratonovich transformation for \texorpdfstring{$\bm{\mathcal{A}}^{\vec{s}} \succeq 0$}{TEXT}}
Next, we consider the case when $\bm{\mathcal{A}}^{\vec{s}}$ has zero eigenvalues, where the other eigenvalues are positive.
Here, we introduce a matrix $\bm{\mathcal{A}}^{\vec{s}}(\varepsilon)$ such that
\begin{align}
    \label{eq:calculation of HS transformation appendix 9}
    \bm{\mathcal{A}}^{\vec{s}}(\varepsilon) = \bm{\mathcal{A}}^{\vec{s}} + \varepsilon
    \begin{bmatrix}
        \bm{1} & \bm{0} \\
        \bm{0} & \bm{1}
    \end{bmatrix},
\end{align}
where $\varepsilon$ is a positive value, and $\bm{0}$ is the $M\times M$ zero matrix.
We can diagonalize the matrices $\bm{\mathcal{A}}^{\vec{s}}$ and $\bm{\mathcal{A}}^{\vec{s}}(\varepsilon)$ by using the same unitary matrix $\bm{\mathcal{U}}^{\vec{s}}$, given in the form of Eq.~\eqref{eq:calculation of HS transformation appendix 4}.
Letting $\sigma_{\mu} \geq 0$ ($\mu=1,2,\cdots, 2M$) be the eigenvalues of $\bm{\mathcal{A}}^{\vec{s}}$, $\bm{\mathcal{A}}^{\vec{s}}(\varepsilon)$ is diagonalized as
\begin{align}
    \bm{\mathcal{U}}^{\vec{s}\dagger}\bm{\mathcal{A}}^{\vec{s}}(\varepsilon)\bm{\mathcal{U}}^{\vec{s}} = \bm{\mathcal{A}}^{\vec{s}}_{\rm diag}(\varepsilon)
    =
    \begin{bmatrix}
        \sigma_1 + \varepsilon & & \text{\huge 0} \\
        & \ddots & \\
        \text{\huge 0} & & \sigma_{2M} + \varepsilon
    \end{bmatrix}.
\end{align}
In the limit of $\varepsilon\to 0$,
the matrices $\bm{\mathcal{A}}^{\vec{s}}(\varepsilon)$ and $\bm{\mathcal{A}}_{\rm diag}^{\vec{s}}(\varepsilon)$ reduce to $\bm{\mathcal{A}}^{\vec{s}}$ and $\bm{\mathcal{A}}^{\vec{s}}_{\rm diag}$, respectively:
\begin{gather}
    \label{eq:calculation of HS transformation appendix 11}
    \lim_{\varepsilon\to 0}\bm{\mathcal{A}}^{\vec{s}}(\varepsilon) = \bm{\mathcal{A}}^{\vec{s}},\\
    \label{eq:calculation of HS transformation appendix 12}
    \lim_{\varepsilon\to 0}\bm{\mathcal{A}}^{\vec{s}}_{\rm diag}(\varepsilon) = \bm{\mathcal{A}}^{\vec{s}}_{\rm diag}.
\end{gather}
Since $\bm{\mathcal{A}}^{\vec{s}}(\varepsilon)$ is a positive-definite matrix, we can follow the same procedures in \ref{appendix:Hubbard-Stratonovich transformation A>0} by replacing $\bm{\mathcal{A}}^{\vec{s}}$ with $\bm{\mathcal{A}}^{\vec{s}}(\varepsilon)$, obtaining
\begin{align}
    \label{eq:calculation of HS transformation appendix 10}
    {\rm exp}\left(-\Delta t 
    \begin{bmatrix}
        \vec{\eta}_{j+1}^{*{\rm T}},\vec{\eta}^{\rm T}_{j+1}
    \end{bmatrix}
    \bm{\mathcal{A}}^{\vec{s}}(\varepsilon)
    \begin{bmatrix}
        \vec{\eta}_{j+1}^* \\
        \vec{\eta}_{j+1}
    \end{bmatrix}
    \right)
    =
    \prod_{\mu = 1}^{2M}\int^{\infty}_{-\infty}d\Delta\mathcal{W}_{\mu}\frac{e^{-\Delta\mathcal{W}^2_{\mu}/(2\Delta t)}}{\sqrt{2\pi\Delta t}}
    \prod_{m=1}^M{\rm exp}\left(\eta^*_{m,j+1}\left[i\bm{\mathcal{U}}^{\vec{s}}\sqrt{\bm{\mathcal{A}}^{\vec{s}}_{\rm diag}(\varepsilon)}\bm{\mathcal{Q}}\Delta\overrightarrow{\mathcal{W}}\right]_m - {\rm c.c.}\right).
\end{align}
Taking the limit of $\varepsilon\to 0$ in both side of Eq.~\eqref{eq:calculation of HS transformation appendix 10} and using Eq.~\eqref{eq:calculation of HS transformation appendix 11}, we obtain
\begin{align}
    \label{eq:calculation of HS transformation appendix 13}
    {\rm exp}\left(-\Delta t 
    \begin{bmatrix}
        \vec{\eta}_{j+1}^{*{\rm T}},\vec{\eta}^{\rm T}_{j+1}
    \end{bmatrix}
    \bm{\mathcal{A}}^{\vec{s}}
    \begin{bmatrix}
        \vec{\eta}_{j+1}^* \\
        \vec{\eta}_{j+1}
    \end{bmatrix}
    \right)
    =
    \lim_{\varepsilon\to 0}
    \prod_{\mu = 1}^{2M}\int^{\infty}_{-\infty}d\Delta\mathcal{W}_{\mu}\frac{e^{-\Delta\mathcal{W}^2_{\mu}/(2\Delta t)}}{\sqrt{2\pi\Delta t}}
    \prod_{m=1}^M{\rm exp}\left(\eta^*_{m,j+1}\left[i\bm{\mathcal{U}}^{\vec{s}}\sqrt{\bm{\mathcal{A}}^{\vec{s}}_{\rm diag}(\varepsilon)}\bm{\mathcal{Q}}\Delta\overrightarrow{\mathcal{W}}\right]_m - {\rm c.c.}\right).
\end{align}

In order to take the limit of $\varepsilon\to 0$ in the right-hand side of Eq.~\eqref{eq:calculation of HS transformation appendix 13}, we introduce the $c$-number functions:
\begin{gather}
    \label{eq:definitnion of the functions F appendix}
    F(\Delta \overrightarrow{\mathcal{W}}) = \prod_{\mu = 1}^{2M}\frac{e^{-\Delta\mathcal{W}^2_{\mu}/(2\Delta t)}}{\sqrt{2\pi\Delta t}}
    \prod_{m=1}^M{\rm exp}\left(\eta^*_{m,j+1}\left[i\bm{\mathcal{U}}^{\vec{s}}\sqrt{\bm{\mathcal{A}}^{\vec{s}}_{\rm diag}}\bm{\mathcal{Q}}\Delta\overrightarrow{\mathcal{W}}\right]_m - {\rm c.c.}\right), \\
    \label{eq:definitnion of the functions widetildeF appendix}
    G(\Delta \overrightarrow{\mathcal{W}},\varepsilon) = \prod_{\mu = 1}^{2M}\frac{e^{-\Delta\mathcal{W}^2_{\mu}/(2\Delta t)}}{\sqrt{2\pi\Delta t}}
    \prod_{m=1}^M{\rm exp}\left(\eta^*_{m,j+1}\left[i\bm{\mathcal{U}}^{\vec{s}}\sqrt{\bm{\mathcal{A}}^{\vec{s}}_{\rm diag}(\varepsilon)}\bm{\mathcal{Q}}\Delta\overrightarrow{\mathcal{W}}\right]_m - {\rm c.c.}\right).
\end{gather}
If these functions satisfy the conditions:
\begin{gather}
    \label{eq:Scheffe's lemma conditnion 1}
    \lim_{\varepsilon\to 0}G(\Delta \overrightarrow{\mathcal{W}},\varepsilon) = F(\Delta \overrightarrow{\mathcal{W}}), \\
    \label{eq:Scheffe's lemma conditnion 2}
    \lim_{\varepsilon\to 0}\prod_{\mu = 1}^{2M}\int_{-\infty}^{\infty}d\Delta\mathcal{W}_{\mu}\left|G(\Delta \overrightarrow{\mathcal{W}},\varepsilon)\right| = \prod_{\mu = 1}^{2M}\int_{-\infty}^{\infty}d\Delta\mathcal{W}_{\mu}\left|F(\Delta \overrightarrow{\mathcal{W}})\right| < \infty,
\end{gather}
which will be proved below, the Scheff\'e's lemma \cite{David} leads to
\begin{align}
    \label{eq:Scheffe's lemma}
    \lim_{\varepsilon\to 0}\prod_{\mu = 1}^{2M}\int_{-\infty}^{\infty}d\Delta\mathcal{W}_{\mu}G(\Delta \overrightarrow{\mathcal{W}},\varepsilon) = \prod_{\mu = 1}^{2M}\int_{-\infty}^{\infty}d\Delta\mathcal{W}_{\mu}F(\Delta \overrightarrow{\mathcal{W}}).
\end{align}
Substituting Eqs.~\eqref{eq:definitnion of the functions F appendix} and \eqref{eq:definitnion of the functions widetildeF appendix} into Eq.~\eqref{eq:Scheffe's lemma}, we can rewrite Eq.~\eqref{eq:calculation of HS transformation appendix 13} as
\begin{align}
    \label{eq:calculation of HS transformation appendix 15}
    {\rm exp}\left(-\Delta t 
    \begin{bmatrix}
        \vec{\eta}_{j+1}^{*{\rm T}},\vec{\eta}^{\rm T}_{j+1}
    \end{bmatrix}
    \bm{\mathcal{A}}^{\vec{s}}
    \begin{bmatrix}
        \vec{\eta}_{j+1}^* \\
        \vec{\eta}_{j+1}
    \end{bmatrix}
    \right)
    =
    \prod_{\mu = 1}^{2M}\int^{\infty}_{-\infty}d\Delta\mathcal{W}_{\mu}\frac{e^{-\Delta\mathcal{W}^2_{\mu}/(2\Delta t)}}{\sqrt{2\pi\Delta t}}
    \prod_{m=1}^M{\rm exp}\left(\eta^*_{m,j+1}\left[i\bm{\mathcal{U}}^{\vec{s}}\sqrt{\bm{\mathcal{A}}^{\vec{s}}_{\rm diag}}\bm{\mathcal{Q}}\Delta\overrightarrow{\mathcal{W}}\right]_m - {\rm c.c.}\right),
\end{align}
and this is the Hubbard-Stratonovich transformation Eq.~\eqref{eq:Hubbard-Stratonovich transformation} for $\bm{\mathcal{A}}^{\vec{s}} \succeq 0$.

Below, we show that $F(\Delta \overrightarrow{\mathcal{W}})$ and $G(\Delta \overrightarrow{\mathcal{W}},\varepsilon)$ actually satisfy the conditions \eqref{eq:Scheffe's lemma conditnion 1} and \eqref{eq:Scheffe's lemma conditnion 2} and completes the derivation of the Hubbard-Stratonovich transformation Eq.~\eqref{eq:Hubbard-Stratonovich transformation} for $\bm{\mathcal{A}}^{\vec{s}} \succeq 0$. 
Taking the limit of $\epsilon\to 0$ in both side of Eq.~\eqref{eq:definitnion of the functions widetildeF appendix} by using Eqs.~\eqref{eq:calculation of HS transformation appendix 12} and \eqref{eq:definitnion of the functions F appendix}, we can obtain the condition Eq.~\eqref{eq:Scheffe's lemma conditnion 1}.
In order to obtain Eq.~\eqref{eq:Scheffe's lemma conditnion 2}, we use the fact that $|G(\Delta \overrightarrow{\mathcal{W}},\varepsilon)|$ is identical to $|F(\Delta \overrightarrow{\mathcal{W}})|$ because the dependence of $\varepsilon$ in Eq.~\eqref{eq:definitnion of the functions widetildeF appendix} can be eliminated by taking the absolute value:
\begin{align}
    \label{eq:determinants of the functions widetildeF and F appendix}
    \left|G(\Delta \overrightarrow{\mathcal{W}},\varepsilon)\right| = \left|F(\Delta \overrightarrow{\mathcal{W}})\right| = \prod_{\mu = 1}^{2M}\frac{e^{-\Delta\mathcal{W}^2_{\mu}/(2\Delta t)}}{\sqrt{2\pi\Delta t}}.
\end{align}
Integrating Eq.~\eqref{eq:determinants of the functions widetildeF and F appendix} with respect to $\Delta\mathcal{W}_{\mu}$ for $\forall \mu$, we obtain
\begin{align}
    \lim_{\varepsilon\to 0}\prod_{\mu = 1}^{2M}\int_{-\infty}^{\infty}d\Delta\mathcal{W}_{\mu}\left|G(\Delta \overrightarrow{\mathcal{W}},\varepsilon)\right| &= \prod_{\mu = 1}^{2M}\int_{-\infty}^{\infty}d\Delta\mathcal{W}_{\mu}\left|F(\Delta \overrightarrow{\mathcal{W}})\right| = \prod_{\mu = 1}^{2M}\int_{-\infty}^{\infty}d\Delta\mathcal{W}_{\mu}\frac{e^{-\Delta\mathcal{W}^2_{\mu}/(2\Delta t)}}{\sqrt{2\pi\Delta t}} = 1 < \infty,
\end{align}
which completes the derivation of Eq.~\eqref{eq:Scheffe's lemma conditnion 2}.


\section{\label{appendix:Non-equal two-time correlation function in the phase space}Non-equal two-time correlation function in the phase space{\rm :} Derivation of Eq.~\texorpdfstring{\eqref{eq:two time correlation function in the phase space}}{TEXT}}

Before deriving Eq.~\eqref{eq:two time correlation function in the phase space}, we derive the phase-space representation of a product of two operators $[\hat{A}\hat{B}]_{\vec{s}}(\vec{\alpha},\vec{\alpha}^*)$.
Using Eq.~\eqref{eq:relation between A0 and As}, we obtain
\begin{align}
    \label{eq:phase-space rep of AB 1}
    [\hat{A}\hat{B}]_{\vec{s}}(\vec{\alpha},\vec{\alpha}^*) = {\rm exp}\left(\sum_{m=1}^{M}\frac{s_m}{2}\frac{\partial^2}{\partial\alpha_m\partial\alpha_m^*}\right)[\hat{A}\hat{B}]_{\vec{0}}(\vec{\alpha},\vec{\alpha}^*),
\end{align}
where $[\hat{A}\hat{B}]_{\vec{0}}(\vec{\alpha},\vec{\alpha}^*)$ is given by the well-known formula \cite{Hillery}:
\begin{align}
    \label{eq:phase-space rep of AB 2}
    [\hat{A}\hat{B}]_{\vec{0}}(\vec{\alpha},\vec{\alpha}^*) = A_{\vec{0}}(\vec{\alpha},\vec{\alpha}^*)\star B_{\vec{0}}(\vec{\alpha},\vec{\alpha}^*)
\end{align}
with $\star$ being the Moyal product defined by Eq.~\eqref{eq:definition of the Moyal product}.
Substituting Eq.~\eqref{eq:phase-space rep of AB 2} into Eq.~\eqref{eq:phase-space rep of AB 1} and using Eq.~\eqref{eq:relation between A stare B and A stars B}, we can rewrite Eq.~\eqref{eq:phase-space rep of AB 1} as
\begin{align}
    [\hat{A}\hat{B}]_{\vec{s}}(\vec{\alpha},\vec{\alpha}^*) = A_{\vec{s}}(\vec{\alpha},\vec{\alpha}^*)\star_{\vec{s}} B_{\vec{s}}(\vec{\alpha},\vec{\alpha}^*).
\end{align}

Following the same procedure to obtain Eq.~\eqref{eq:Kraus representation in the phase space} (see \ref{appendix:The propagator in the Kraus representation}), we can rewrite Eq.~\eqref{eq:frist line of two time correlation function in the phase space} as
\begin{align}
    \braket{\hat{A}(t)\hat{B}(t_0)} = \int\frac{d^2\vec{\alpha}_{\rm f}d^2\vec{\alpha}_0}{\pi^{2M}}A_{\vec{s}}(\vec{\alpha}_{\rm f},\vec{\alpha}^*_{\rm f})\varUpsilon_{\vec{s}}(\vec{\alpha}_{\rm f},t;\vec{\alpha}_0,t_0)[\hat{B}\hat{\rho}(t_0)]_{-\vec{s}}(\vec{\alpha}_0,\vec{\alpha}_0^*).
\end{align}
Substituting $[\hat{B}\hat{\rho}(t_0)]_{-\vec{s}}(\vec{\alpha}_0,\vec{\alpha}_0^*) = B_{-\vec{s}}(\vec{\alpha}_0,\vec{\alpha}^*_0)\star_{-\vec{s}} W_{\vec{s}}(\vec{\alpha}_0,\vec{\alpha}^*_0,t_0)$ into the above equation, we obtain
\begin{align}
    \braket{\hat{A}(t)\hat{B}(t_0)} &= \int\frac{d^2\vec{\alpha}_{\rm f}d^2\vec{\alpha}_0}{\pi^{2M}}A_{\vec{s}}(\vec{\alpha}_{\rm f},\vec{\alpha}^*_{\rm f})\varUpsilon_{\vec{s}}(\vec{\alpha}_{\rm f},t;\vec{\alpha}_0,t_0)\left[B_{-\vec{s}}(\vec{\alpha}_0,\vec{\alpha}^*_0)\star_{-\vec{s}} W_{\vec{s}}(\vec{\alpha}_0,\vec{\alpha}^*_0,t_0)\right].
\end{align}
This completes the derivation of Eq.~\eqref{eq:two time correlation function in the phase space}.
















\bibliographystyle{elsarticle-num_new}
\bibliography{elsarticle-num_new}




\end{document}